\pdfoutput=1

\documentclass[11pt]{article}
\usepackage{jheppub}

\usepackage{amsmath}
\usepackage{amssymb}
\usepackage{graphicx,color,slashed}
\usepackage{ifpdf}
\usepackage[english]{babel}
\usepackage{url}
\usepackage{booktabs}

\usepackage{float}

\hypersetup{bookmarksnumbered}

\allowdisplaybreaks[1]

\DeclareFontFamily{U}{mathx}{\hyphenchar\font45}
\DeclareFontShape{U}{mathx}{m}{n}{
      <5> <6> <7> <8> <9> <10>
      <10.95> <12> <14.4> <17.28> <20.74> <24.88>
      mathx10
      }{}
\DeclareSymbolFont{mathx}{U}{mathx}{m}{n}
\DeclareFontSubstitution{U}{mathx}{m}{n}
\DeclareMathAccent{\widecheck}{0}{mathx}{"71}

\restylefloat{figure}
\definecolor{dgreen}{rgb}{0,0.70,0.30}
\definecolor{gold}{rgb}{0.85,.66,0}
\definecolor{purple}{rgb}{1.0,0.3,0.6}

\usepackage{tikz}
\usetikzlibrary{calc} \usetikzlibrary{patterns} \usetikzlibrary{decorations.pathreplacing} \usetikzlibrary{decorations.markings} \usetikzlibrary{decorations.pathmorphing} \usetikzlibrary{positioning} \usetikzlibrary{bending}

\def\be{\begin{equation}}
\def\ee{\end{equation}}
\def\ba{\begin{array}}
\def\ea{\end{array}}

\newcommand{\bea}{\begin{eqnarray}}
\newcommand{\eea}{\end{eqnarray}}

\def\beq{\begin{equation}}
\def\eeq{\end{equation}}
\let\Re\relax
\let\Im\relax
\DeclareMathOperator{\Re}{Re}
\DeclareMathOperator{\Im}{Im}

\newcommand{\dd}{\mathrm{d}}
\newcommand{\te}{\textrm}
\newcommand{\ap}{{\alpha'}}

\DeclareMathOperator*{\esumsym}{\hspace{0.5em}\sum\raisebox{-0.5em}{\makebox[0.2em]{$\scriptstyle \mathrm{E}$}}\hspace{0.3em}}
\newcommand{\esum}{\hspace{-0.5em}\esumsym}

\newcommand{\cform}[1]{\,{\cal C}[\protect\begin{smallmatrix}#1\protect\end{smallmatrix}]}
\newcommand{\cformtri}[3]{\,{\cal C}\!\left[\protect\begin{smallmatrix}#1\protect\end{smallmatrix}\middle|\protect\begin{smallmatrix}#2\protect\end{smallmatrix}\middle|\protect\begin{smallmatrix}#3\protect\end{smallmatrix}\right]}

\newcommand{\cregform}[1]{\,{\cal C}_{\rm reg}[\protect\begin{smallmatrix}#1\protect\end{smallmatrix}]}

\newcommand{\RR}{\mathbb R}
\newcommand{\CC}{\mathbb C}
\newcommand{\NN}{\mathbb N}
\newcommand{\ZZ}{\mathbb Z}

\usepackage[format=plain,
            labelfont=bf,
            textfont=it]{caption}

\title{Heterotic-string amplitudes at one loop: modular graph forms
and relations to open strings}
\author[a]{Jan E. Gerken,}
\author[a,b]{Axel Kleinschmidt,}
\author[a,c]{Oliver Schlotterer}

\affiliation[a]{Max--Planck--Institut f\"ur Gravitationsphysik,
Albert--Einstein--Institut,
14476 Potsdam, Germany}
\affiliation[b]{International Solvay Institutes ULB-Campus Plaine CP231, BE-1050 Brussels, Belgium}
\affiliation[c]{Perimeter Institute for Theoretical Physics,
Waterloo, ON N2L 2Y5, Canada}

\emailAdd{jan.gerken@aei.mpg.de}
\emailAdd{axel.kleinschmidt@aei.mpg.de}
\emailAdd{olivers@aei.mpg.de}

\date{\today}

\abstract{We investigate one-loop four-point scattering of non-abelian gauge bosons
in heterotic string theory and identify new connections with the corresponding open-string 
amplitude. In the low-energy expansion of the heterotic-string amplitude,
the integrals over torus punctures are systematically evaluated in terms of 
modular graph forms, certain non-holomorphic modular forms. 
For a specific torus integral, the modular graph forms in the low-energy expansion
are related to the elliptic multiple zeta values from the analogous open-string integrations over 
cylinder boundaries. The detailed correspondence between these modular graph forms
and elliptic multiple zeta values supports a recent proposal for an elliptic generalization of the 
single-valued map at genus zero.}

\begin{document}
\maketitle{}

\setcounter{tocdepth}{2}

\numberwithin{equation}{section}



\section{Introduction}
\label{sec:1}

Scattering amplitudes in string theories have become a rewarding laboratory to encounter
modern number-theoretic concepts in a simple setup. String amplitudes are
derived from integrating over moduli spaces of punctured Riemann surfaces
whose genus matches the loop order in perturbation theory. Accordingly, the
low-energy expansion of string amplitudes provides generating functions for 
the periods of the relevant moduli spaces -- multiple zeta values at tree level
\cite{Terasoma, Brown:2009qja, Schlotterer:2012ny, Broedel:2013aza} and 
various elliptic generalizations at loop level.

At genus one, a variety of mathematical structures have recently been revealed in maximally
supersymmetric open- and closed-string amplitudes. For one-loop amplitudes of the open
superstring, elliptic multiple zeta values (eMZVs) \cite{Enriquez:Emzv} were identified as the natural language to
capture the expansion in the inverse string tension $\ap$ \cite{Broedel:2014vla, Broedel:2017jdo}. Closed-string 
amplitudes of type-II superstrings in turn introduce an intriguing system of non-holomorphic modular functions
known as modular graph functions \cite{DHoker:2015wxz}. 

One-loop closed-string worldsheets are toroidal with complex modulus $\tau$ on which the modular group acts. Modular graph functions are invariant under this action and depend on a graph on the worldsheet describing the contractions of the punctures related to external states. The properties and relations of modular graph functions have been
studied from various perspectives \cite{Green:1999pv, Green:2008uj, DHoker:2015foa, DHoker:2015sve,
Basu:2015ayg, Zerbini:2015rss, DHoker:2016mwo, Basu:2016xrt, Basu:2016kli, Basu:2016mmk, 
DHoker:2016quv, Kleinschmidt:2017ege, DHoker:2017zhq, Broedel:2018izr}. For open strings at genus one, eMZVs result from integration over the punctures on the boundary of a cylindrical open-string worldsheet at one-loop level, where $\tau$ is treated as the modulus of the cylinder.

This work is dedicated to one-loop amplitudes of the heterotic string, where the moduli-space
integrals are constrained by only half of the supersymmetries as compared to the type-II setup.
Accordingly, the $\ap$-expansion of heterotic-string amplitudes admits a larger class of integrals 
over torus punctures which we will evaluate in terms of modular graph \textit{forms} in representative
four-point examples. Modular graph forms have been introduced as generalizations of modular 
graph functions \cite{DHoker:2016mwo} by allowing a non-trivial transformation under the modular 
group action that is expressed through non-vanishing modular weights $(w,\bar{w})$. They can 
often be related to modular invariant functions by differential equations in $\tau$.

In this article, we will simplify the four-point amplitude among gauge bosons in a way such that the 
integrals over the punctures at any order of the $\ap$-expansions manifestly evaluate to modular graph forms. 
Explicit results are given for both the single-trace and the double-trace sector up to the third subleading 
order in $\ap$. Moreover, we give general arguments that modular graph forms capture
the integrations over the torus punctures in the $n$-point heterotic-string amplitude involving 
any combination of gauge bosons and gravitons. Earlier work on the 
correlation functions and integrations over the punctures in multi-particle amplitudes of the heterotic 
string includes \cite{Lerche:1987qk, Stieberger:2002wk, Dolan:2007eh}. Furthermore, the subleading 
order in the $\ap$-expansion of four-point amplitudes involving gravitons has been recently studied 
in~\cite{Basu:2017nhs, Basu:2017zvt}.

A major motivation for this work concerns the connection between open- and closed-string amplitudes at the 
level of their low-energy expansion. At genus zero, the multiple zeta values (MZVs) from open- and closed-string
scattering on a disk and a sphere, respectively, are related by the single-valued map \cite{Schlotterer:2012ny, Stieberger:2013wea, Stieberger:2014hba, BrownDupont, Schlotterer:2018zce, Brown:2018omk}\footnote{The conjectures
of \cite{Schlotterer:2012ny, Stieberger:2013wea, Stieberger:2014hba} have been proven recently in~\cite{BrownDupont, Brown:2018omk}; see also~\cite{Schlotterer:2018zce} for a derivation
where certain transcendentality conjectures on MZVs are assumed.}.
The latter refers to the origin of MZVs from polylogarithms, where the defining property of single-valued MZVs is
their descent from single-valued polylogarithms \cite{Schnetz:2013hqa, Brown:2013gia}. The notion of a 
single-valued map applies to a variety of periods \cite{Francislecture}, and it is natural to expect loop-level relations between 
open- and closed strings on these grounds. An ambitious long-term goal is to completely sidestep the moduli-space 
integrations of closed-string amplitudes at various genera and to infer their results from suitable maps acting on 
open-string quantities.

In particular, by comparing the eMZVs and modular graph functions in one-loop amplitudes of open and closed
strings, a conjecture for the explicit form of an elliptic single-valued map has been made in \cite{Broedel:2018izr}. This conjecture
is based on a graphical organization of the open-string $\ap$-expansion, but it has been limited to the scattering of
abelian gauge bosons on the open-string side and modular invariant functions of $\tau$ on the closed-string side. 
As one of our main results, we extend the proposal of \cite{Broedel:2018izr} to non-abelian open-string 
states, where the conjectural elliptic single-valued map reproduces substantial parts of modular graph 
forms of modular weight $(2,0)$.
The integration cycles for four open-string punctures on a cylinder boundary are proposed to relate to certain
elliptic functions of the closed-string punctures through a Betti--deRham duality \cite{betti1, betti2}.

In the mathematics literature, a general construction of single-valued eMZVs has been given by 
Brown \cite{Brown:2017qwo, Brown:2017qwo2}. So far, it remains conjectural that modular 
graph functions or modular graph forms are contained in the image of this elliptic single-valued map. 
The real-analytic functions in Brown's construction \cite{Brown:2017qwo, Brown:2017qwo2} include 
modular forms of various holomorphic and antiholomorphic weights. Hence, our extension of the string-theory 
motivated conjecture for an elliptic single-valued map to modular graph forms should be helpful to 
identify the missing link to the setup of~\cite{Brown:2017qwo, Brown:2017qwo2}.
Given that the heterotic string relaxes the maximal supersymmetry of type--II superstrings, its
extended set of moduli-space integrals (as investigated here at four points)
is hoped to give a more general picture of an elliptic single-valued map.

As another consequence of the half-maximal supersymmetry of the heterotic string,
the coefficients of a given modular graph form in massless one-loop\footnote{See 
\cite{Huang:2016tag, Azevedo:2018dgo} for the
analogous phenomenon in tree-level amplitudes of the heterotic string, where the coefficients
of a given multiple zeta value are usually geometric series in $\ap$.} 
amplitudes usually mix different orders in $\ap$. 
In superstring amplitudes in turn, the order in the $\ap$-expansion correlates with the 
transcendental weights of the accompanying iterated integrals -- (elliptic) multiple zeta values 
or modular graph functions \cite{Green:2008uj, Schlotterer:2012ny, Broedel:2013aza, Broedel:2014vla}. 
This property known as {\it uniform transcendentality} can also be found in the context of dimensionally 
regularized Feynman integrals with the regularization parameter $\varepsilon$
taking the r$\hat{\rm o}$le of $\alpha'$ \cite{Kotikov:1990kg, ArkaniHamed:2010gh, Henn:2013pwa, 
Adams:2018yfj, Broedel:2018qkq}.
We decompose the four-point gauge amplitude of the heterotic string into integrals 
that are individually believed to be uniformly transcendental -- both in the single-trace and the 
double-trace sector. The non-uniform transcendentality of the overall amplitude is
then reflected by the coefficients of these basis integrals. The classification of uniformly transcendental 
moduli-space integrals is expected to give important clues about the mathematical properties of the
underlying twisted cohomologies \cite{Mizera:2017cqs}.

In summary, the main results of this work are the following:
\begin{itemize}
\item a general argument and explicit four-point examples that modular graph forms
capture the integrals over the punctures in massless one-loop amplitudes of the heterotic string 
\item relating an integral of modular weight $(2,0)$ in the four-point gauge amplitude of
the heterotic string to a cylinder integral of the open superstring by applying the tentative
elliptic single-valued map of \cite{Broedel:2018izr} order by order in the $\ap$-expansion
\item an explicit decomposition of the four-point one-loop gauge amplitude of the heterotic string
into integrals of conjecturally uniform transcendentality
\end{itemize}
This paper is structured as follows. We first review the worldsheet building blocks for constructing one-loop heterotic-string amplitudes in section~\ref{sec:2}. Then we develop the notion of modular graph forms and review their salient properties in section~\ref{sec:8}. In section~\ref{sec:3}, we bring the two concepts together and show how the integrands of one-loop amplitudes of the heterotic string can be expressed as modular graph forms. In particular, we explicitly perform the analysis of the leading planar and non-planar contributions to four-point gauge amplitudes in the $\alpha'$-expansion. We also reorganize the integrals in terms of representatives of uniform transcendentality, determine the integrated amplitude to second order in $\ap$ and comment on general $n$-point amplitudes and their relation to modular graph forms. In section~\ref{sec:4}, we investigate the relation between open-string amplitudes and heterotic strings and develop the conjectural notion of single-valued map relevant in this context. Section~\ref{sec:concl} contains concluding remarks.  Several technical details needed in the analysis have been relegated to a number of appendices.


\section{Basics of heterotic-string amplitudes}
\label{sec:2}


In this section, we introduce the basic building blocks for correlation functions in heterotic-string 
amplitudes at one loop and how they combine to produce modular forms after integration over the punctures corresponding to the external states.

\subsection{Kronecker--Eisenstein series and elliptic functions}
\label{sec:2.1}

\begin{figure}
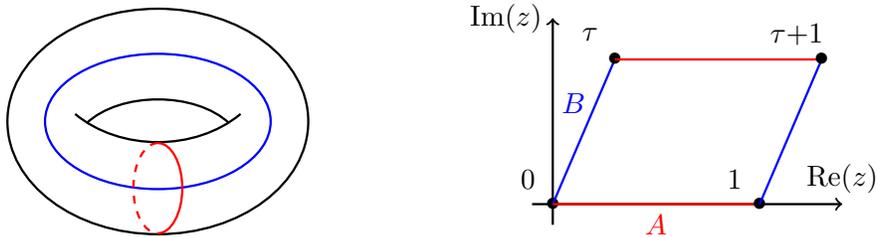

\begin{center}
\tikzpicture[scale=0.5,line width=0.30mm]
\draw(0,0) ellipse  (4cm and 3cm);
\draw(-2.2,0.2) .. controls (-1,-0.8) and (1,-0.8) .. (2.2,0.2);
\draw(-1.9,-0.05) .. controls (-1,0.8) and (1,0.8) .. (1.9,-0.05);
\draw[blue](0,0) ellipse  (3cm and 1.8cm);
\draw[red] (0,-2.975) arc (-90:90:0.65cm and 1.2cm);
\draw[red,dashed] (0,-0.575) arc (90:270:0.65cm and 1.2cm);
\scope[xshift=10.5cm,yshift=-2.2cm,scale=1.1]
\draw[->](-0.5,0) -- (7,0) node[above]{${\rm Re}(z)$};
\draw[->](0,-0.5) -- (0,4.5) node[left]{${\rm Im}(z)$};
\draw(0,0)node{$\bullet$};
\draw(-0.6,0.6)node{$0$};
\draw[blue](0,0) -- node[left,pos=0.7]{$\textcolor{blue}{B}$}(1.5,3.5);
\draw (1.5,3.5)node{$\bullet$} ;
\draw(0.9,4.1)node{$\tau$};
\draw[red](0,0) -- node[below]{$\textcolor{red}{A}$}(5,0);
\draw (5,0)node{$\bullet$};
\draw (4.4,0.6)node{$1$};
\draw[red](1.5,3.5) -- (6.5,3.5);
\draw[blue](5,0) -- (6.5,3.5);
\draw(6.5,3.5)node{$\bullet$};
\draw(5.9,4.1)node{$\tau{+}1$};
\endscope
\endtikzpicture
\caption{Throughout this work, toroidal genus-one worldsheets are parametrized through the
depicted parallelogram with discrete identifications $z \cong z{+}1$ and $z \cong z{+}\tau$ corresponding
to the $A$-cycle and $B$-cycle, respectively.}
\label{figureone}
\end{center}
\end{figure}

The relevant heterotic-string correlation functions are defined to live on a worldsheet of torus topology 
that will be parametrized by the parallelogram in figure \ref{figureone}. The homology cycles of the torus are mapped
to the periodicities $z \cong z+1$ and $z\cong z+\tau$ for the torus coordinate $z\in \CC$, and the
modular parameter $\tau\in \CC$ is taken to be in the upper half plane $\Im \tau >0$.
 
In this setup, a universal starting point for constructing elliptic functions and iterated integrals is the
following definition of the Kronecker--Eisenstein series \cite{Kronecker, BrownLev}
\begin{equation}
F(z,\beta,\tau) := \frac{ \theta_1'(0,\tau) \theta_1(z+\beta,\tau) }{\theta_1(z,\tau)  \theta_1(\beta,\tau)}
\label{1.2}
\end{equation}
involving the odd Jacobi theta function
\begin{align}
\theta_1(z,\tau) := 2 q^{1/8} \sin(\pi z) \prod_{n=1}^{\infty} (1-q^n) (1-e^{2\pi i z} q^n) (1-e^{-2\pi i z} q^n) 
\label{1.0}
\end{align}
and its $z$-derivative. The formal parameter $\beta$ appearing in definition~\eqref{1.2} will be used below
to define an infinite family of doubly-periodic functions. The Kronecker--Eisenstein series itself is quasi-periodic with~\cite{ZagierF}  
\begin{align}
\label{KEmod}
F(z{+}1,\beta,\tau)= F(z,\beta,\tau)\,,\quad  F(z{+}\tau,\beta,\tau)= e^{-2\pi i \beta}F(z,\beta,\tau)\,,
\end{align}
but its monodromies around the $B$-cycle cancel from the following doubly-periodic completion
\begin{equation}
\Omega(z,\beta,\tau) := \exp\Big( 
2\pi i \beta \, \frac{ \Im \, z}{\Im \, \tau} \Big)
\, F(z,\beta,\tau) \ .
\label{1.1}
\end{equation}
When viewing $\beta$ as a bookkeeping variable, a Laurent expansion of (\ref{1.1})
introduces doubly-periodic but non-meromorphic functions $f^{(w)}$ \cite{Broedel:2014vla}
\begin{equation}
\Omega(z,\beta,\tau) =: \sum_{w=0}^{\infty} \beta^{w-1} f^{(w)}(z,\tau)\ ,
\label{1.1b}
\end{equation}
starting with $f^{(0)}(z,\tau)=1$ and  $f^{(1)}(z,\tau) = \partial_z \log \theta_1(z,\tau) + 2\pi i \frac{ \Im  z}{\Im  \tau}$.
From the Kronecker--Eisenstein series (\ref{1.2}),
one can see that the only singularity among the $f^{(w)}$ functions is the simple 
pole $f^{(1)}(z,\tau) = \frac{1}{z}+{\cal O}(z,\bar z)$ at the origin (and its translates by $\ZZ + \tau \ZZ$).

The non-holomorphic exponentials of (\ref{1.1}) drop out if several Kronecker--Eisenstein series are combined
to cyclic products with first arguments $z_{ij}=z_i-z_j$: A formal expansion in the bookkeeping variable $\beta$
defines a series of elliptic functions $V_w$ \cite{Dolan:2007eh}\footnote{Since (\ref{1.1a}) is at the same time an elliptic function of
$\beta$, the coefficients $V_w(1,\ldots,n), \ w <n$ of the singular terms in $\beta$ determine the coefficients
$V_w(1,\ldots,n), \ w >n$ of the regular terms $\beta^{\geq 1}$ \cite{Dolan:2007eh}.} 
\begin{align}
&F(z_{12},\beta,\tau) F(z_{23},\beta,\tau) \ldots F(z_{n1},\beta,\tau) = \Omega(z_{12},\beta,\tau) \Omega(z_{23},\beta,\tau) \ldots \Omega(z_{n1},\beta,\tau)
\notag \\
&=: \beta^{-n} \sum_{w=0}^\infty \beta^w V_w(1,2,\ldots,n)  \ . \label{1.1a}
\end{align}
Here and in later places of this work, the dependence of $V_w(1,2,\ldots,n)=V_w(1,2,\ldots,n|\tau)$ 
on the modular parameter is left implicit. The definition (\ref{1.1a}) and
$F(-z,-\beta,\tau) = - F(z,\beta,\tau)$ immediately manifest the following dihedral symmetry properties
\begin{equation}
V_w(1,2,\ldots,n) = V_w(2,\ldots,n,1) \, , \ \ \ \ \ \
V_w(n,n{-}1,\ldots,2,1) = (-1)^w V_w(1,2,\ldots,n)\label{1.1symm}
\end{equation}
under cyclic shifts and reversal. The $V_w$ functions can be conveniently expressed in terms of the
doubly-periodic coefficients $f^{(w)}$ in (\ref{1.1b}), for instance (with cyclic identification $z_{n+1}=z_1$)
\begin{align}
V_0(1,2,\ldots,n) &=1 \ , \ \ \ \ \ \  V_1(1,2,\ldots,n) = \sum_{j=1}^n f^{(1)}(z_{j}{-}z_{j+1},\tau) \label{1.1c} \\
V_2(1,2,\ldots,n) &= \sum_{j=1}^n f^{(2)}(z_{j}{-}z_{j+1},\tau)
+\sum_{i=1}^n \sum_{j=i{+}1}^n f^{(1)}(z_{i}{-}z_{i+1},\tau)f^{(1)}(z_{j}{-}z_{j+1},\tau) \ .
\notag
\end{align}
The modular properties of the Kronecker--Eisenstein series~\eqref{KEmod}
\cite{ZagierF} give rise to the ${\rm SL}_2(\ZZ)$ transformations
\begin{align}
f^{(w)}\Big( \frac{ z}{\gamma\tau + \delta} , \frac{\alpha \tau + \beta }{\gamma\tau + \delta} \Big) &= (\gamma\tau + \delta)^{w} f^{(w)}(z,\tau) \label{1.1mod} \\
V_w(1,2,\ldots,n) \, \Big|^{z_j \rightarrow \frac{ z_j }{\gamma\tau + \delta}}_{\; \tau \rightarrow \frac{\alpha \tau + \beta }{\gamma\tau + \delta}}
&=(\gamma\tau +\delta)^w V_w(1,2,\ldots,n) \, , \notag
\end{align}
where $\left(\begin{smallmatrix}  \alpha&\beta\\\gamma&\delta\end{smallmatrix}\right)\in{\rm SL}_2(\ZZ)$. The purely holomorphic modular weight $(w,0)$ of the $V_w$ functions will be later on 
seen to resonate with modular invariance of the heterotic string.


\subsection{One-loop gauge amplitudes of the heterotic string}
\label{sec:2.2}

In this work, we will be interested in one-loop scattering of gauge bosons in the heterotic string.
Up to an overall normalization factor, the prescription for the four-point function 
reads \cite{Gross:1985rr, Sakai:1986bi, Ellis:1987dc, Abe:1988cq} 
\begin{align}
{\cal M}_{4}= \int_{{\cal F}} \frac{ \dd^2 \tau }{(\Im \tau)^5} \, \frac{1}{\eta^{24}(\tau)}  
\int_{T(\tau)} \dd^2 z_2
\int_{T(\tau)} \dd^2 z_3
\int_{T(\tau)} \dd^2 z_4\
\langle \prod_{j=1}^4 {\cal V}^{a_j}(z_j,\epsilon_j,k_j) \rangle^\tau\, ,
\label{presc1}
\end{align}
where ${\cal F}$ denotes the fundamental domain of the modular group ${\rm SL}_2(\mathbb Z)$,
the integration domain $T(\tau)$ for $z_2,z_3,z_4$ is the torus in the parametrization 
of figure \ref{figureone}, and translation invariance has been used to fix $z_1=0$. Moreover, the inverse factors
of the Dedekind eta function
\begin{align}
\eta(\tau) := q^{1/24} \prod_{n=1}^{\infty}(1-q^n) \, , \ \ \ \ \ \ q:=e^{2\pi i \tau}
\label{presc2}
\end{align}
arise as the partition function of the 26 worldsheet bosons in the non-supersymmetric sector, 
and ${\cal V}^{a}(z,\epsilon,k)$ denotes the vertex operator for an external gauge boson \cite{Gross:1985rr}
\begin{align}
{\cal V}^{a}(z,\epsilon,k) = J^a(z) V_{\rm SUSY}(\bar z,\epsilon,k) e^{ik\cdot X(z,\bar z)} 
\label{presc3}
\end{align}
with polarization vector $\epsilon$, lightlike momentum $k$ and adjoint index $a$.
The correlation function $\langle \ldots \rangle^{\tau}$ in (\ref{presc1}) is
evaluated on a torus of modular parameter $\tau$ and allows factoring out the contribution
from the Kac--Moody currents $J^a(z)$. The leftover correlator involving $V_{\rm SUSY}(\bar z,\epsilon,k)$
and $e^{ik\cdot X(z,\bar z)} $ matches a chiral half of type-II superstrings, and its four-point instance is
completely determined by maximal supersymmetry \cite{Green:1982sw}, 
\begin{align}
\langle \prod_{j=1}^4  V_{\rm SUSY}(\bar z_j,\epsilon_j,k_j) e^{ik_j \cdot X(z_j,\bar z_j)} \rangle^\tau = (k_1\cdot k_2)(k_2\cdot k_3) A_{\rm SYM}^{\rm tree}(1,2,3,4) \langle \prod_{j=1}^4 e^{ik_j \cdot X(z_j,\bar z_j)} \rangle^\tau\ .
\label{presc4}
\end{align}
The permutation invariant combination of
polarizations $\epsilon$ and momenta $k$ on the right-hand side has been expressed through
a color-ordered tree-level amplitude $A_{\rm SYM}^{\rm tree}(1,2,3,4)$ of ten-dimensional super-Yang--Mills. The manifestly 
supersymmetric calculation in the pure-spinor formalism \cite{Berkovits:2004px}
leads to the same conclusion for any combination of gauge bosons and gauginos.
The left-over correlator over the plane waves
\begin{align}
\Big\langle  
\prod_{j=1}^n e^{ik_j\cdot X(z_j,\bar z_j)}
\Big \rangle^{\tau} &= \exp \Big( \sum_{1\leq i<j}^n s_{ij} G_{ij}(\tau) \Big) 
\label{mgf0}
\end{align}
will be referred to as the Koba--Nielsen factor and
involves the bosonic Green function $G_{ij}(\tau) :=G(z_i{-}z_j,\tau) $ as well as 
dimensionless Mandelstam invariants in the exponent,
\begin{align}
G(z,\tau) := - \log \left| \frac{ \theta_1(z,\tau)}{\eta(\tau)} \right|^2 - \frac{ \pi (z{-}\bar z)^2}{2 \, \Im \tau}  \, , \ \ \ \ \ \ 
s_{ij} := -\frac{\ap}{2} k_i \cdot k_j \, .
\label{mgf2}
\end{align}
Note that this representation of the Green function relates it to the function $f^{(1)}(z,\tau)$ defined in \eqref{1.1b} via
\begin{align}
  f^{(1)}(z,\tau)=-\partial_{z}G(z,\tau)\ .\label{eq:27}
\end{align}
We shall now focus on the correlation function of the Kac--Moody currents $J^a$, see 
(\ref{presc1}) and (\ref{presc3}), that carries all the dependence on the adjoint indices $a_1,\ldots,a_4$
of the external gauge bosons. In a fermionic representation 
$J^a(z) = t^a_{ij} \psi^i \psi^j(z)$ of the currents,
the correlators receive contributions from different spin structures -- the
boundary conditions for the worldsheet spinors under $\psi^j(z+1)=\pm \psi^j(z)$
and $\psi^j(z+\tau)=\pm \psi^j(z)$. We will be mostly interested in the gauge group
 $\mathrm{Spin}(32)/\ZZ_2$ with Lie-algebra generators  $t^a_{ij}$ and fundamental 
 indices $i,j=1,2,\ldots,32$.

For four-point functions, the only contributions come from the even spin structures that we label with an integer $\nu=2,3,4$, and the corresponding fermionic two-point function or ``Szeg\"o kernel'' \cite{DHoker:1988pdl} can be brought into the universal form
\begin{equation}
S_\nu(z,\tau) := \frac{ \theta_1'(0,\tau) \theta_\nu(z,\tau) }{\theta_\nu(0,\tau) \theta_1(z,\tau) }\, ,
\label{1.01}
\end{equation}
where the even theta functions analogous to (\ref{1.0}) read
\begin{align}
\theta_2(z,\tau) &:= 2q^{1/8} \cos(\pi z)\prod_{n=1}^{\infty} (1-q^n) (1+e^{2\pi i z} q^n) (1+e^{-2\pi i z} q^n)  \notag \\
\theta_3(z,\tau) &:=  \prod_{n=1}^{\infty} (1-q^n) (1+e^{2\pi i z} q^{n-1/2}) (1+e^{-2\pi i z} q^{n-1/2}) \label{1.01ev}
\\
\theta_4(z,\tau) &:=  \prod_{n=1}^{\infty} (1-q^n) (1-e^{2\pi i z} q^{n-1/2}) (1-e^{-2\pi i z} q^{n-1/2}) \, .  \notag
\end{align}
In the fermionic realization of Kac--Moody currents, the contribution of a given spin structure to
the correlation function reduces to (sums of products of) Szeg\"o kernels (\ref{1.01}), e.g.\ \cite{Dolan:2007eh}
\begin{align}
\langle J^{a_1}(z_1) J^{a_2}(z_2) \rangle^{\tau}_\nu &= {\rm Tr}(t^{a_1} t^{a_2}) S_\nu(z_{12},\tau) S_\nu(z_{21},\tau)
\label{1.02} \\
\langle J^{a_1}(z_1) J^{a_2}(z_2) J^{a_3}(z_3) \rangle^{\tau}_\nu &= \  \buildrel{\leftrightarrow}\over{{\rm Tr}} \hspace{-0.11cm} (t^{a_1} t^{a_2} t^{a_3} ) S_\nu(z_{12},\tau) S_\nu(z_{23},\tau) S_\nu(z_{31},\tau)
\label{1.03} \\
\langle J^{a_1}(z_1) J^{a_2}(z_2) J^{a_3}(z_3) J^{a_4}(z_4) \rangle^{\tau}_\nu &= \  \buildrel{\leftrightarrow}\over{{\rm Tr}} \hspace{-0.11cm} (t^{a_1} t^{a_2} t^{a_3} t^{a_4}  ) S_\nu(z_{12},\tau) S_\nu(z_{23},\tau) S_\nu(z_{34},\tau) S_\nu(z_{41},\tau) \notag \\
& \! \! \!  \! \! \!  \! \! \!  \! \! \! \! \! \! \! \! \!\! \! \! \!\! \! \! \! \! \! \! \! \! \! \! \! \!\! \! \! \!\! \! \! \!  \! \! \! \! \! \! \!\! \! \! \!\! \! \! \!  \!
+ {\rm Tr}(t^{a_1} t^{a_2})  {\rm Tr}(t^{a_3} t^{a_4}) S_\nu(z_{12},\tau) S_\nu(z_{21},\tau) S_\nu(z_{34},\tau) S_\nu(z_{43},\tau) 
 + {\rm cyc}(2,3,4)\, , \label{1.04} 
\end{align}
where we use the following shorthand for parity-weighted traces
relevant for $n\geq 3$ currents,
\begin{align}
\buildrel{\leftrightarrow}\over{{\rm Tr}} \hspace{-0.11cm}(t^{a_1}t^{a_2} \ldots  t^{a_{n}})
:= {\rm Tr}(t^{a_1}t^{a_2} \ldots  t^{a_{n}}) + (-1)^n{\rm Tr}(t^{a_n} \ldots t^{a_2} t^{a_{1}}) \, .
\label{trarrow}
\end{align}
The sum over cyclic permutations refers to both lines of (\ref{1.04}), and it acts on
both the adjoint indices $a_2,a_3,a_4$ and the punctures $z_2,z_3,z_4$.
Furthermore, each of the spin-structure dependent current correlators is weighted by
the fermionic partition function of the $\mathrm{Spin}(32)/\ZZ_2$ model,
\begin{equation}
Z^{\rm het}_{\nu}(\tau) := 2\zeta_4^2 \theta_\nu^{16}(0,\tau)\, .
\label{1.00}
\end{equation}
Since we do not track the overall normalization of the amplitude in (\ref{presc1}), 
the prefactor $2\zeta_4^2$ is introduced along the way for later convenience. 
The end results for the current correlators in (\ref{presc1})
are proportional to the spin-summed expressions,
\begin{equation}
\langle J^{a_1}(z_1) J^{a_2}(z_2) \ldots J^{a_n}(z_n) \rangle^\tau = \sum_{\nu=2}^4 
Z^{\rm het}_{\nu}(\tau) \langle J^{a_1}(z_1) J^{a_2}(z_2) \ldots J^{a_n}(z_n) \rangle_\nu^\tau \ ,
\label{1.06} 
\end{equation}
and we will next construct convenient representations of (\ref{1.06}) from the elliptic functions of section \ref{sec:2.1}.


\subsection{Spin-summed current correlators}
\label{sec:2.20}

By the form of the spin-structure dependent correlators (\ref{1.02}) and (\ref{1.04}), 
we will decompose the spin sums (\ref{1.06}) according to the traces of Lie-algebra generators. For $n$ gauge currents we get in general both single- and multi-trace contributions
\begin{align}
{\cal H}_{12\ldots n}(\tau) &:= \langle J^{a_1}(z_1) J^{a_2}(z_2) \ldots J^{a_n}(z_n) \rangle^\tau \, \big|_{{\rm Tr}(t^{a_1}t^{a_2}\ldots t^{a_{n-1}}t^{a_n} )} \label{1.07}  \\
&\phantom{:}=  \sum_{\nu=2}^4 
Z^{\rm het}_{\nu}(\tau) S_\nu(z_{12},\tau) S_\nu(z_{23},\tau) \ldots S_\nu(z_{n-1,n},\tau) S_\nu(z_{n1},\tau) \notag \\
{\cal H}_{12\ldots p|p{+}1\ldots n}(\tau) &:= \langle J^{a_1}(z_1) J^{a_2}(z_2) \ldots J^{a_n}(z_n) \rangle^\tau \, \big|_{{\rm Tr}(t^{a_1}t^{a_2}\ldots  t^{a_p} ){\rm Tr}(t^{a_{p{+}1}}\ldots  t^{a_n} )} \label{1.08} \\
&\phantom{:}=  \sum_{\nu=2}^4 
Z^{\rm het}_{\nu}(\tau) S_\nu(z_{12},\tau) S_\nu(z_{23},\tau) \ldots S_\nu(z_{p{-}1,p},\tau)S_\nu(z_{p1},\tau) \notag \\
& \ \ \times  S_\nu(z_{p{+}1,p{+}2},\tau) S_\nu(z_{p{+}2,p{+}3},\tau) \ldots S_\nu(z_{n{-}1,n},\tau)S_\nu(z_{n,p{+}1},\tau)\, . \notag
\end{align}
The dependence of the Szeg\"o kernels (\ref{1.01}) on the spin structure $\nu$ can be simplified by
relating them to Kronecker--Eisenstein series (\ref{1.2}) with one of the half-periods
\begin{align}
\omega_2 = \frac{1}{2} \, , \ \ \ \ \ \ \omega_3 = -\frac{1+\tau}{2} \, , \ \ \ \ \ \ \omega_4 = \frac{ \tau }{2}
\label{halfper}
\end{align}
in the second argument $\beta$. Given that the $\theta_{\nu=1,2,3,4}$ functions can be mapped into each 
other by a half-period shift in the first argument\footnote{More explicitly, the theta functions defined in
(\ref{1.0}) and (\ref{1.01ev}) are related by $\theta_2(z+\tfrac{1}{2},\tau) =  - \theta_1(z,\tau)$ and
\[
\theta_4(z+\tfrac{\tau}{2},\tau) =  i e^{-i\pi z} q^{-1/8} \theta_1(z,\tau) \, , \ \ \ \
 \theta_4(z+\tfrac{1}{2},\tau) =  \theta_3(z,\tau) \, , \ \ \ \
\theta_3(z+\tfrac{\tau}{2},\tau) =   e^{-i\pi z} q^{-1/8} \theta_2(z,\tau) 
\, .\]}, we have $S_\nu(z_{ij},\tau) \sim F(z_{ij},\omega_\nu,\tau)$ up to phase factors
that drop out from the cyclic products of Szeg\"o kernels in the spin sums (\ref{1.07}) and (\ref{1.08}) 
\cite{Tsuchiya:2017joo}
\begin{align}
S_\nu(z_{12},\tau) S_\nu(z_{23},\tau) \ldots  S_\nu(z_{n1},\tau) =
F(z_{12},\omega_\nu,\tau)F(z_{23},\omega_\nu,\tau)\ldots F(z_{n1},\omega_\nu,\tau)\, .
\label{halfper2}
\end{align}
This naturally introduces the elliptic functions $V_w$ generated by the cycles of 
Kronecker--Eisenstein series in (\ref{1.1a}). Given that the right-hand side of (\ref{halfper2})
defines an elliptic function of $\omega_\nu$, all the spin-structure dependence
can be absorbed into Weierstrass invariants
\begin{align}
e_\nu(\tau) := \wp(\omega_\nu,\tau) \, , \ \ \ \ \ \ \wp(z,\tau):= - \partial_z^2 \log \theta_1(z,\tau) + \frac{ \partial_z^3\theta_1(0,\tau) }{3 \partial_z \theta_1(0,\tau)} \, .
\label{halfper3}
\end{align}
The vanishing of $\partial_z \wp(z,\tau)$ at $z=\omega_\nu$ and the differential equation $\partial_z^2 \wp(z,\tau)=6 ( \wp(z,\tau))^2 - 30 {\rm G}_4$ then lead to a polynomial appearance of the Weierstrass invariants which carry the entire $\nu$-dependence \cite{Tsuchiya:2012nf, Tsuchiya:2017joo}
\begin{align}
S_\nu(z_{12},\tau) S_\nu(z_{21},\tau) &= V_2(1,2) + e_\nu \notag \\
S_\nu(z_{12},\tau) S_\nu(z_{23},\tau) S_\nu(z_{31},\tau)&= V_3(1,2,3) + e_\nu V_1(1,2,3)
\label{halfper4} \\
S_\nu(z_{12},\tau) S_\nu(z_{23},\tau) S_\nu(z_{34},\tau) S_\nu(z_{41},\tau) &= V_4(1,2,3,4) + e_\nu V_2(1,2,3,4) + e_\nu^2 - 6 {\rm G}_4 \, .\notag
\end{align}
We are using the following normalization conventions for holomorphic Eisenstein series\footnote{Note
that our normalization conventions in the expression (\ref{1.6eis}) for holomorphic Eisenstein series ${\rm G}_{2m}$
differ from (3.11) in \cite{DHoker:2016mwo} by a factor of $\pi^m$. The style of the letter ${\rm G}_k$ and the number of
subscripts distinguishes the holomorphic Eisenstein series ${\rm G}_k$ in (\ref{1.6eis}) from the bosonic Green functions
$G_{ij}(\tau)=G(z_{i}-z_j,\tau)$ in (\ref{mgf2}).},
\begin{equation}
{\rm G}_k(\tau) := \sum_{(m,n) \neq (0,0)} \frac{1}{(m\tau + n)^{k}} 
= 2\zeta_k + \frac{2 (2\pi i)^k}{(k{-}1)!} \sum_{m,n=1}^\infty m^{k-1} q^{mn}
,\qquad k\geq 4 \ \te{and even}   \ ,
\label{1.6eis}
\end{equation}
where $(m,n) \neq (0,0)$ is a shorthand for the summation domain $(m,n) \in \ZZ^2 \setminus \{(0,0)\}$.
Since the Weierstrass invariants are furthermore related by $e_\nu^3 - 15 {\rm G}_4 e_\nu - 35 {\rm G}_6=0$, 
the spin sums of the $n$-point current correlators in (\ref{1.07}) and (\ref{1.08}) can be reduced to the three
inequivalent cases\footnote{These identities are a consequence of the relations $e_2+e_3+e_4=0$, $\ e_2e_3+e_3 e_4+e_4 e_2= - 15 {\rm G}_4$ and $e_2 e_3 e_4 = 35 {\rm G}_6$ among the Weierstrass invariants as well as the
connection with theta functions via $\pi^2 (\theta_4(0,\tau))^4=e_2-e_3$, $\ \pi^2 (\theta_2(0,\tau))^4=e_3-e_4$
and $\pi^2 (\theta_3(0,\tau))^4=e_2-e_4$.} \cite{Ellis:1987dc}
\begin{align}
\sum_{\nu=2}^4 \!Z^{\rm het}_{\nu} =  {\rm G}_4^2 \, , \ \ \ \ \
\sum_{\nu=2}^4 \! Z^{\rm het}_{\nu} e_\nu= - \frac{7}{2} {\rm G}_4 {\rm G}_6 \, , \ \ \ \ \
\sum_{\nu=2}^4 \!Z^{\rm het}_{\nu} e^2_\nu = \frac{49}{6} {\rm G}_6^2  + \frac{5}{3} {\rm G}_4^3\, .
\label{halfper5} 
\end{align}
As a bottom line, (\ref{halfper4}) and (\ref{halfper5}) lead us to the following representations
\begin{align}
  {\cal H}_{12}  &= {\rm G}_4^2 V_2(1,2) - \frac{7}{2} {\rm G}_4 {\rm G}_6 
\label{halfper6} \\
  {\cal H}_{123} &= {\rm G}_4^2 V_3(1,2,3) - \frac{7}{2} {\rm G}_4{\rm G}_6 V_1(1,2,3)
\label{halfper7}\\
  {\cal H}_{1234} &= {\rm G}_4^2 V_4(1,2,3,4) - \frac{7}{2} {\rm G}_4{\rm G}_6 V_2(1,2,3,4) - \frac{13}{3} {\rm G}_{4}^3 + \frac{49}{6} {\rm G}_{6}^2\label{halfper8}\\
 {\cal H}_{12|34}  &= {\rm G}_4^2  V_2(1,2) V_2(3,4) - \frac{7}{2} {\rm G}_4{\rm G}_6\big[V_2(1,2) + V_2(3,4)\big] + \frac{5}{3} {\rm G}_{4}^3 + \frac{ 49  }{6} {\rm G}_6^2
\label{halfper9}
\end{align}
for the spin sums in (\ref{1.07}) and (\ref{1.08}) which enter the
trace decomposition of the four-point correlator,
\begin{align}
\langle J^{a_1}(z_1) J^{a_2}(z_2)  &J^{a_3}(z_3)  J^{a_4}(z_4) \rangle^\tau = \ 
\buildrel{\leftrightarrow}\over{{\rm Tr}}  \hspace{-0.11cm} (t^{a_1}t^{a_2}t^{a_3}t^{a_4}) {\cal H}_{1234}   \label{halfper99}  \\
& \ \ \  
+ {\rm Tr}(t^{a_1}t^{a_2}) {\rm Tr}(t^{a_3}t^{a_4}) {\cal H}_{12|34}  + {\rm cyc}(2,3,4) \, .
\notag
\end{align}
As before, the sum over cyclic permutations refers to both lines of (\ref{halfper99}).


\subsection{The key integrals over torus punctures}
\label{sec:2.new}

We shall now pinpoint the integrals over torus punctures that need to be performed
in the four-point gauge amplitude (\ref{presc1}) and whose low-energy expansion will be the main
topic of the later sections. After factoring out the polarization dependent parts (\ref{presc4}),\footnote{\label{fn:norm}In order to reproduce the normalization conventions for the four-point gauge amplitude in \cite{Ellis:1987dc}, the right-hand
side of (\ref{3.0}) needs to be multiplied by $\frac{1}{(2\zeta_4)^2} \big( \frac{ \ap g }{64 \pi^5} \big)^2$,
where $g$ denotes the gauge coupling. The inverse factor of $2\zeta_4^2$ compensates for
our choice of normalization of the partition function in (\ref{1.00}).}
\begin{align}
{\cal M}_4 =  (k_1\cdot k_2)(k_2\cdot k_3) A_{\rm SYM}^{\rm tree}(1,2,3,4) \,   \int_{\cal F} \frac{\dd^2 \tau }{(\Im \tau)^2\, \eta^{24}(\tau) }   \, M_4(\tau)
\, ,
\label{3.0}
\end{align}
we will be interested in the following integral over the punctures,
\begin{align}
M_4(\tau)&= 
 \int \dd \mu_4 \, \langle J^{a_1}(0) J^{a_2}(z_2) J^{a_3}(z_3) J^{a_4}(z_4) \rangle^{\tau} \,  \exp \Big( \sum_{1\leq i<j}^4 s_{ij} G_{ij}(\tau) \Big)
\, .\label{3.1}
\end{align}
We use the following notation for the modular invariant integration measure,
\begin{align}
\int \dd \mu_4 &:= \prod_{j=2}^4 \int_{T(\tau)} \frac{ \dd^2 z_j}{\Im \tau} \, ,
\label{meas}
\end{align}
and have used translation invariance on the torus to fix the first coordinate $z_1\to0$ such that the integral is only over the remaining three punctures. The dimensionless Mandelstam invariants $s_{ij}$ obey the following constraints
by momentum conservation and the mass-shell condition $k_j^2=0$ for external gauge bosons $j=1,2,3,4$,
\begin{align}
s_{34}=s_{12} \, , \ \ \ \ \ \
s_{14}= s_{23} \, , \ \ \ \ \ \
s_{13}=s_{24}=-s_{12}-s_{23}\, .
\label{mgf3}
\end{align}
With the results for the current correlators (\ref{halfper99}) in terms of the spin sums
(\ref{halfper8}) and (\ref{halfper9}), the right-hand side of (\ref{3.1}) boils down
to five inequivalent Koba--Nielsen integrals ${\cal I}^{(w,0)}_{\ldots}$ over elliptic functions,
\begin{align}
M_4(\tau) &= \  \buildrel{\leftrightarrow}\over{{\rm Tr}} \hspace{-0.11cm} (t^{a_1}t^{a_2}t^{a_3}t^{a_4}) 
\Big[ {\rm G}_4^2 {\cal I}^{(4,0)}_{1234} - \frac{7}{2} {\rm G}_4{\rm G}_6 {\cal I}^{(2,0)}_{1234}
- \frac{13}{3} {\rm G}_{4}^3  {\cal I}^{(0,0)} + \frac{49}{6} {\rm G}_{6}^2 {\cal I}^{(0,0)} \Big]  \label{3.6} \\
&\! \! \! \! \! \! \! \! \! +{\rm Tr}(t^{a_1}t^{a_2}) {\rm Tr}(t^{a_3}t^{a_4}) \Big[
 {\rm G}_4^2   {\cal I}^{(4,0)}_{12|34} - \frac{7}{2} {\rm G}_4{\rm G}_6 
{\cal I}^{(2,0)}_{12|34}  + \frac{5}{3} {\rm G}_{4}^3 {\cal I}^{(0,0)} + \frac{ 49  }{6} {\rm G}_6^2 {\cal I}^{(0,0)} \Big]
+ {\rm cyc}(2,3,4)  \, .\notag 
\end{align}
The notation `$+ {\rm cyc}(2,3,4)$' refers to cyclic permutations of both lines w.r.t.\
the adjoint indices $a_{2},a_3,a_4$ and the Mandelstam invariants $s_{ij}$. Two 
of the integrals in the single-trace or `planar' sector of (\ref{3.6}) are defined by
a cyclic ordering in the subscript
\begin{align}
{\cal I}^{(4,0)}_{1234}(s_{ij},\tau) &:= \int \dd \mu_4 \,
V_4(1,2,3,4)\, \exp \Big( \sum_{1\leq i<j}^4 s_{ij} G_{ij}(\tau) \Big)   \label{3.2}\\
{\cal I}^{(2,0)}_{1234}(s_{ij},\tau) &:= \int \dd \mu_4 \,
 V_2(1,2,3,4)\, \exp \Big( \sum_{1\leq i<j}^4 s_{ij} G_{ij}(\tau) \Big) \, . \label{3.3}
 \end{align}
Furthermore, the permutation-invariant integral
\begin{align}
{\cal I}^{(0,0)}(s_{ij},\tau) := \int \dd \mu_4 \, \exp \Big( \sum_{1\leq i<j}^4 s_{ij} G_{ij}(\tau) \Big)
\label{mgf1}
\end{align}
is universal to the single- and double-trace sectors of (\ref{3.6}), and it furthermore occurs 
in the four-point one-loop amplitude of type-II superstrings \cite{Green:1982sw}.
In the double-trace or `non-planar' sector of (\ref{3.6}), we have further instances of ${\cal I}^{(0,0)}$ and
\begin{align}
{\cal I}^{(4,0)}_{12|34}(s_{ij},\tau)  &:= \int \dd \mu_4 \,
V_2(1,2) V_2(3,4)\, \exp \Big( \sum_{1\leq i<j}^4 s_{ij} G_{ij}(\tau) \Big)  \label{3.4}\\
{\cal I}^{(2,0)}_{12|34}(s_{ij},\tau)  &:= \int \dd \mu_4 \,
 \big[V_2(1,2) + V_2(3,4)\big]\, \exp \Big( \sum_{1\leq i<j}^4 s_{ij} G_{ij}(\tau) \Big)\,.  \label{3.5}
 \end{align}
In all of (\ref{3.2}) to (\ref{3.5}), translation invariance has been used to fix $z_1=0$. 

The superscripts in the notation for the integrals keep track of their modular weights: 
By the modular properties (\ref{1.1mod}) of the elliptic $V_w$-functions,
the integrals are easily checked to transform as modular forms of holomorphic 
weights $(w,0)$,
\begin{align}
{\cal I}^{(w,0)}_{\ldots}\Big(s_{ij}, \frac{ \alpha \tau + \beta }{\gamma \tau + \delta} \Big) = (\gamma \tau+\delta)^w {\cal I}^{(w,0)}_{\ldots}(s_{ij},\tau)  \, ,
\label{3.7}
\end{align}
where the ellipsis may represent any permutation of $1234$ and $12|34$ (or be
empty to incorporate modular invariance of ${\cal I}^{(0,0)}$).
With the modular weight $(k,0)$ of ${\rm G}_k$, each term in the four-point integral
(\ref{3.6}) is a form of weight $(12,0)$. In the integrated amplitude (\ref{3.0}), this
compensates the weight $(-12,0)$ of the bosonic partition function $\eta^{-24}$ in agreement with modular invariance.

In the rest of this work, we will develop and apply methods for a systematic low-energy expansion of the integrals 
(\ref{3.2}) to (\ref{3.5}) over the punctures. This amounts to a simultaneous Taylor expansion in all the
Mandelstam invariants (\ref{mgf2}) and will also be referred to as $\ap$-expansion. 
The modular weights (\ref{3.7}) apply to each order in the $\ap$-expansion of the integrals
${\cal I}^{(w,0)}_{\ldots}$. For all of (\ref{3.2}) to (\ref{3.5}),
the coefficients of any monomial in $s_{ij}$ will be shown to line up with modular graph 
forms \cite{DHoker:2016mwo} which we will review in the next section. Hence, 
the ${\cal I}^{(w,0)}_{\ldots}$ are generating series for 
modular graph forms of weight $(w,0)$, and the modular invariants obtained from ${\cal I}^{(0,0)}$ 
are known as modular graph functions \cite{DHoker:2015wxz}.


\section{Basics of modular graph forms}
\label{sec:8}

In this section, we introduce some basic material and notation for modular graph functions 
(that are invariant under the modular group) and modular graph forms (that transform covariantly under the modular group).


\subsection{Modular graph functions}
\label{sec:2.3}

One of the main goals in this paper is to perform the integrals over the punctures $z_2,z_3,z_4$
in the low-energy expansion of the $\tau$-integrand (\ref{3.6}) of the heterotic-string amplitude (\ref{3.0}). 
We will start with a review of the techniques that apply to the simplest integral ${\cal I}^{(0,0)}(s_{ij},\tau)$ in 
(\ref{mgf1}) known from the type-II superstring \cite{Green:1982sw}. While a closed-form evaluation of 
${\cal I}^{(0,0)}(s_{ij},\tau)$ is currently out of reach, its Taylor expansion in the $s_{ij}$ has been thoroughly 
investigated in \cite{Green:1999pv, Green:2008uj, DHoker:2015foa, 
DHoker:2015wxz, Kleinschmidt:2017ege}. As explained in the references, the $\ap$-expansion of
${\cal I}^{(0,0)}(s_{ij},\tau)$ probes low-energy effective interactions at the one-loop order of type-II superstrings and has
important implications for S-duality of the type-IIB theory \cite{Hull:1994ys, 
Green:1997tv, Green:1999pu, Green:2005ba, Green:2014yxa}.

However, by keeping $\tau$ fixed in this calculation, one cannot detect the logarithmic 
dependence of the full amplitude on $s_{ij}$ which arises 
from the $\tau \rightarrow i\infty$ region of the later integration over modular parameters, cf.\ (\ref{3.0}). 
As explained in \cite{Green:2008uj}, Taylor expansion of (\ref{mgf1}) amounts to isolating the analytic momentum 
dependence of the one-loop amplitude as opposed to non-analytic threshold contributions including factors 
of $\log s_{ij}$ that are determined from the tree amplitude via unitarity.
The analytic momentum dependence of graviton and gauge-boson amplitudes in turn can be used 
to identify curvature and field-strength operators in the low-energy effective action. Still, a subtle 
interplay between the analytic and non-analytic sectors has to be taken into 
account \cite{Green:2008uj, DHoker:2015foa}.

After Taylor expanding the exponentials in the integrand of (\ref{mgf1}), the leftover challenge is to integrate
monomials in Green functions $\prod_{1\leq i<j}^4 G_{ij}^{n_{ij}}$ with $n_{ij}\in \NN_0$ over the torus.
This kind of order-by-order integration is most conveniently performed by means of the lattice-sum 
representation \cite{Green:1999pv}
\begin{align}
G(z,\tau) 
= \frac{ \Im \tau}{\pi} \sum_{(m,n) \neq (0,0)} \frac{  e^{2\pi i (mv - nu)} }{| m\tau + n|^{2}}\, , \ \ \ \ 
z=u\tau + v \, ,
\label{mgf4}
\end{align}
where $(m,n) \neq (0,0)$ is again a shorthand for $(m,n) \in \ZZ^2 \setminus \{(0,0)\}$,
and we use the following real coordinates $u,v\in [0,1)$ for the torus puncture $z = u\tau + v$
in (\ref{mgf4}) and later equations,
\begin{equation}
u = \frac{ \Im  z}{\Im \tau} \ , \ \ \ \ \ \ 
v = \Re  z -  \frac{ \Im z \ \Re \tau}{\Im  \tau} \ .
\label{1.3}
\end{equation}
Note that the lattice-sum representation (\ref{mgf4}) of the Green function requires a summation prescription as the sum is not absolutely convergent. We shall use the Eisenstein summation prescription defined in \eqref{eq:19}.
The lattice sum exhibits the modular invariance of the Green function, i.e.\ the integral ${\cal I}^{(0,0)}(s_{ij},\tau)$ is modular invariant term by term
in its $\ap$-expansion.

By the absence of the zero mode $(m,n)=(0,0)$ in (\ref{mgf4}), a single Green function integrates
to zero under the measure $\dd \mu_4$ in (\ref{meas}). A large class of integrals over 
$\prod_{1\leq i<j}^4 G_{ij}^{n_{ij}}$ vanishes for the same reason, and the conditions on the 
exponents $n_{ij} \in \NN_0$ can be characterized by the following graphical organization: Assign
a vertex for each puncture $z_{j}, \ j=1,2,3,4$ and draw an undirected edge between vertices $i$ and $j$ for each
factor of $G_{ij}=G_{ji}$ in the integrand, see figure \ref{figdefGij}. The contributions from disconnected graphs 
factorize in this setup, and trivial graphs with just one zero-valent vertex evaluate to one by 
$\int_{T(\tau)} \frac{ \dd^2 z }{\Im \tau}=1$. Furthermore, a monomial in $G_{ij}$ integrates 
to zero whenever the associated graph is one-particle reducible, i.e.\ whenever it can be disconnected by 
cutting a single line.

\begin{figure}
  \begin{center}
    \tikzpicture[scale=1,line width=0.30mm]
    \draw(-1.7,0)node{$G_{ij}(\tau) \ \ \leftrightarrow$};
    \draw(0,0)node{$\bullet$}node[left]{$i$} -- (1.5,0)node{$\bullet$}node[right]{$j$};
    \draw(2.2,0)node{,};
    \draw(5.3,0)node{$\displaystyle \int \dd \mu_4 \, G_{23}G_{24}G^2_{34} \ \ \leftrightarrow$};
    \scope[xshift=8cm]
    \draw(0,0.75)node{$\bullet$}node[left]{2};
    \draw(0,-0.75)node{$\bullet$}node[left]{1};
    \draw(1.5,0.75)node{$\bullet$}node[right]{3};
    \draw(1.5,-0.75)node{$\bullet$}node[right]{4};
    \draw(0,0.75)--(1.5,-0.75);
    \draw(0,0.75)--(1.5,0.75);
    \draw(1.5,0.75) .. controls (1.8,0) .. (1.5,-0.75);
    \draw(1.5,0.75) .. controls (1.2,0) .. (1.5,-0.75);
    \endscope
    \endtikzpicture
    \caption{Graphical representation of a bosonic Green function $G_{ij}(\tau)$ and a sample
      integral in the low-energy expansion of ${\cal I}^{(0,0)}$.}
    \label{figdefGij}
  \end{center}
\end{figure}

The simplest non-vanishing integrals admitted by this graphical selection rule turn out to be
non-holomorphic Eisenstein series
\begin{align}
  {\rm E}_{k}(\tau) := \left(\frac{\Im \tau}{\pi}\right)^{k}\sum_{(m,n)\neq(0,0)}\frac{1}{|m\tau+n|^{2k}}\ ,\qquad k\geq 2 \, .
  \label{mgf5}
\end{align}
One can check from the lattice-sum representation (\ref{mgf4}) of the Green functions that closed
cycles $G_{12}G_{23}\ldots G_{k-1,k}G_{k,1}$ associated with one-loop graphs
as depicted in figure \ref{figeisen} integrate to ${\rm E}_{k}(\tau)$, e.g.
\begin{align}
\int \dd \mu_4 \, G_{12}^2 = {\rm E}_2 \, , \ \ \ \ \ 
\int \dd \mu_4 \, G_{12}G_{23}G_{31} = {\rm E}_3 \, , \ \ \ \ \ 
\int \dd \mu_4 \, G_{12}G_{23}G_{34} G_{41} = {\rm E}_4\, .
  \label{mgf6}
\end{align}

\begin{figure}
  \begin{center}
    \tikzpicture[scale=1,line width=0.30mm]
    \draw (0.5,0)--(-0.5,0);
    \draw (-0.5,0)--(-0.85,-0.35);
    \draw [dashed](-0.85,-0.35)--(-1.2,-0.7);
    \draw (0.5,0)--(1.2,-0.7);
    \draw[dashed] (-1.2,-1.7)--(-1.2,-0.7);
    \draw (1.2,-1.7)--(1.2,-0.7);
    \draw (1.2,-1.7)--(0.85,-2.05);
    \draw[dashed] (0.85,-2.05)--(0.5,-2.4);
    \draw[dashed] (-0.5,-2.4)--(0.5,-2.4);
    \draw[dashed] (-0.5,-2.4)--(-1.2,-1.7);
    \draw(0.5,0)node{$\bullet$}node[above]{1};
    \draw(-0.5,0)node{$\bullet$}node[above]{$k$};
    \draw(1.2,-0.7)node{$\bullet$}node[right]{2};
    \draw(1.2,-1.7)node{$\bullet$}node[right]{3};
    \draw(2.7,-1.2)node{$\leftrightarrow \ \ {\rm E}_k$};
    \endtikzpicture
    \caption{One-loop graphs due to a length-$k$ cycle of Green functions 
      $G_{12}G_{23}\ldots G_{k-1,k}G_{k1}$ correspond to non-holomorphic Eisenstein series ${\rm E}_k$
      defined in (\ref{mgf5}).}
    \label{figeisen}
  \end{center}
\end{figure}

In the graphical organization of the $\ap$-expansion of ${\cal I}^{(0,0)}(s_{ij},\tau)$,
the denominators in (\ref{mgf5}) can be interpreted as Feynman propagators of a scalar field on a torus.
The discrete momenta $m\tau + n$ arise from the Fourier coefficients of the $G_{ij}$ associated with the edges,
see (\ref{mgf4}), and integration over the $z_j$ imposes momentum conservation at each vertex.
The results of integrating more general monomials
$\prod_{1\leq i<j}^N G_{ij}^{n_{ij}}$ over the measure $\prod_{j=2}^N \frac{ \dd^2 z_j}{\Im \tau}$ for $N$ punctures
are dubbed modular graph functions \cite{DHoker:2015wxz}. Modular invariance of modular graph functions is a direct consequence of the modular invariance of the integration measure and the Green function.

\begin{figure}
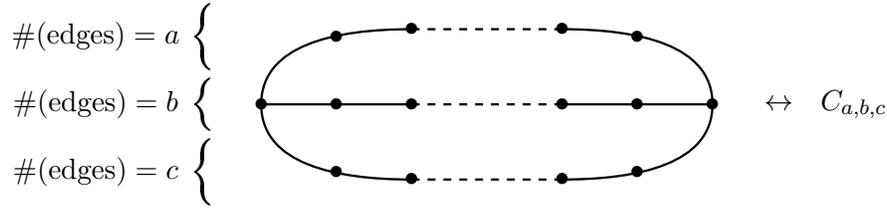

  \begin{center}
    \tikzpicture[scale=1,line width=0.30mm]
    \draw (-1.7,0.9)node[rotate=-90]{$\underbrace{\phantom{xxxx}}$};
    \draw (-1.7,0)node[rotate=-90]{$\underbrace{\phantom{xx}}$};
    \draw (-1.7,-0.9)node[rotate=-90]{$\underbrace{\phantom{xxxx}}$};
    \draw(-3.2,0.9)node{$\#(\text{edges}) = a$};
    \draw(-3.2,0)node{$\#(\text{edges}) = b$};
    \draw(-3.2,-0.9)node{$\#(\text{edges}) = c$};
    \draw (-1,0) node{$\bullet$};
    \draw (0,0.9)node{$\bullet$};
    \draw (0,0)node{$\bullet$};
    \draw (0,-0.9)node{$\bullet$};
    \draw (1,1)node{$\bullet$};
    \draw (1,0)node{$\bullet$};
    \draw (1,-1)node{$\bullet$};
    \draw[dashed](1,1) -- (3,1);
    \draw[dashed](1,0) -- (3,0);
    \draw[dashed](1,-1) -- (3,-1);
    \draw (3,1)node{$\bullet$};
    \draw (3,0)node{$\bullet$};
    \draw (3,-1)node{$\bullet$};
    \draw (4,0.9)node{$\bullet$};
    \draw (4,0)node{$\bullet$};
    \draw (4,-0.9)node{$\bullet$};
    \draw (5,0)node{$\bullet$};
    \draw (0,0)--(-1,0);
    \draw(-1,0) .. controls (-1,0.8) and (0.0,1) .. (1,1);
    \draw(-1,0) .. controls (-1,-0.8) and (0.0,-1) .. (1,-1);
    \draw (0,0)--(1,0);
    \draw (3,0)--(5,0);
    \draw(5,0) .. controls (5,0.8) and (4.0,1) .. (3,1);
    \draw(5,0) .. controls (5,-0.8) and (4.0,-1) .. (3,-1);
    \draw(6.5,0)node{$\leftrightarrow \ \ C_{a,b,c}$};
    \endtikzpicture
    \caption{Two loop graphs associated with the modular graph functions $C_{a,b,c}$ defined in~\eqref{mgf7}.}
    \label{fig:Cabc}
  \end{center}
\end{figure}

Modular graph functions as the ones depicted in figure \ref{figeisen} with one loop on the toroidal worldsheet involve only a single lattice sum. Modular graph functions with more than one loop on the toroidal worldsheet as e.g.\ the ones depicted in figure \ref{fig:Cabc} involve several lattice sums. In order to avoid cluttering notation, we shall use the shorthand
$\sum_{p \neq 0}$ for the lattice sum $\sum_{(m,n)\neq (0,0)}$ over
the discrete momenta $p=m\tau + n$ (instead of writing out $p \in( \ZZ\tau +\ZZ )\setminus \{0\}$).
In this way, the expression (\ref{mgf5}) for one-loop graph functions condenses to ${\rm E}_k= ( \frac{ \Im \tau }{\pi} )^k \sum_{p\neq 0} \frac{1}{|p|^{2k}}$, and the 
two-loop graphs depicted in figure \ref{fig:Cabc} give rise to double lattice sums\footnote{Without the notation for discrete momenta $p_i=m_i \tau + n_i$,
(\ref{mgf7}) takes the lengthier form
$$
C_{a,b,c}(\tau) = \left(\frac{\Im \tau}{\pi}\right)^{a+b+c} \! \!  \! \!  \! \!  \! \!  \! \! \sum_{(m_1,n_1),(m_2,n_2)\neq(0,0) \atop{(m_1,n_1)+(m_2,n_2)\neq(0,0)}}   \frac{1}{|m_1\tau+n_1|^{2a}
  |m_2\tau+n_2|^{2b} |(m_1{+}m_2)\tau+(n_1{+}n_2)|^{2c}}\, .
$$}
\begin{align}
  C_{a,b,c}(\tau) := \left(\frac{\Im \tau}{\pi}\right)^{a+b+c} \sum_{p_1,p_2 \neq 0 \atop{p_1+p_2 \neq 0}}   \frac{1}{|p_1|^{2a}
  |p_2|^{2b} |p_1{+}p_2|^{2c}} \, .
  \label{mgf7}
\end{align}

The modular invariant functions $C_{a,b,c}$ can be viewed as a generalization of 
non-holo\-morphic Eisenstein series, and many of their properties including Laplace eigenvalue equations 
and Fourier zero modes w.r.t.\ $\Re \tau$ have been studied for generic $a,b,c\geq 1$
\cite{DHoker:2015foa, DHoker:2017zhq}. Their simplest examples in the $\ap$-expansion 
of the integral (\ref{mgf1}) are
\begin{align}
\int \dd \mu_4 \, G_{12}^3 = C_{1,1,1} \, , \ \ \ \ \ \
\int \dd \mu_4 \, G_{12}^2G_{23}G_{31} = C_{2,1,1}   \, .
  \label{mgf8}
\end{align}
One of the fascinating properties of modular graph functions is their multitude of relations 
which often involve multiple zeta values (MZVs) and mix different loop orders \cite{DHoker:2015foa, DHoker:2015sve, DHoker:2016quv}. 
For instance, the simplest two- and three-loop modular graph functions turn out to reduce to 
lower-complexity objects \cite{DHoker:2015foa, DHoker:2015sve}
\begin{align}
C_{1,1,1} = {\rm E}_3 + \zeta_3 \, , \ \ \ \ \ \
D_4 := \int \dd \mu_4 \, G_{12}^4  = 24 C_{2,1,1}  - 18 {\rm E}_4 + 3 {\rm E}_2^2  \, .
  \label{mgf9}
\end{align}
In general MZVs $\zeta_{n_{1},n_{2},\ldots,n_{r}}$ are defined by the conical sums
\begin{align}
\zeta_{n_1,n_2,\ldots,n_r} := \sum_{0<k_1<k_2<\ldots <k_r} k_1^{-n_1} k_2^{-n_2}\ldots k_r^{-n_r} \, , \ \ \ \ n_i \in \NN \, , \ \ \ \ n_r \geq 2\, ,
\label{defMZV}
\end{align}
where $r$ and $n_1{+}n_2{+}\ldots{+}n_r$ are referred to as their depth and weight, respectively.

With the above input, the leading orders in the $\ap$-expansion of the integral (\ref{mgf1}) read\footnote{Note that the relation $s_{12}+s_{13}+s_{23}=0$ among the Mandelstam variables in (\ref{mgf10}) has been
used to attain compact expressions in the $\ap$-expansion. Elimination of $s_{13}$ leaves $\mathcal{I}^{(0,0)}$ to be function of $s_{12}$, $s_{23}$ and $\tau$.}
\begin{align}
{\cal I}^{(0,0)}(s_{ij},\tau) = 1 &+ 2 {\rm E}_2(s_{13}^{2} -s_{12}s_{23}) + (5 {\rm E}_3 + \zeta_3) s_{12} s_{23} s_{13}   \label{mgf10}  \\
&+ 2(2C_{2,1,1} + {\rm E}_2^2 - {\rm E}_4) (s_{13}^{2} -s_{12}s_{23})^2 + {\cal O}(\ap^5)\, ,
\notag
\end{align}
and we will later on give similar expansions for the additional integrals that enter the heterotic-string
amplitude (\ref{3.6}). 


\subsection{Modular graph forms}
\label{sec:2.4}

\begin{figure}
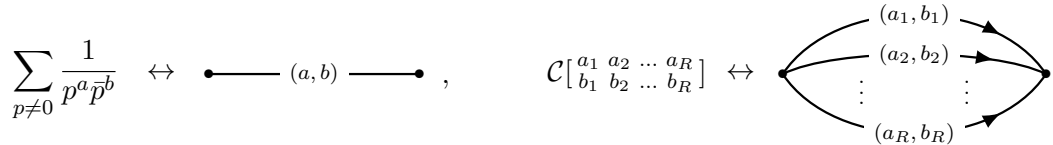

  \begin{center}
     \tikzpicture[scale=0.7,line width=0.3mm,decoration={markings,mark=at position 0.8 with {\arrow[scale=1.3]{latex}}}]
    \draw(-1.62,2.17)node{$\displaystyle \sum_{p\neq0} \frac{1}{p^a \bar p^b}$};
    \draw(0.2,2.3)node{$\displaystyle \leftrightarrow$};
    \draw(1.1,2.3)node{\scriptsize $\bullet$} --node[fill=white]{\scriptsize $(a,b)$} (5.1,2.3)node{\scriptsize $\bullet$};
    \draw(5.6,2.1)node{$,$};
    \scope[xshift=11cm, yshift=2.3cm]
    \draw(-2,0)node{$\cform{a_1&a_2&\ldots &a_R \protect\\b_1&b_2&\ldots &b_R}$};
    \draw(0.2,0)node{$\displaystyle \leftrightarrow$};
    \draw[postaction={decorate}] (1,0)node{\scriptsize $\bullet$} .. controls (2,1.5) and (5,1.5) .. node[fill=white]{\scriptsize $(a_1,b_1)$} (6,0)node{\scriptsize $\bullet$};
    \draw[postaction={decorate}] (1,0) .. controls (2,0.5) and (5,0.5) .. node[fill=white]{\scriptsize $(a_2,b_2)$} (6,0);
    \draw (2.5,-0.2) node{\scriptsize $\vdots$};
    \draw (4.5,-0.2) node{\scriptsize $\vdots$};
    \draw[postaction={decorate}] (1,0) .. controls (2,-1.5) and (5,-1.5) .. node[fill=white]{\scriptsize $(a_R,b_R)$} (6,0);
    \endscope
    \endtikzpicture 
    \vspace*{-1.3em}
    \caption{Graph associated to a single lattice sum (left panel) and the general dihedral modular graph form $\cform{a_1&a_2&\ldots &a_R \protect\\b_1&b_2&\ldots &b_R}$ (right panel). The holomorphic and antiholomorphic decorations of the edges are denoted by $(a_j,b_j)$. The arrows indicate the directions of the momenta whose relative orientation is important for the momentum-conserving delta function.}
    \label{fig:dihedralMGF}
  \end{center}
\end{figure}

The lattice-sum representations of holomorphic Eisenstein series
${\rm G}_k$ in (\ref{1.6eis}) and their non-holomorphic counterparts ${\rm E}_k$ in (\ref{mgf5})
can be unified in the framework of modular graph forms  \cite{DHoker:2016mwo}. The idea is to study
lattice sums of the type $\sum_{p\neq 0} \frac{1}{p^{a}\bar p^{b}}$, where the holomorphic and antiholomorphic 
momenta $p=m\tau+n$ and $\bar p = m \bar \tau+n$ may have different integer 
powers $a\neq b$ with $a+b\geq 2$.
Generalizations to multiple sums $\sum_{p_1,p_2,\ldots \neq 0}$ arise naturally from the following dictionary
to one-particle irreducible graphs: Associate powers of momenta 
$p^{-a}\bar p^{-b}$ with
an edge decorated by the two integers $(a,b)$ as depicted in figure \ref{fig:dihedralMGF}
and take vertices to impose momentum conservation as before.
Under a change of direction $p \rightarrow -p$, the term $p^{-a}\bar p^{-b}$ changes sign if $a{+}b$ is odd, hence odd values of $a{+}b$ give rise to directed edges, while even values of $a{+}b$ give rise to undirected edges. Flipping the direction of a directed edge flips the sign of the associated modular graph form.
It will be shown in section \ref{sec:3.0}
that the $\ap$-expansions of the integrals ${\cal I}_{\ldots}^{(w,0)}$ defined in 
(\ref{3.2}) to (\ref{3.5}) are expressible in terms of modular graph forms.

For the \emph{dihedral} graphs with two vertices and $R$ decorated edges $(a_1,b_1),(a_2,b_2),\ldots,(a_R,b_R)$ 
depicted in figure \ref{fig:dihedralMGF}, the associated modular graph form is
\begin{align}
  \cform{a_1&a_2&\ldots &a_R \\b_1&b_2&\ldots &b_R}:=\left(\frac{\Im\tau}{\pi}\right)^{\sum_{i=1}^{R}b_{i}}\sum_{p_1,p_2,\ldots,p_R\neq 0} \frac{\delta(p_1+p_2+\ldots+p_R)}{p_1^{a_1} \bar p_1^{b_1} \, p_2^{a_2} \bar p_2^{b_2} \, \ldots \, p_R^{a_R} \bar p_R^{b_R} }\, ,
   \label{mgf11} 
\end{align}
which matches the notation of \cite{DHoker:2016mwo} up to overall powers of $ \pi$ and $\Im \tau$. The 
normalization conventions in (\ref{mgf11}) with holomorphic and antiholomorphic momenta entering on asymmetric 
footing are chosen for later convenience\footnote{The 
quantities which are denoted by $\cform{a_1&a_2&\ldots &a_R \\b_1&b_2&\ldots &b_R}$,
$\ {\cal C}^+ \!  \! \left[ \begin{smallmatrix} a_1&a_2&\ldots &a_R \\b_1&b_2&\ldots &b_R\end{smallmatrix} \right]$ and
${\cal C}^- \!  \! \left[ \begin{smallmatrix} a_1&a_2&\ldots &a_R \\b_1&b_2&\ldots &b_R\end{smallmatrix} \right]$
in \cite{DHoker:2016mwo} are obtained
by multiplying (\ref{mgf11}) by factors of $( \frac{ \Im \tau}{\pi} )^{\frac{1}{2} (a-b)}$, 
$\ \left(\frac{ (\Im \tau )^{2}}{\pi}\right)^{\frac{1}{2}(a-b)} $ and $\left(\frac{1}{\pi}\right)^{\frac{1}{2}(a-b)} $,
respectively, where $a=\sum_{i=1}^{R}a_{i}$ and $b=\sum_{i=1}^{R}{b_{i}}$.}. The 
momentum-conserving delta function is introduced to manifest the invariance of (\ref{mgf11}) under permutations 
$\sigma \in S_R$ of the edges and their decorations $(a_i,b_i)$, 
\begin{align}
\cform{a_1&a_2&\ldots &a_R \\b_1&b_2&\ldots &b_R} = \cform{a_{\sigma(1)}&a_{\sigma(2)}&\ldots &a_{\sigma(R)} \\b_{\sigma(1)}&b_{\sigma(2)}&\ldots &b_{\sigma(R)}} \quad \forall \ \sigma \in S_R\, .
\label{mgf11perm}
\end{align}
After resolving the delta functions, the remaining summation variables $p_j\neq 0$ in
\begin{align}
  \cform{a_1&\ldots &a_R \\b_1&\ldots &b_R}\hspace{-0.2em}=\hspace{-0.2em}\left(\frac{\Im\tau}{\pi}\right)^{\sum_{i=1}^{R}\hspace{-0.2em}b_{i}}\hspace{-2em} \sum_{p_1,\ldots,p_{R-1}\neq 0 \atop{p_1+\ldots+p_{R-1} \neq 0}} \hspace{-0.5em}\frac{(-1)^{a_R+b_R}}{p_1^{a_1} \bar p_1^{b_1}  \, \ldots \, p_{R-1}^{a_{R-1}} \bar p_{R-1}^{b_{R-1}} \, (p_1{+}\ldots{+}p_{R-1})^{a_R} (\bar p_1{+}\ldots{+}\bar p_{R-1})^{b_R}  }
   \label{mgf12} 
\end{align}
are additionally constrained by $p_1+p_2+\ldots+p_{R-1} \neq 0$, and similar expressions are obtained
upon eliminating a different momentum. Hence, a single summation variable is attained
from (\ref{mgf12}) by formally starting with a two-edge graph,
\begin{align}
  \cform{a&0 \\b&0}=\left(\frac{\Im\tau}{\pi}\right)^{b}   \sum_{p\neq 0 } \frac{1}{p^a \bar p^b}\, ,
   \label{mgf13} 
\end{align}
and the two classes of Eisenstein series ${\rm G}_k$ and ${\rm E}_k$ are recovered via
\begin{align}
{\rm G}_k =  \cform{k&0 \\0&0}  \, , \ \ \ \ \ \ 
 {\rm E}_k =   \cform{k&0 \\k&0} \,  .
   \label{mgf14} 
\end{align}
Note that we require $a_{i}+b_{i}+a_{R}+b_{R}>2$ for all $i=1,\ldots, R{-}1$ in order to ensure absolute convergence of the series\footnote{Since the columns in the matrix $
  (\begin{smallmatrix}
    a_{1}&a_2&\ldots&a_{R}\\
    b_{1}&b_2&\ldots&b_{R}
  \end{smallmatrix})
  $ can be rearranged arbitrarily by (\ref{mgf11perm}), this is equivalent to the criterion that $a_{i}+b_{i}+a_{j}+b_{j}>2$ for any $i,j=1,2,\ldots, R$.}. If $a_{i}+b_{i}+a_{R}+b_{R}=2$ for some $i$, the sum is only conditionally convergent, as e.g.\ in the case of
  $ {\rm G}_{2}$. We define ${\rm G}_{2}$ using the Eisenstein summation prescription $\esum$ (cf.\ appendix \ref{app:Eisreg} and in particular \eqref{eq:19}),
\begin{align}
  {\rm G}_{2}(\tau):=\esum_{p\neq 0} \frac{1}{p^{2}}=\sum_{n\in\ZZ\backslash \{0\}}\frac{1}{n^{2}}+\sum_{m\in\ZZ\backslash \{0\}}\sum_{n\in\ZZ}\frac{1}{(m\tau+n)^{2}}\ ,
  \label{1.6AA}
\end{align}
which gives rise to a holomorphic but non-modular expression. One can also obtain a modular but non-holomorphic version ${\rm \hat G}_{2}$ via \cite{Schoen:1974}
 \begin{equation}
\hat {\rm G}_2(\tau):=  \lim_{s\rightarrow 0}\sum_{(m,n) \neq (0,0)} \frac{1}{(m\tau + n)^{2} \,|m\tau + n|^s} = {\rm G}_2(\tau)  - \frac{\pi}{\Im  \tau}\, .
  \label{1.6BB}
\end{equation}
For some of these conditionally convergent lattice sums, regularized values 
$\cregform{a_1&a_2&\ldots &a_R \\b_1&b_2&\ldots &b_R}$ will be given in section \ref{sec:3.3}.

By definition, modular graph functions are special cases of modular graph forms, where
the edges have decorations of the form $(a,a)$, i.e.\ where the holomorphic and antiholomorphic
momenta always have the same exponents. For instance, the two-loop graph function in (\ref{mgf7}) and the three-loop 
graph function in (\ref{mgf9}) line up as follows with the notation of (\ref{mgf12}):
\begin{align}
C_{a,b,c} = \cform{a&b&c \\a&b&c}\, , \ \ \ \ \ \ 
D_4 = \cform{1&1&1 &1 \\ 1&1&1 &1}
\label{mgf14A}
\end{align}

\subsubsection{Relations among modular graph forms}

The momentum-conserving delta function in (\ref{mgf11}) gives rise to a
variety of algebraic relations among absolutely convergent modular graph forms \cite{DHoker:2016mwo}.
By inserting the vanishing momenta $\sum_{j=1}^R p_j$ or $\sum_{j=1}^R \bar p_j$
into the numerators for dihedral graphs, one finds
\begin{align}
0=\sum_{j=1}^R \cform{a_1&a_2&\ldots &a_j{-}1 &\ldots &a_R \\b_1&b_2&\ldots &b_j &\ldots &b_R}
=\sum_{j=1}^R \cform{a_1&a_2&\ldots &a_j &\ldots &a_R \\b_1&b_2&\ldots &b_j{-}1 &\ldots &b_R} \, .
\label{mgfrel1}
\end{align}
Whenever an edge in a dihedral graph carries the trivial decoration $(a_j,b_j)=(0,0)$, one
can reduce the number of summation variables via
\begin{align}
\cform{a_1&a_2&\ldots &a_{R-1} &0 \\b_1&b_2&\ldots &b_{R-1} &0} 
= \prod_{j=1}^{R-1} \cform{a_j &0 \\b_j &0} - \cform{a_1&a_2&\ldots &a_{R-1}  \\b_1&b_2&\ldots &b_{R-1} }  \, ,
\label{mgfrel2}
\end{align}
where the last term stems from the constraint $\sum_{j=1}^{R-1} p_j\neq0$ on the left-hand side.
With $R=3$ edges, one can use the $R=2$ identity $\cform{a &c \\ b &d} = (-1)^{c+d} \cform{a+c &0 \\ b+d &0}$ to 
reduce the right-hand side to the simplest lattice sums (\ref{mgf13}),
\begin{align}
\cform{a &c &0 \\ b &d &0} = \cform{a &0 \\ b &0}\cform{c &0 \\ d &0} - (-1)^{c+d} \cform{a+c &0 \\ b+d &0} \, .
\label{mgfrel3}
\end{align}
By combinations of (\ref{mgfrel1}) and (\ref{mgfrel2}), one can often reduce the number 
of summation variables as in
\begin{align}
\cform{2 &1 &1 \\ 0 &1 &1} = \cform{4 &0 \\ 2 &0}  \, , \ \ \ \ \ \ 
\cform{2 &1 &1 \\ 1 &1 &0} = - \frac{1}{2} \cform{4 &0 \\ 2 &0} \ ,
\label{mgfrel4}
\end{align}
and a variety of similar identities is spelled out in appendix \ref{app:B1}.

\subsubsection{Holomorphic subgraph reduction}
\label{sec:hsr}

Another opportunity to reduce the number of summation variables in a modular graph form
arises for graphs that have closed subgraphs of only holomorphic momenta \cite{DHoker:2016mwo, Gerken:2018zcy}.
By the methods of \cite{DHoker:2016mwo}, in the dihedral case, such ``holomorphic subgraphs'' can be expressed in terms
of subgraphs without closed loops with coefficients ${\rm G}_{k\geq4}$ or $\hat {\rm G}_2$. 
Given collective labels $A=(a_1,a_2,\ldots,a_{R-2})$ and $B=(b_1,b_2,\ldots,b_{R-2})$ for the decorations
and $a_+{+}a_-\geq 3$, we have
\begin{align}
  \cform{a_{+}&a_{-}&A\\0&0&B}&=\cform{a_{+}&a_{-}\\0&0}\cform{A\\B}-\binom{a_{+} + a_{-}}{a_{-}}\cform{a_{+}+a_{-}&A\\0&B}\notag\\
              &+\sum_{k=4}^{a_{+}}\binom{a_{+}+a_{-}-1-k}{a_{+}-k}
              {\rm G}_k\cform{a_{+}+a_{-}-k&A\\0&B} \label{eq:42} \\
              &+\sum_{k=4}^{a_{-}}\binom{a_{+}+a_{-}-1-k}{a_{-}-k}
              {\rm G}_k\cform{a_{+}+a_{-}-k&A\\0&B}\notag\\
              &+\binom{a_{+}+a_{-}-2}{a_{+}-1}\left\{\hat{\rm G}_{2}\cform{a_{+}+a_{-}-2&A\\0&B}+\cform{a_{+}+a_{-}-1&A\\-1&B}\right\} \, ,\notag
\end{align}
where the entry $-1$ of the last term can be resolved via momentum conservation (\ref{mgfrel1}).
As detailed in \cite{DHoker:2016mwo} and reviewed in appendix \ref{app:Eisreg}, the 
contributions of $\hat {\rm G}_2$ or $\cform{a_{+}+a_{-}-1&A\\-1&B}$ in
the last line stem from Eisenstein-regularized sums which cancel from certain 
linear combination of terms with the same value of $a_+ + a_-$, e.g.
\begin{align}
\cform{2 &2&A\\ 0 &0 &B} - 2 \cform{3 &1&A \\ 0 &0 &B}  &= 2 \cform{4 &A\\0 &B} + 3 {\rm G}_4 \cform{A\\B} \label{eq:42more} \\
\cform{3 &2&A\\ 0 &0 &B} - 3 \cform{4 &1&A \\ 0 &0 &B}  &= 5 \cform{5 &A\\0 &B} - 3 {\rm G}_4 \cform{1 &A\\0 &B} \, . \notag
\end{align}
In fact, some of the integrals in the heterotic-string amplitude will involve the
Eisenstein-regularized contributions, namely
\begin{align}
  \cform{2&1&2\\0&0&1}&=3 \cform{ 5 & 0 \\ 1 & 0 }-{\rm \hat G}_{2} \cform{ 3 & 0 \\ 1 & 0 }-{\rm G}_{4}\notag\\
  \cform{2&2&1\\0&0&1}&=-6 \cform{5&0\\1&0}+2 {\rm \hat G}_{2}\cform{3&0\\1&0}+2{\rm G}_{4} \label{eq:28} \\
  \cform{3&1&1\\0&0&1}&=-4 \cform{5&0\\1&0}+{\rm \hat G}_{2}\cform{3&0\\1&0}+{\rm G}_{4} \notag
\end{align}
and
\begin{align}
  \cform{3&1&1&1\\0&0&1&1}&={\rm G}_{4}{\rm E}_{2} -4 \cform{4&1&1\\0&1&1}+{\rm \hat G}_{2}\cform{2&1&1\\0&1&1}-2\cform{3&1&1\\0&0&1}\notag\\
  \cform{2&2&1&1\\0&0&1&1}&={\rm G}_{4}{\rm E}_{2}-6 \cform{4&1&1\\0&1&1}+2 {\rm \hat G}_{2} \cform{2&1&1\\0&1&1}-4\cform{3&1&1\\0&0&1}\ .\label{eq:20}
\end{align}
The right-hand sides of the identities~\eqref{eq:20} can be further simplified using the relations \eqref{mgfrel4} and \eqref{manyrels} as well as the holomorphic subgraph reductions \eqref{eq:28}.

\subsubsection{Trihedral modular graph forms}

\begin{figure}
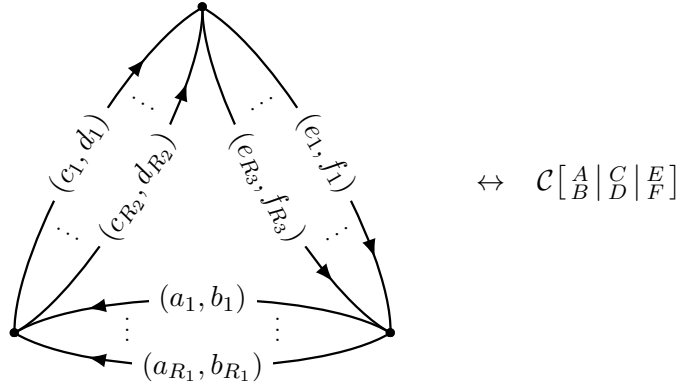

  \begin{center}
    \tikzpicture[scale=1,line width=0.30mm]
    \scope[shift={(-1,0)},decoration={markings,mark=at position 0.8 with {\arrow[scale=1.3]{latex}}}]
    \draw[postaction={decorate}] (5,0)node{\scriptsize $\bullet$} .. controls (4,0.6) and (1,0.6) ..node[fill=white,transform shape]{$(a_1,b_1)$} (0,0)node{\scriptsize $\bullet$};
    \draw (1.5,0.15) node{\scriptsize $\vdots$};
    \draw (3.5,0.15) node{\scriptsize $\vdots$};
    \draw[postaction={decorate}] (5,0)node{\scriptsize $\bullet$} .. controls (4,-0.6) and (1,-0.6) ..node[fill=white,transform shape]{$(a_{R_{1}},b_{R_{1}})$} (0,0);

    \scope[rotate=60]
    \draw[postaction={decorate}] (0,0)node{\scriptsize $\bullet$} .. controls (1,0.6) and (4,0.6) ..node[fill=white,transform shape,sloped]{$(c_1,d_1)$} (5,0)node{\scriptsize $\bullet$};
    \draw (1.5,0.15) node[transform shape]{\scriptsize $\vdots$};
    \draw (3.5,0.15) node[transform shape]{\scriptsize $\vdots$};
    \draw[postaction={decorate}] (0,0)node{\scriptsize $\bullet$} .. controls (1,-0.6) and (4,-0.6) ..node[transform shape,fill=white]{$(c_{R_{2}},d_{R_{2}})$} (5,0);
    \endscope

    \scope[shift={(60:5)},rotate=-60]
    \draw[postaction={decorate}] (0,0)node{\scriptsize $\bullet$} .. controls (1,0.6) and (4,0.6) ..node[fill=white,transform shape]{$(e_1,f_1)$} (5,0)node{\scriptsize $\bullet$};
    \draw (1.5,0.15) node[transform shape]{\scriptsize $\vdots$};
    \draw (3.5,0.15) node[transform shape]{\scriptsize $\vdots$};
    \draw[postaction={decorate}] (0,0)node{\scriptsize $\bullet$} .. controls (1,-0.6) and (4,-0.6) ..node[transform shape,fill=white]{$(e_{R_{3}},f_{R_{3}})$} (5,0);
    \endscope

    \endscope
    \draw(6.5,2)node{$\leftrightarrow \ \ \cformtri{A\\B}{C\\D}{E\\F}$};
    \endtikzpicture
    \caption{Graph associated to a trihedral modular graph form.}
    \label{fig:trihedralMGF}
  \end{center}
\end{figure}

We will also need modular graph forms descending from the trihedral topology.
As depicted in figure \ref{fig:trihedralMGF}, trihedral graphs are characterized by three vertices
and three sets of edges each of which connects a different pair of vertices.
We will denote the decorations for the $R_1,R_2$ and $R_3$ edges in the first,
second and third set by $(a_i,b_i)$, $(c_i,d_i)$ and $(e_i,f_i)$, respectively.
By gathering the decorations $a_1,a_2,\ldots,a_{R_1}$ in a collective label $A$ 
(and a similar convention for $B,C,\ldots,F$), the most general trihedral
modular graph form can be written as
\begin{align}
\cformtri{A\\B}{C\\D}{E\\F} :=\left(\frac{\Im\tau}{\pi}\right)^{|B|+|D|+|F|}\hspace{-1.8em} \sum_{p_1,p_2,\ldots ,p_{R_1} \neq 0 \atop{
k_1,k_2,\ldots ,k_{R_2} \neq 0 \atop{ \ell_1,\ell_2,\ldots ,\ell_{R_3} \neq 0
}}} 
\frac{ \delta\Big(\sum_{i=1}^{R_1} p_i - \sum_{i=1}^{R_2} k_i   \Big)
\delta\Big(\sum_{i=1}^{R_2} k_i - \sum_{i=1}^{R_3} \ell_i   \Big) }{ \Big( \prod_{j=1}^{R_1} p_j^{a_j}\bar p_j^{b_j} \Big)
\Big( \prod_{j=1}^{R_2} k_j^{c_j}\bar k_j^{d_j} \Big)
\Big( \prod_{j=1}^{R_3} \ell_j^{e_j}\bar \ell_j^{f_j} \Big)
 } \, ,
\label{mgf15}
\end{align}
where $|B|:=\sum_{i=1}^{R_{1}}b_{i}$ and similarly for $|D|$ and $|F|$, and the delta 
functions in the numerator enforce that each of the three sets of edges
carries the same overall momentum, $\sum_{i=1}^{R_1} p_i = \sum_{i=1}^{R_2} k_i = \sum_{i=1}^{R_3} \ell_i $. 
Each vertex is taken to connect with more than two edges since the graph 
would otherwise be dihedral rather than trihedral.
Accordingly, the simplest trihedral modular graph function contributing to (\ref{mgf1}) requires integration over five Green functions,
\begin{align}
D_{2,2,1} := \int \dd \mu_4\, G_{12}^2 G_{23}^2 G_{31} =  \cformtri{1&1\\1&1}{1&1\\1&1}{1\\1}\, ,
\label{mgf16}
\end{align}
whereas in the remaining integrals ${\cal I}^{(w,0)}_{\ldots}$ in (\ref{3.6}), trihedral graphs arise from fewer 
Green functions.
The ideas of momentum conservation and holomorphic subgraph reduction can be applied
to generate identities between trihedral modular graph functions and more general graph topologies
\cite{Gerken:2018zcy}. Examples and further details can be found in appendix \ref{app:B3}.

\subsubsection{Modular properties}

Under ${\rm SL}_2(\ZZ)$ transformations, each factor of $p^{-a}$ and $\bar p^{-b}$ in the summand of
modular graph forms contributes modular weight $(a,0)$ and $(0,b)$, respectively, and $\Im\tau$ contributes modular weight $(-1,-1)$, i.e.\ for dihedral graphs
\begin{align}
\cform{A\\B}\Big( \frac{\alpha\tau + \beta}{\gamma\tau + \delta} \Big) 
=(\gamma\tau+\delta)^{|A|-|B|} \cform{A\\B}(\tau)
\label{mgf17}
\end{align}
is a non-holomorphic modular form of weight $(|A|{-}|B|,0)$. Here, we again use the 
notation $|A|=\sum_{i=1}^R a_i$ of \eqref{mgf15}. In the trihedral case, 
the analogous transformation is,
\begin{align}
\cformtri{A\\B}{C\\D}{E\\F}\Big(\frac{\alpha\tau + \beta}{\gamma\tau + \delta} \Big) =
(\gamma\tau+\delta)^{|A|+|C|+|E|-|B|-|D|-|F|}
\cformtri{A\\B}{C\\D}{E\\F}(\tau)\, .
\label{mgf18}
\end{align}
%

\subsection{Differential equations and iterated Eisenstein integrals}
\label{sec:2.5}

A major motivation for the introduction of modular graph forms in \cite{DHoker:2016mwo}
is their descent from modular graph functions under Cauchy--Riemann derivatives
\begin{align}
\nabla := 2i (\Im \tau)^2 \frac{\partial}{\partial \tau} \, , \ \ \ \ \ \ \overline{\nabla} := - 2i (\Im \tau)^2 \frac{\partial}{\partial \bar \tau} \, .
\label{crder1}
\end{align}
These operators shift the weights of a non-holomorphic modular form\footnote{One can extend $\nabla$ to a map between
modular forms of weights $(w_1,w_2) \rightarrow (w_1,w_2{-}2)$ by adding weight-dependent connection terms.} according to $\nabla: \, (0,w) \rightarrow (0,w{-}2)$ and 
$\overline \nabla: \, (w,0) \rightarrow (w{-}2,0)$. By repeated action of $\nabla$ on modular graph functions,
their relations including (\ref{mgf9}) can be elegantly proven based on properties 
of holomorphic modular forms up to integration constants
\cite{DHoker:2016mwo}. Moreover, Cauchy--Riemann equations of 
this type are instrumental to generate the expansions of various modular graph forms around the 
cusp $\tau \rightarrow i\infty$ \cite{Broedel:2018izr}. 

\subsubsection{Cauchy--Riemann derivatives of modular graph forms}

Cauchy--Riemann derivatives of the above lattice sums can be computed from
\begin{align}
\nabla \Big( \frac{ \Im \tau }{p} \Big) = (\Im \tau)^2 \frac{ \bar p}{p^2 } \, , \ \ \ \ \ \ 
\overline\nabla \Big( \frac{ \Im \tau }{\bar p} \Big) = (\Im \tau)^2 \frac{ p}{ \bar p^2 } \, ,
\label{crder2}
\end{align}
which for instance relate non-holomorphic Eisenstein series  to
\begin{align}
\nabla^k {\rm E}_a = \frac{(a{+}k{-}1)! }{ (a{-}1)!}  \frac{ (\Im\tau)^{a+k} }{\pi^a}\sum_{p\neq 0} \frac{1}{p^{a+k} \bar p^{a-k}}
=  \frac{(a{+}k{-}1)!}{(a{-}1)!} \frac{(\Im \tau)^{2k}}{\pi^{k}} \cform{a+k &0 \\ a-k &0} \, .
\label{crder3}
\end{align}
In particular, setting $k=a$ reproduces holomorphic Eisenstein series by (\ref{mgf14}),
\begin{align}
 (\pi \nabla)^k {\rm E}_k = \frac{ (2k{-}1)! }{(k{-}1)!} (\Im \tau)^{2k} {\rm G}_{2k}\, .
 \label{crder4}
\end{align}
The simplest examples of these identities include
\begin{align}
\pi \nabla {\rm E}_2 =   2 (\Im \tau)^2  \cform{3 &0 \\ 1 &0} \, , \ \ \ \ 
\pi \nabla {\rm E}_3 =  3 (\Im \tau)^2  \cform{4 &0 \\ 2 &0} \, , \ \ \ \ 
(\pi \nabla)^2 {\rm E}_3 =  12 (\Im \tau)^4  \cform{5 &0 \\ 1 &0} 
\label{toinv}
\end{align}
as well as
\begin{align}
(\pi \nabla)^2 {\rm E}_2 = 6 (\Im \tau)^4 {\rm G}_4  \, , \ \ \ \ \ \ 
(\pi \nabla)^3 {\rm E}_3 = 60 (\Im \tau)^6 {\rm G}_6 \, .\label{toinvB}
\end{align}
For more general modular graph functions, iterated Cauchy--Riemann derivatives introduce
combinations of holomorphic Eisenstein series and modular graph forms of lower complexity.
For instance, the simplest irreducible two-loop graph function (\ref{mgf7}) obeys
\begin{align}
 {\rm E}_{2,2} := C_{2,1,1} - \frac{9}{10}{\rm E}_4
\, , \ \ \ \ \ \
(\pi \nabla)^3 {\rm E}_{2,2} =  - 6 ( \Im \tau)^4 {\rm G}_4 \, \pi \nabla {\rm E}_2\, ,
 \label{crder5}
\end{align}
where ${\rm G}_4$ on the right-hand side stems from holomorphic subgraph reduction \cite{DHoker:2016mwo},
and the subtraction of ${\rm E}_4$ has been tailored to simplify the differential equation \cite{Broedel:2018izr}.

More generally, the Cauchy--Riemann derivatives of dihedral modular graph forms as defined in (\ref{mgf11}) are
given by \cite{DHoker:2016mwo}
\begin{align}
\pi \nabla \big( (\Im \tau)^{|A|-|B|} \cform{A\\B} \big)&=
(\Im \tau)^{|A|-|B|+2}  \sum_{j=1}^R a_j \cform{a_1&a_2&\ldots &a_j{+}1 &\ldots &a_R \\b_1&b_2&\ldots 
&b_j{-}1 & \ldots &b_R}
\notag \\
\bar \nabla  \cform{A\\B} &=\pi
\sum_{j=1}^R b_j \cform{a_1&a_2&\ldots &a_j{-}1 &\ldots &a_R \\b_1&b_2&\ldots 
&b_j{+}1 & \ldots &b_R}  \, ,  \label{mgf11nabla} 
\end{align}
where the appearance of $\pi$ and $\Im \tau$ is due to our normalization convention (\ref{mgf11})
with holomorphic and antiholomorphic momenta entering on asymmetric footing.
Negative entries $b_j{-}1<0$ on the right-hand side of (\ref{mgf11nabla}) can be avoided using 
momentum conservation
(\ref{mgfrel1}) if the left-hand side has already been simplified using holomorphic subgraph 
reduction (\ref{eq:42}) \cite{DHoker:2016mwo}. Similar differential equations for trihedral and more 
general modular graph forms are a straightforward consequence of (\ref{crder2}).

\subsubsection{Iterated Eisenstein integrals}

Given that repeated $\tau$-derivatives of modular graph forms introduce factors 
of ${\rm G}_k$, it is natural to attempt to express modular graph forms in terms of iterated 
Eisenstein integrals with $k_1,k_2,\ldots,k_r  \in 2\NN_0$,
\begin{align}
{\cal E}_0(k_1,k_2,\ldots,k_r;\tau) &:= 2\pi i \int_\tau^{i \infty} \dd \tau_r \, \frac{ {\rm G}^0_{k_r}(\tau_r) }{(2\pi i)^{k_r}} \, {\cal E}_0(k_1,k_2,\ldots,k_{r-1};\tau_r)  \label{crder6} \\
&\phantom{:}=(-1)^r \! \! \! \! \!  \int \limits_{0\leq q_1\leq q_2\leq \ldots \leq q_r \leq q} \! \! \! \! \! 
 \frac{ \dd q_1}{q_1} \, \frac{ \dd q_2}{q_2} \, \ldots \, \frac{ \dd q_r }{q_r} \, 
  \frac{ {\rm G}^0_{k_1}(\tau_1) }{(2\pi i)^{k_1}} \,  \frac{ {\rm G}^0_{k_2}(\tau_2) }{(2\pi i)^{k_2}}  \,\ldots \,  \frac{ {\rm G}^0_{k_r}(\tau_r) }{(2\pi i)^{k_r}} \, .\notag
\end{align}
The recursive definition in the first line is based on ${\cal E}_0(;\tau) := 1$, and
we have subtracted the zero modes in the $q$-expansion of the Eisenstein 
series ${\rm G}_{k}$ with even weight $k\neq 0$,
\begin{align}
{\rm G}^0_k := {\rm G}_k - 2 \zeta_k = \frac{2 (2\pi i)^k}{(k{-}1)!} \sum_{m,n=1}^\infty m^{k-1} q^{mn}
\, ,  \ k \neq 0 \, , \ \ \ \ \ \ 
{\rm G}_0^0 := -1 \, .
 \label{crder6a}
\end{align}
In this way, iterated Eisenstein integrals (\ref{crder6}) are convergent if $k_1\neq 0$,
and their expansion around the cusp can be inferred from straightforward integration \cite{Broedel:2015hia}
\begin{align}
&{\cal E}_0(k_1,0^{p_1-1},k_2,0^{p_2-1},\ldots,k_r,0^{p_r-1};\tau) = (-2)^r \Big( \prod_{j=1}^r \frac{ 1}{(k_j{-}1)!} \Big) 
 \label{crder7} \\
 & \ \ \ \ \times \sum_{m_i,n_i=1}^\infty \frac{ m_1^{k_1-1} m_2^{k_2 -1} \ldots m_r^{k_r-1} \, q^{m_1n_1+m_2n_2+\ldots +m_r n_r} }{(m_1 n_1)^{p_1} \,(m_1n_1{+}m_2n_2)^{p_2} \ldots (m_1n_1{+}m_2n_2{+}\ldots{+}m_rn_r)^{p_r}} \ ,
 \notag
\end{align}
where $k_j\neq 0 \ \forall \ j=1,2,\ldots,r$, and $0^p$ is a shorthand for $p$ successive zeros. The 
number $r$ of non-zero entries on the left-hand side
of (\ref{crder7}) is also referred to as the {\it depth} of an iterated Eisenstein integral as in~\eqref{crder6}. With these definitions, (\ref{crder4}) can be
integrated\footnote{In performing these integrations, it is useful to note that \cite{Broedel:2018izr}
$$
\pi \nabla(y^n) = n y^{n+1} \, , \ \ \ \ \ \ \pi \nabla\big( {\cal E}_0(k_1,k_2,\ldots,k_r) \big) = \frac{ 4 y^2 \, {\rm G}_{k_r}^0 }{(2\pi i)^{k_r}} \, {\cal E}_0(k_1,k_2,\ldots,k_r)
\, .$$} to the following representations of non-holomorphic Eisenstein 
series \cite{Ganglzagier, DHoker:2015wxz, Broedel:2018izr}
\begin{align}
{\rm E}_2&=  \frac{y^2}{45} + \frac{ \zeta_3}{y}   - 12 \Re[ {\cal E}_0({4, 0}) ] - \frac{6}{y} \Re[{\cal E}_0({4, 0, 0}) ] \,,
 \notag\\
{\rm E}_3&= \frac{2y^3}{945} + \frac{3 \zeta_5}{4y^2}  
  -120 \Re[ {\cal E}_0({6, 0, 0}) ]- \frac{180 }{y} \Re[ {\cal E}_0({6, 0, 0, 0})] - \frac{ 90}{y^2} \Re[ {\cal E}_0({6, 0, 0, 0, 0}) ]
\notag \\
{\rm E}_4&=  \frac{y^4}{4725} + \frac{5 \zeta_7}{8 y^3}
  -1680 \Re [{\cal E}_0({8, 0, 0, 0}) ]- \frac{5040 }{y} \Re [{\cal E}_0({8, 0, 0, 0, 0})]   \label{nholo3} \\
& \ \ \ \ \ - 
\frac{ 6300 }{y^2} \Re [ {\cal E}_0({8, 0, 0, 0, 0, 0}) ] - \frac{ 3150 }{y^3} \Re [ {\cal E}_0({8, 0, 0, 0, 0, 0, 0}) ] \, .
\notag
\end{align}
Here and in later equations, we are using the shorthand
\begin{align}
y := \pi \, \Im \tau \, ,
\label{crder8}
\end{align}
and the Riemann zeta values in (\ref{nholo3}) arise as integration constants 
whose coefficients can be determined from modular invariance. The modular transformations of the iterated Eisenstein integrals~\eqref{crder6} has been discussed in~\cite{Brown:mmv, Broedel:2018izr}.

For instance, integration of (\ref{crder5})
yields the following representation of the simplest irreducible two-loop graph function \cite{Broedel:2018izr}
\begin{align} 
{\rm E}_{2,2} &= -\frac{ y^4}{20250} + \frac{y \zeta_3}{45}+ \frac{ 5 \zeta_5}{12 y}  - \frac{ \zeta_3^2}{4 y^2} -
\Big(\frac{ 2y}{15}  - \frac{ 3  \zeta_3}{y^2} \Big) \Re[ {\cal E}_0(4, 0, 0) ]   \notag \\
&- \frac{9 \Re[ {\cal E}_0(4, 0, 0) ] ^2}{y^2}
 - 72 \Re[ {\cal E}_0(4, 4, 0, 0) ]    -  \frac{1}{5} \Re[ {\cal E}_0(4, 0, 0, 0) ]   \label{newCR8}\\
& - \frac{36 \Re[ {\cal E}_0(4, 0, 4, 0, 0) ] }{y} 
 - \frac{ 108 \Re[ {\cal E}_0(4, 4, 0, 0, 0) ] }{y} 
 - \frac{ \Re[ {\cal E}_0(4, 0, 0, 0, 0) ] }{10 y} \, , \notag
\end{align}
where the coefficients of the zeta values are again fixed by modular invariance. 
The corresponding ${\cal E}_0$-representations of the first Cauchy--Riemann 
derivatives $\pi \nabla {\rm E}_{\ldots}$ are spelled out in (\ref{service4}) below.

All the modular graph forms ${\cal C}[\Gamma] $ that will be encountered in the $\ap$-expansions
of section~\ref{sec:3} have been reduced to iterated Eisenstein integrals 
with Laurent polynomials in $y$ as their coefficients. In these cases, the
$q$-series representations (\ref{crder7}) of iterated Eisenstein integrals give rise to an
expansion
\begin{align}
{\cal C}[\Gamma] = \sum_{m,n=0}^{\infty} c^\Gamma_{m,n}(y) q^m \bar q^n \, ,
\label{crder9}
\end{align}
where $\Gamma$ may represent the labels of the dihedral and trihedral cases 
(\ref{mgf11}) and (\ref{mgf15}) or a more general graph. The coefficients $c^\Gamma_{m,n}(y)$ 
are Laurent polynomials in $y$, and the terms in (\ref{nholo3}) and (\ref{newCR8})
without any factor of $\Re {\cal E}_0$ reproduce the zero modes $c^\Gamma_{0,0}(y)$ of 
the simplest modular graph functions \cite{Green:2008uj}. 

From the methods of \cite{DHoker:2016mwo}, an expansion of the form (\ref{crder9}) is 
expected to exist for {\it any} modular graph form, and a proof has been given in 
\cite{Zerbini:2015rss, Zerbini:2018sox} for the special case of modular graph functions.
Indeed, the proof of the references can be adapted to general modular graph forms\footnote{We
are grateful to Federico Zerbini for discussions on this point.},
including global bounds (which only depend on the graph $\Gamma$ but not
on $m$ and $n$) on the highest and lowest powers of $y$ in the above 
$c^\Gamma_{m,n}(y) $. Note that, for modular graph functions,
the Laurent coefficients within $c^\Gamma_{m,n}(y) $ are proven 
to be $\mathbb Q$-linear combinations of cyclotomic MZVs
and conjectured to be single-valued MZVs \cite{Zerbini:2015rss, Zerbini:2018sox}.

A major motivation for expressing modular graph forms in terms of iterated Eisenstein integrals
is the connection with open-string amplitudes. As will be detailed in section \ref{sec:4}, the $\ap$-expansion 
of the iterated integrals over open-string punctures yields elliptic multiple zeta 
values \cite{Broedel:2014vla, Broedel:2017jdo} which are in turn expressible in terms 
of the ${\cal E}_0$ in (\ref{crder6}) \cite{Enriquez:Emzv, Broedel:2015hia}.
The representations (\ref{nholo3}) and (\ref{newCR8}) of modular graph functions have been 
connected with open-string quantities through a candidate prescription for an elliptic
single-valued map \cite{Broedel:2018izr}. In section~\ref{sec:4}, we will find a similar 
correspondence between open strings and the integral ${\cal I}^{(2,0)}_{1234}$ in (\ref{3.3})
which enters the $\tau$-integrand (\ref{3.6}) of the heterotic-string amplitude.


\section{$\alpha'$-expansion of the heterotic-string amplitude and modular graph forms}
\label{sec:3}

This section is dedicated to the low-energy expansion of the four-point integrals ${\cal I}^{(w,0)}_{\ldots}$
defined in (\ref{3.2}) to (\ref{3.5}) which arise in the gauge sector of the heterotic string. We will give general arguments
in section \ref{sec:3.0} that each order in their $\alpha'$-expansion is expressible via modular graph forms.
The low-energy expansion of the modular invariant integral ${\cal I}^{(0,0)}$ is well-studied in the literature, see 
(\ref{mgf10}) for the leading orders. The integrals ${\cal I}^{(w,0)}_{1234}$ and ${\cal I}^{(w,0)}_{12|34}$
from the planar and the non-planar sector of (\ref{3.6}) are expanded in sections \ref{sec:3.1} and \ref{sec:3.3}, respectively. We will elaborate on their transcendentality properties in 
section \ref{sec:unif-transc} and
perform the integrals over $\tau$ in the amplitude (\ref{3.0}) up to the
order of $\ap^2$ in section \ref{sec:tauint}.
Finally, it will be explained in section \ref{sec:3.8} why the integrals over the punctures 
in all massless $n$-point amplitudes boil down
to modular graph forms.


\subsection{$\ap$-expansions from modular graph forms}
\label{sec:3.0}

Our constructive method to extract modular graph forms from the $\ap$-expansions of
(\ref{3.2}) to (\ref{3.5}) starts from the double Fourier expansion of the
non-holomorphic Kronecker--Eisenstein series (\ref{1.1}),
\begin{equation}
\Omega(z,\beta,\tau) = \sum_{m,n\in \ZZ} \frac{ e^{2\pi i (mv - nu)} }{m\tau + n + \beta} \ ,
\label{1.4}
\end{equation}
where again $z=u\tau+v$.
After removing the contribution $\beta^{-1}$ due the origin $(m,n)=(0,0)$, a geometric-series expansion in $\beta$ implies the lattice-sum representation of the
doubly-periodic functions in (\ref{1.1b}),
\begin{equation}
f^{(w)}(z,\tau) = (-1)^{w-1}   \sum_{(m,n)\neq (0,0)} \frac{ e^{2\pi i (mv - nu)} }{(m\tau + n)^w}
\ , \ \ \ \ \ \ w\geq 1\, ,
\label{1.5}
\end{equation}
which exposes their modular properties (\ref{1.1mod}).
The lattice sums at $w=1,2$ are not absolutely convergent on the entire torus, and
we will comment on regularization prescriptions in section~\ref{sec:3.3}. Still, the 
sum at $w=1$ is formally consistent with \eqref{eq:27} by taking the holomorphic derivative
\begin{align}
  -\frac{\Im \tau}{ \pi}\partial_z \frac{ e^{2\pi i (mv - nu)} }{|m\tau + n|^{2}}=\frac{e^{2\pi i (mv - nu)}}{m\tau+n}
\end{align}
of the summand of the Green function \eqref{mgf4}. The single pole of $f^{(1)}(z,\tau)$ at $z=0$ mentioned below \eqref{1.1b} is not obvious in the lattice-sum representation \eqref{1.5} since the limit $z\rightarrow0$ does not commute with the sum.

The lattice-sum representation (\ref{1.5}) manifests that the holomorphic
Eisenstein series ${\rm G}_k$ in (\ref{1.6eis}) are recovered when evaluating 
the even-weight $f^{(w)}$ at the origin,
\begin{equation}
 f^{(k)}(0,\tau)  =- \sum_{(m,n) \neq (0,0)} \frac{1}{(m\tau + n)^{k}} =-{\rm G}_k(\tau) ,\qquad k\geq 4   \ .
\label{1.6}
\end{equation}
At $w=2$, however, the limit $u,v\rightarrow 0$ introduces conditionally convergent sums,
and one can obtain a holomorphic but non-modular expression (\ref{1.6AA}).

\subsubsection{Fourier expansion of the $V_w$ functions}

The elliptic integrands $V_w$ of ${\cal I}^{(w,0)}_{\ldots}$ are defined by cyclic products of
Kronecker--Eisenstein series, see (\ref{1.1a}). Hence, the dependence of  $V_w$ on the punctures 
can be brought into the same plane-wave form (\ref{1.4}) as the Green function (\ref{mgf4}). In the planar cases 
${\cal I}^{(w,0)}_{1234}$ with $w=2,4$, for instance, the Fourier expansion (\ref{1.4}) propagates to
\begin{align}
&V_w(1,2,3,4) = \! \! \! \! \! \! \! \! \sum_{m_1,m_2,m_3,m_4 \in \ZZ \atop{n_1,n_2,n_3,n_4 \in \ZZ}} \! \frac{ 1 }{(m_1\tau{+}n_1{+}\beta)
(m_2\tau{+}n_2{+}\beta)(m_3\tau{+}n_3{+}\beta)(m_4\tau{+}n_4{+}\beta)} \Big|_{\beta^{w-4}}
 \label{2.13} \\
& \  \times e^{2\pi i  \big[(m_1{-}m_4) v_{12}{+}(m_2{-}m_4) v_{23} {+} (m_3{-}m_4) v_{34}{-}
(n_1{-}n_4) u_{12}{-}(n_2{-}n_4) u_{23} {-} (n_3{-}n_4) u_{34} \big]} \notag
\end{align}
with $z_j = u_j \tau + v_j$ as well as $u_{ij}= u_i-u_j$ and $v_{ij}= v_i-v_j$. With
the following notation for the scalar product involving discrete momenta $p_i=m_i\tau +n_i$,
\begin{align}
\langle p_i,z_j \rangle := m_i v_j - n_i u_j \, ,
\label{notA}
\end{align}
the representation (\ref{2.13}) of the $V_w$ functions shares the lattice sums
of modular graph forms, see section \ref{sec:2.4},
\begin{align}
V_w(1,2,3,4) = \sum_{p_1,p_2,p_3,p_4} \frac{e^{2\pi i(  
\langle p_1{-}p_4, z_{12} \rangle + \langle p_2{-}p_4, z_{23} \rangle+\langle p_3{-}p_4, z_{34} \rangle
) }}{(p_1{+}\beta) (p_2{+}\beta) (p_3{+}\beta) (p_4{+}\beta)}  \Big|_{\beta^{w-4}} \, .
\label{notB}
\end{align}
Laurent expansion in $\beta$ will generate arbitrary powers of inverse holomorphic momenta,
\begin{align}
\sum_{p} \frac{1}{p+\beta}
=\frac{1}{\beta}+\sum_{p\neq 0} \frac{1}{p+\beta}
=\frac{1}{\beta}+\sum_{p\neq 0}  \Big(\frac{1}{p} -\frac{ \beta}{p^2} + \frac{\beta^2}{p^3} - \frac{\beta^3}{p^4}+\ldots  \Big)
=\frac{1}{\beta}+\sum_{p\neq 0}  \sum_{a=1}^\infty \frac{ (-\beta)^{a-1}  }{p^a}\, ,
\label{notC}
\end{align}
which correspond to decorated edges $(a,0)$ in the
discussion of modular graph forms. However, in contrast to the lattice-sum definition
of modular graph forms, the sums over $p_j$ in (\ref{notB}) include the origin and must
be decomposed according to $\sum_{p_j} f(p_j)= f(0)+\sum_{p_j\neq 0} f(p_j)$ before
making contact with modular graph forms.

The integrands in the non-planar sector of (\ref{3.6}) boil down to $V_2(i,j)$ with the even simpler Fourier expansion
\begin{align}
  V_{2}(i,j)&=-\wp(z_{ij},\tau)={\rm \hat G}_{2} -\partial_{z_{ij}}^{2}G(z_{ij},\tau)  \notag \\
  &={\rm \hat G}_{2}   -\frac{\pi}{\Im \tau}\sum_{(m,n)\neq(0,0)}\frac{m\bar\tau+n}{m\tau+n} \, e^{2\pi i(mv_{ij}-nu_{ij})}
  \label{2.999} \\
&={\rm \hat G}_{2}   -\frac{\pi}{\Im \tau}  \,  \sum_{p \neq 0}  \,  \frac{\bar p}{p} \, e^{2\pi i \langle p,z_{ij} \rangle}
\, .  \notag
\end{align}

\subsubsection{Heterotic graph forms}

By (\ref{2.999}) and (\ref{notB}), each term in the Taylor expansion of the integrals ${\cal I}^{(w,0)}_{\ldots}$
 in (\ref{3.2}) to (\ref{3.5}) boils down to straightforward Fourier integrals over $z_j = u_j \tau + v_j$,
\begin{align}
\int \dd \mu_4 &=  \prod_{j=1}^3 \int_0^1 \dd u_{j,j+1} \int_0^1 \dd v_{j,j+1}  \ \ \Longrightarrow \ \
\int \dd \mu_4 \, e^{2\pi i (  \langle p, z_{12} \rangle + \langle k, z_{23} \rangle+\langle \ell, z_{34} \rangle )}=\delta(p)\delta(k) \delta(\ell)
\label{2.14}
\end{align}
which collapse some of the above sums over $p_i$ through the delta functions on the right-hand side. 
Hence, integrals of this type over (\ref{2.999}) and each term in the 
Laurent expansion of (\ref{notB}) w.r.t.\ $\beta $ line up with the framework of modular 
graph forms: Different types of decorated edges arise from the Green functions $G_{ij}$ of the 
Koba--Nielsen factor, the Laurent-expanded $\frac{1}{p_j{+}\beta}$ 
in (\ref{notB}) and the ratio $\frac{\bar p}{p}$ in (\ref{2.999}).
Moreover, the integrations (\ref{2.14}) yield momentum conserving vertices.

In order to track the different contributions from the expanded Koba--Nielsen factor to ${\cal I}^{(w,0)}_{1234}$,
we introduce the following function of $\beta$ for each monomial in Green functions $G_{ij}$,
\begin{align}
H_{1234}\Big[\prod_{i<j} G_{ij}^{n_{ij}} ;\beta\Big] 
&=  \int \dd \mu_4\,  \Omega(z_{12},\beta,\tau)  \Omega(z_{23},\beta,\tau) 
 \Omega(z_{34},\beta,\tau)  \Omega(z_{41},\beta,\tau) \, \prod_{i<j} G_{ij}^{n_{ij}}
\notag \\
&= \! \!   \sum_{p_1,p_2,p_3,p_4} \int \dd \mu_4 \frac{   e^{2\pi i(  
\langle p_1{-}p_4, z_{12} \rangle + \langle p_2{-}p_4, z_{23} \rangle+\langle p_3{-}p_4, z_{34} \rangle
) } }{  (p_1{+}\beta) (p_2{+}\beta) (p_3{+}\beta) (p_4{+}\beta)  }  \, \prod_{i<j} G_{ij}^{n_{ij}} 
\label{hetgraph1} \\
&= H^{(0,0)}_{1234}\Big[\prod_{i<j} G_{ij}^{n_{ij}} \Big] \frac{1}{\beta^4} 
+ H^{(2,0)}_{1234}\Big[\prod_{i<j} G_{ij}^{n_{ij}} \Big] \frac{1}{\beta^2} 
+  H^{(4,0)}_{1234}\Big[\prod_{i<j} G_{ij}^{n_{ij}} \Big]   +{\cal O}(\beta^{2}) \, . \notag
\end{align}
Since $H_{1234}[\ldots;\beta]$ is an even function of $\beta$, odd powers of $\beta$ are absent in the Laurent expansions in
the third line. The latter define the 
contributions $H^{(w,0)}_{1234}[\ldots]$ relevant to ${\cal I}^{(w,0)}_{1234}$ 
(with $w=0,2,4$ and $V_{0}(1,2,3,4)=1$),
\begin{align}
  H^{(w,0)}_{1234}\Big[\prod_{i<j} G_{ij}^{n_{ij}} \Big] &= \int \dd \mu_4  \, V_w(1,2,3,4)  \, \prod_{i<j} G_{ij}^{n_{ij}} \ .\label{eq:1}
\end{align}
With the above definitions, the leading orders of the $\ap$-expansions of ${\cal I}^{(w,0)}_{1234}$ read
\begin{align}
{\cal I}^{(w,0)}_{1234}(s_{ij},\tau) &= H^{(w,0)}_{1234}[\emptyset]+\! \! \sum_{1\leq i<j}^4 \! \! s_{ij}  H^{(w,0)}_{1234}[G_{ij}]  +\frac{1}{2}  \sum_{1\leq i<j \atop{1\leq k<l}}^4 \! \! s_{ij} s_{kl}  H^{(w,0)}_{1234}[G_{ij} G_{kl}]+{\cal O}(\ap^3) \, . \label{hetgraph2}
\end{align}
Analogous definitions can be made in the non-planar sector, where the bookkeeping variable
$\beta$ can be bypassed from the simple form (\ref{2.999}) of $V_2(i,j)$,
\begin{align}
H^{(4,0)}_{12|34}\Big[\prod_{i<j} G_{ij}^{n_{ij}} \Big] &= \int \dd \mu_4  \, V_2(1,2) V_2(3,4)  \, \prod_{i<j} G_{ij}^{n_{ij}} 
\label{hetgraph3} \\
H^{(2,0)}_{12|34}\Big[\prod_{i<j} G_{ij}^{n_{ij}} \Big] &=  \int \dd \mu_4  \, \big[ V_2(1,2)+V_2(3,4)\big]   \, \prod_{i<j}  G_{ij}^{n_{ij}} \, ,
\notag
\end{align}
and (\ref{hetgraph2}) with $1234\rightarrow 12|34$ applies to
the $\ap$-expansions of the non-planar integrals ${\cal I}^{(w,0)}_{12|34} $.

By analogy with the definition of modular graph functions through the $\ap$-expansion of 
${\cal I}^{(0,0)}$ in (\ref{mgf1}), we will refer to the above $H^{(w,0)}_{1234}[\ldots]$ 
and $H^{(w,0)}_{12|34}[\ldots]$ as {\it heterotic graph forms}.
They are modular graph forms of weight $(w,0)$, as one can see from modular 
invariance of $G_{ij}$ and $\dd \mu_4$ and the weight $(w,0)$ of $V_w$. Modular graph 
functions are recovered from the weight-zero instances $H^{(0,0)}_{1234}[\ldots]$.

Note that similar techniques have been applied in \cite{Stieberger:2002wk}\footnote{The integrations in \cite{Stieberger:2002wk} are performed before summing over
the spin structure (\ref{1.06}), based on a double Fourier expansion of the Szeg\"o kernel similar to (\ref{1.4}).}
to evaluate certain integrals 
over the punctures in five- and six-point gauge correlators at $\exp\left(\sum_{i<j}s_{ij}G_{ij}\right)\rightarrow1$ 
that demonstrate the absence of ${\rm Tr}(F^5)$ and 
${\rm Tr}(F^6)$ operators in the one-loop effective action.


\subsection{Planar heterotic graph forms}
\label{sec:3.1}

In this section, we compute and simplify the planar heterotic graph forms 
that arise from the Koba--Nielsen integrals over $V_w(1,2,3,4)$ at the leading orders in $\ap$ and $w=2,4$.
These results follow from Laurent expansion of (\ref{hetgraph1}), but one can equivalently
employ the representation of the $V_w$ functions in terms of the doubly-periodic
$f^{(w)}$ with Fourier expansion in (\ref{1.5}). In a shorthand with $f^{(w)}_{ij}:= f^{(w)}(z_i{-}z_j,\tau)$, 
the relevant integrands are
\begin{align}
V_2(1,2,3,4) &=  f^{(1)}_{12}f^{(1)}_{34}+ f^{(1)}_{23}f^{(1)}_{41}+\big[ f^{(2)}_{12}+  f^{(1)}_{12}f^{(1)}_{23}+ {\rm cyc}(1,2,3,4) \big]
\label{hetgraph4}\\
V_4(1,2,3,4) &= f^{(1)}_{12}f^{(1)}_{23}f^{(1)}_{34}f^{(1)}_{41} +
f^{(1)}_{12}f^{(3)}_{34}+f^{(2)}_{12}f^{(2)}_{34}+f^{(3)}_{12}f^{(1)}_{34}+
f^{(1)}_{23}f^{(3)}_{41}+f^{(2)}_{23}f^{(2)}_{41}+f^{(3)}_{23}f^{(1)}_{41} \notag\\
& \hspace{-2cm}+ \big[f^{(1)}_{12}f^{(1)}_{23}f^{(2)}_{34}{+}f^{(1)}_{12}f^{(2)}_{23}f^{(1)}_{34}{+}f^{(2)}_{12}f^{(1)}_{23}f^{(1)}_{34}
{+} f^{(1)}_{12}f^{(3)}_{23}{+} f^{(2)}_{12}f^{(2)}_{23}{+} f^{(3)}_{12}f^{(1)}_{23} {+} f^{(4)}_{12} 
+ {\rm cyc}(1,2,3,4) \big] \, . \notag
\end{align}
By permutation symmetry of the Koba--Nielsen factor, the dihedral symmetry (\ref{1.1symm})
of the $V_w$ functions propagates to the integrals ($w=2,4$),
\begin{align}
{\cal I}^{(w,0)}_{1234}={\cal I}^{(w,0)}_{4321} \, , \ \ \ \ \ \ {\cal I}^{(w,0)}_{1234}={\cal I}^{(w,0)}_{2341} \, .
\label{hetgraph5}
\end{align}
On these grounds, only a small fraction of heterotic graph forms at a given order in $\ap$ are inequivalent
under dihedral symmetry, see table \ref{tableplan}.
\begin{table}
  \begin{center}
    \begin{tabular}{cc}  
      \toprule
      Order &Inequivalent planar heterotic graph forms\\
      \midrule
      $\ap^0$ &$H_{1234}^{(w,0)}[\emptyset]$\\
      \addlinespace
      $\ap^1$ &$H_{1234}^{(w,0)}[G_{12}] ,\  H_{1234}^{(w,0)}[G_{13}]$\\
      \addlinespace
      $\ap^2$ &$H_{1234}^{(w,0)}[G^2_{12}] , \  H_{1234}^{(w,0)}[G_{13}G_{24}]  , \   H_{1234}^{(w,0)}[G_{12}G_{13}]$ \\
               & $H_{1234}^{(w,0)}[G^2_{13}] ,\    H_{1234}^{(w,0)}[G_{12}G_{34}]  , \   H_{1234}^{(w,0)}[G_{12}G_{23}]$\\
      \addlinespace
      $\ap^3$ &$H_{1234}^{(w,0)}[G^3_{12}], \ H_{1234}^{(w,0)}[G^2_{13}G_{24}]  , \  H_{1234}^{(w,0)}[G_{12}G_{23} G_{34}]   ,\ H_{1234}^{(w,0)}[G_{12}G_{23} G_{31}]$ \\
               &$H_{1234}^{(w,0)}[G^3_{13}] ,\ H_{1234}^{(w,0)}[G^2_{12}G_{34}]  , \ H_{1234}^{(w,0)}[G_{12}G_{13} G_{14}]  , \ H_{1234}^{(w,0)}[G_{12}G_{13} G_{34}]  $\\
               &$H_{1234}^{(w,0)}[G^2_{12}G_{13}] , \, \, H_{1234}^{(w,0)}[G_{12}G^2_{13}] , \; \, H_{1234}^{(w,0)}[G^2_{12}G_{23}] , \, \, H_{1234}^{(w,0)}[G_{12}G_{13} G_{24}]$\\
      \bottomrule
    \end{tabular}
    \caption{Inequivalent planar heterotic graph forms with respect to the dihedral symmetry.}
    \label{tableplan}
  \end{center}
\end{table}
In the remainder of this subsection, we will restrict our attention to the heterotic graph forms in the table.

\subsubsection{Leading orders $\ap^0$ and $\ap^1$}

In the absence of any $G_{ij}$ in the integrand, we obtain the simplest heterotic graph forms
\begin{align}
  H_{1234}^{(2,0)}[\emptyset] = 0\, , \ \ \ \ \ \ 
    H_{1234}^{(4,0)}[\emptyset] = {\rm G}_4\, ,
\end{align}
where the Eisenstein series ${\rm G}_4$ on the right-hand side can be traced back to the contribution 
of $f^{(1)}_{12}f^{(1)}_{23}f^{(1)}_{34}f^{(1)}_{41}$ in (\ref{hetgraph4}), see \cite{Lerche:1987qk}. 
At first order in $\ap$, the two inequivalent heterotic graph forms in table \ref{tableplan} are found to be
\begin{align}
  H_{1234}^{(2,0)}[G_{12}]  = - \cform{3&0\\1&0} \, , \ \ \ \ \ \
  H_{1234}^{(2,0)}[G_{13}]  = 2\cform{3&0\\1&0}  \ ,\label{eq:4}
  \end{align}
as well as
\begin{align}
  H_{1234}^{(4,0)}[G_{12}]  = - \cform{3&1&1\\0&0&1} - 4\cform{5&0\\1&0} \, , \ \ \ \ \ \
  H_{1234}^{(4,0)}[G_{13}]  =  \cform{2&2 &1\\0&0&1} + 6\cform{5&0\\1&0}  \ , \label{eq:4a}
\end{align}
see (\ref{mgf11}) for the dihedral modular graph forms ${\cal C}[\ldots]$. 
The modular graph forms associated with three-edge graphs can be simplified via
holomorphic subgraph reduction (\ref{eq:28}), and the $H_{1234}^{(4,0)}[G_{ij}]$
can be expressed solely in terms of single lattice sums (just like the $H_{1234}^{(2,0)}[G_{ij}]$),
\begin{align}
  H_{1234}^{(4,0)}[G_{12}]  = -{\rm \hat G}_{2}\cform{3&0\\1&0}-{\rm G}_{4}\, , \ \ \ \ \ \
  H_{1234}^{(4,0)}[G_{13}]  =  2 {\rm \hat G}_{2}\cform{3&0\\1&0}+2{\rm G}_{4} \, .  \label{eq:4b}
\end{align}

\subsubsection{Subleading orders $\ap^2$ and $\ap^3$}\label{sec:subl-orders-ap2}

At the order $\ap^2$, the
six inequivalent heterotic graph forms in table~\ref{tableplan} evaluate to
\begin{align}
  H_{1234}^{(2,0)}[G_{12}^2]  &= - \cform{2& 1&1\\0 &1&1} \notag \\
  H_{1234}^{(2,0)}[G_{13}^2]  &=  2\cform{2&1&1\\0&1&1} \notag\\
  H_{1234}^{(2,0)}[G_{13}G_{24}]  &= - 2\cform{4&0\\2&0} \label{eq:6} \\
  H_{1234}^{(2,0)}[G_{13}G_{12}]  &= - \cform{4&0\\2&0}-\cform{2&1&1\\1&1 &0} \notag \\
  H_{1234}^{(2,0)}[G_{12}G_{34}]  &=  \cform{4&0\\2&0} \notag \\
  H_{1234}^{(2,0)}[G_{12}G_{23}]  &=  \cform{4&0\\2&0}   \ , \notag
\end{align}
as well as
\begin{align}
  H_{1234}^{(4,0)}[G_{12}^2]  &= - 4\cform{4&1&1\\0&1&1}- \cform{3&1&1&1\\0 &0& 1&1} \notag\\
  H_{1234}^{(4,0)}[G_{13}^2]  &=  6\cform{4&1&1\\0&1&1}+\cform{2&2&1&1\\0&0&1&1} \notag \\
  H_{1234}^{(4,0)}[G_{13}G_{24}]  &=  -2\cform{6&0\\2&0}-4\cform{2&2&2\\1&1&0}+8\cform{3&2&1\\1&1&0}+ {\cal C}_{\rm tet}  \label{eq:11}\\
  H_{1234}^{(4,0)}[G_{13}G_{12}]  &=  -\cform{6&0\\2&0}+\cform{3&2&1\\0&1&1}-\cform{4&1&1\\1&1&0}+\cformtri{2&1\\0&1}{1&1\\1&0}{1\\0}-\left(\cform{3&0\\1&0} \,\right)^2   \notag \\
  H_{1234}^{(4,0)}[G_{12}G_{34}]  &=  3\cform{6&0\\2&0}+4\cform{4&1&1\\1&1&0}-2\cform{3&2&1\\1&0&1}+\left(\cform{3&0\\1&0} \,\right)^{2}+\cformtri{1&1\\1&0}{1&1\\1&0}{2\\0}  \notag \\
  H_{1234}^{(4,0)}[G_{12}G_{23}]  &= 3\cform{6&0\\2&0}+4\cform{4&1&1\\1&1&0}-2\cform{3&2&1\\1&0&1}+\left(\cform{3&0\\1&0} \,\right)^{2}+\cformtri{1&1\\1&0}{1&1\\1&0}{2\\0}\ .  \notag
\end{align}
At this order, the contribution of $f^{(1)}_{12}f^{(1)}_{23}f^{(1)}_{34}f^{(1)}_{41}$ to $V_4(1,2,3,4)$
introduces trihedral modular graph forms (\ref{mgf15}) as well as the following tetrahedral representative in 
$H_{1234}^{(4,0)}[G_{13}G_{24}] $
\begin{equation}
 {\cal C}_{\rm tet}  := \left(\frac{\Im\tau}{\pi}\right)^{2}\hspace{-1em}
\sum_{\substack{p_1,p_2,p_3\neq0\\p_{1}+p_{3}\neq0\\p_{2}+p_{3}\neq0\\p_{1}+p_{2}+p_{3}\neq0}}\frac{1}{|p_1|^2|p_{2}|^{2}p_{3}(p_{1}+p_{3})(p_{2}+p_{3})(p_{1}+p_{2}+p_{3})}  \ .\label{tetgraph}
\end{equation}
Similar to (\ref{eq:4b}), the complexity of the lattice sums in $H_{1234}^{(w,0)}[G_{ij}G_{kl}]$ can be considerably reduced:
The three-edge sums at $w=2$ boil down to $\cform{4 &0 \\2 &0}$ via (\ref{mgfrel4}), e.g.
\begin{align}
 H_{1234}^{(2,0)}[G_{12}^2]  &= - \cform{4 &0 \\2 &0} \, ,&
 H_{1234}^{(2,0)}[G_{13}G_{12}] &= -\frac{1}{2} \cform{4 &0 \\2 &0}\, .
\end{align}
Using the techniques of the appendices \ref{app:B3} and \ref{sec:tet-hsr}, the trihedral and tetrahedral topologies at $w=4$ can be 
expressed in terms of single lattice sums as well,
\begin{align}
\cformtri{2&1\\0&1}{1&1\\1&0}{1\\0} &= \frac{3}{2}\left(\cform{3&0\\1&0} \, \right)^{2}-\frac{1}{2}\cform{6&0\\2&0}-\frac{1}{2}{\rm\hat G}_{2}\cform{4&0\\2&0}+ 3\cform{5&0\\1&0}-{\rm \hat G}_{2}\cform{3&0\\1&0}- {\rm G}_{4}
\notag \\
\cformtri{1&1\\1&0}{1&1\\1&0}{2\\0} &= -2\left(\cform{3&0\\1&0} \, \right)^{2}+2\cform{6&0\\2&0}+{\rm\hat G}_{2}\cform{4&0\\2&0}
 - 6\cform{5&0\\1&0}+2{\rm \hat G}_{2}\cform{3&0\\1&0}+ 2 {\rm G}_{4}
\label{greatrels} \\
{\cal C}_{\rm tet} &= 4 \left(\cform{3&0\\1&0} \, \right)^{2}-2 \cform{6&0\\2&0}-2{\rm \hat G}_{2}\cform{4&0\\2&0}+ 12\cform{5&0\\1&0}- 4 {\rm \hat G}_{2}\cform{3&0\\1&0}- 4{\rm G}_{4} \, .
\notag
\end{align}
Furthermore, the identities among $\cform{a&b&c \\ 1 &1 &0}$ in appendix \ref{app:B1} and holomorphic 
subgraph reduction as in \eqref{eq:20} reduce all the dihedral modular graph forms in 
$H_{1234}^{(4,0)}[G_{ij}G_{kl}]$ to the simplest class $\cform{a&0 \\ b &0}$ of lattice sums. For instance,
\begin{align}
H_{1234}^{(4,0)}[G_{12}^2] &= {\rm G}_4 {\rm E}_2 - \hat{\rm G}_2 \cform{4 &0 \\ 2 &0} - 8 \cform{5 &0 \\ 1 &0} + 2 {\rm \hat G}_{2}  \cform{3 &0 \\ 1 &0} +  2 {\rm G}_4\label{hetgraph23}\\
H_{1234}^{(4,0)}[G_{12}G_{34}] &= \hat{\rm G}_2 \cform{4 &0 \\ 2 &0}- 6 \cform{5 &0 \\ 1 &0} + 2 {\rm \hat G}_{2}  \cform{3 &0 \\ 1 &0} +  2 {\rm G}_4\ .\label{eq:3}
\end{align}
Once this simplification is performed for all heterotic graph forms in \eqref{eq:11}, 
the remaining $H_{1234}^{(4,0)}[G_{ij}G_{kl}]$ are found to be related to \eqref{hetgraph23} and \eqref{eq:3} via
\begin{align}
  H_{1234}^{(4,0)}[G_{13}^2]  &= 3 {\rm G}_{4}{\rm E}_{2}-2H_{1234}^{(4,0)}[G_{12}^2]  \notag\\
  H_{1234}^{(4,0)}[G_{13}G_{24}]  &=-2 H_{1234}^{(4,0)}[G_{12}G_{34}]   \label{eq:5}\\
  H_{1234}^{(4,0)}[G_{13}G_{12}]  &=  -\frac{1}{2}H_{1234}^{(4,0)}[G_{12}G_{34}]   \notag \\
  H_{1234}^{(4,0)}[G_{12}G_{23}]  &= H_{1234}^{(4,0)}[G_{12}G_{34}]\ .  \notag
\end{align}
Furthermore, the contributions of ${\rm \hat G}_{2}\cform{3 &0 \\ 1 &0}$ and
${\rm G}_4 $ turn out to cancel when assembling
the complete $\ap^2$ order of ${\cal I}_{1234}^{(4,0)}$, see (\ref{eq:123}) below.

We have also evaluated the $\ap^3$-order of the integral ${\cal I}^{(2,0)}_{1234}$, see appendix \ref{sec:Z.1}
for the occurring modular graph forms and appendix \ref{sec:Z.2} for their simplifications. 
Their specific combination selected by applying momentum conservation to ${\cal I}^{(2,0)}_{1234}$ will be given
below.

\subsubsection{Checking and assembling the results}

The symmetry properties of the $V_w$ functions \cite{Dolan:2007eh}
\begin{align}
V_2(1,2,3,4) + V_2(1,3,4,2) + V_2(1,4,2,3) &= 0 \label{hetgraph22}\\
V_4(1,2,3,4) + V_4(1,3,4,2) + V_4(1,4,2,3) &= 3 {\rm G}_4 \notag 
\end{align}
impose the following constraints on heterotic graph forms
\begin{align}
H_{1234}^{(2,0)}[\ldots]+H_{1342}^{(2,0)}[\ldots]+H_{1423}^{(2,0)}[\ldots]&=0 \label{hetgraph21}\\
H_{1234}^{(4,0)}[\ldots]+H_{1342}^{(4,0)}[\ldots]+H_{1423}^{(4,0)}[\ldots]&=3 {\rm G}_4H_{1234}^{(0,0)}[\ldots] \, ,
\notag
\end{align}
where the ellipses represent arbitrary monomials in $G_{ij}$ (the same ones in each term of the respective equation),
and all of our results for $H_{1234}^{(w,0)}[\ldots]$ have been checked to satisfy these consistency conditions.
Examples for the modular graph functions on the right-hand side of (\ref{hetgraph21}) include
\begin{align}
H_{1234}^{(0,0)}[\emptyset]=1\, , \ \ \ \ \ \
H_{1234}^{(0,0)}[G_{ij}]=0\, , \ \ \ \ \ \
H_{1234}^{(0,0)}[G_{ij}^2]={\rm E}_2\, .
\end{align}
With the results for the planar heterotic graph forms $H_{1234}^{(2,0)}[\ldots]$ at orders $\ap^{\leq 3}$
and $H_{1234}^{(4,0)}[\ldots]$ at orders $\ap^{\leq 2}$, we arrive at the low-energy expansions (cf.\ (\ref{hetgraph2}))
\begin{align}
  {\cal I}^{(2,0)}_{1234}(s_{ij},\tau) &= 6s_{13}\cform{3&0\\1&0}+2(s_{13}^{2}+2s_{12}s_{23})\cform{4&0\\2&0}    \label{goodrep}\\
  &+ 4s_{13} (s_{13}^{2} -s_{12}s_{23})\left(3\cform{3&1&1\\1&1&1}-4 \cform{5&0\\3&0}+3 {\rm E}_{2}\cform{3&0\\1&0}\right)+{\cal O}(\ap^{4}) \notag\\
   {\cal I}^{(4,0)}_{1234}(s_{ij},\tau) &= {\rm G}_{4}+6s_{13}\left({\rm G}_{4}+{\rm \hat G}_{2}\cform{3&0\\1&0}\right)
  +2(s_{13}^{2} -s_{12}s_{23}) {\rm G}_{4}{\rm E}_{2}    \label{eq:123} \\
  &+2(s_{13}^{2}+2s_{12}s_{23})\left(\cform{5&0\\1&0}+ {\rm \hat G}_{2}\cform{4&0\\2&0}\right)+{\cal O}(\ap^{3})  \notag
\end{align}
after applying the Mandelstam identities (\ref{mgf3}). One can use the equivalent form of (\ref{toinv})\footnote{We have also employed $\pi\nabla C_{2,1,1}=2 (\Im \tau)^{2} \Big(\cform{3&1&1\\1&1&1} + \cform{2&2&1\\2&0&1} \Big)$ and $ \cform{3&1&1\\1&1&1} +2 \cform{2&2&1\\2&0&1} - 2 \cform{5&0\\3&0}=0$ in deriving \eqref{eq:2}.}
\begin{align}
\cform{3 &0 \\1&0} &= \frac{ \pi \nabla {\rm E}_2 }{2 (\Im \tau)^2} \, , \ \ \ \ 
\cform{4 &0 \\2&0} = \frac{ \pi \nabla {\rm E}_3 }{3 (\Im \tau)^2}  \, , \ \ \ \ 
\cform{5 &0 \\1&0} = \frac{( \pi \nabla)^2 {\rm E}_3 }{12 (\Im \tau)^4} \notag  \\
\cform{5 &0 \\3&0} &= \frac{ \pi \nabla {\rm E}_4 }{4 (\Im \tau)^2}  \, , \ \ \ \ 
\cform{3 &1 &1 \\1 &1 &1} - \frac{8}{5} \cform{5 &0 \\3&0}= \frac{ \pi \nabla {\rm E}_{2,2} }{ (\Im \tau)^2}\label{eq:2}
\end{align}
to express these expansions via Cauchy--Riemann derivatives of modular invariants ${\rm E}_{\ldots}$,
\begin{align}
 (\Im \tau)^2 {\cal I}^{(2,0)}_{1234}(s_{ij},\tau) &= 3s_{13}  \pi \nabla {\rm E}_2+\frac{2}{3}(s_{13}^{2}+2s_{12}s_{23}) \pi \nabla {\rm E}_3  \label{goodrepB}\\
  &+ s_{13} (s_{13}^{2} -s_{12}s_{23})\left( \frac{4}{5} \pi\nabla {\rm E}_{4}+12 \pi\nabla {\rm E}_{2,2}+6{\rm E}_{2} \pi\nabla {\rm E}_{2} \right)+{\cal O}(\ap^{4}) \notag \\
   {\cal I}^{(4,0)}_{1234}(s_{ij},\tau) &= {\rm G}_{4}+3s_{13}\Big(2{\rm G}_{4}+\frac{{\rm \hat G}_{2} \, \pi\nabla {\rm E}_2}{(\Im \tau)^2} \Big)  
  +2(s_{13}^{2} -s_{12}s_{23}){\rm G}_{4}{\rm E}_{2}    \label{eq:123B} \\
&+(s_{13}^{2}+2s_{12}s_{23})     \Big(\frac{(\pi \nabla)^2 {\rm E}_3}{6(\Im \tau)^4} 
                          + \frac{2 {\rm \hat G}_{2} \, \pi \nabla {\rm E}_3}{3 (\Im \tau)^2}  \Big)+{\cal O}(\ap^{3})\ ,\notag
\end{align}
see (\ref{crder5}) for the definition of ${\rm E}_{2,2}$ and (\ref{mgf10}) for the analogous expansion
of ${\cal I}^{(0,0)}$.


\subsection{Non-planar heterotic graph forms}
\label{sec:3.3}

We will now adapt the strategy of the previous section to 
the low-energy expansion of the Koba--Nielsen integrals ${\cal I}_{12|34}^{(w,0)}$
in the double-trace sector.
The non-planar heterotic graph forms (\ref{hetgraph3}) in the $\ap$-expansion take the most
compact form when we compute them from the Fourier expansion (\ref{2.999}) of $V_2(i,j)$.
However, some of the resulting lattice sums turn out to be conditionally convergent 
or even divergent. We define a regularization scheme by demanding that (\ref{2.999}) and the
alternative Fourier expansion due to insertion of (\ref{1.5}) into
\begin{align}
V_2(i,j) = 2 f^{(2)}_{ij} - (f^{(1)}_{ij})^2 
\label{npreg1}
\end{align}
yield the same expression for each heterotic graph form. Again, we exploit the symmetries
\begin{align}
{\cal I}_{12|34}^{(w,0)}={\cal I}_{21|34}^{(w,0)}={\cal I}_{34|12}^{(w,0)}
\label{npreg2}
\end{align}
of the integrals at each order in $\ap$ to reduce the number of heterotic graph forms
that need to be calculated independently, see table \ref{npreg3}.
\begin{table}
  \begin{center}
    \begin{tabular}{cc}  
      \toprule
      Order &Inequivalent non-planar heterotic graph forms\\
      \midrule
      $\ap^0$ &$H_{12|34}^{(w,0)}[\emptyset]$\\
      \addlinespace
      $\ap^1$ &$H_{12|34}^{(w,0)}[G_{12}] ,\  H_{12|34}^{(w,0)}[G_{13}]$\\
      \addlinespace
      $\ap^2$ &$H_{12|34}^{(w,0)}[G^2_{12}] , \  H_{12|34}^{(w,0)}[G_{12}G_{34}]  , \   H_{12|34}^{(w,0)}[G_{13}G_{24}]$ \\
               & $H_{12|34}^{(w,0)}[G^2_{13}] ,\    H_{12|34}^{(w,0)}[G_{12}G_{13}]  , \   H_{12|34}^{(w,0)}[G_{13}G_{23}]$\\
      \addlinespace
      $\ap^3$ &$H_{12|34}^{(w,0)}[G^3_{12}], \ H_{12|34}^{(w,0)}[G^2_{12}G_{13}]  , \  H_{12|34}^{(w,0)}[G_{12}G_{13} G_{34}]   , \  H_{12|34}^{(w,0)}[G_{12}G_{13} G_{24}]$ \\
               &$H_{12|34}^{(w,0)}[G^3_{13}], \ H_{12|34}^{(w,0)}[G_{12}G^2_{13}]  , \  H_{12|34}^{(w,0)}[G_{12}G_{13} G_{14}]   , \  H_{12|34}^{(w,0)}[G_{13}G_{14} G_{34}] $  \\
               &$H_{12|34}^{(w,0)}[G^2_{12}G_{34}], \ H_{12|34}^{(w,0)}[G^2_{13}G_{14}]  , \  H_{12|34}^{(w,0)}[G^2_{13} G_{24}]   , \  H_{12|34}^{(w,0)}[G_{13}G_{14} G_{23}]$ \\
      \bottomrule
    \end{tabular}
    \caption{Inequivalent non-planar heterotic graph forms with respect to the symmetries \mbox{$H_{12|34}^{(w,0)}=H_{21|34}^{(w,0)}=H_{34|12}^{(w,0)}$}.}
    \label{npreg3}
  \end{center}
\end{table}

\subsubsection{Leading orders $\ap^0$ and $\ap^1$}

In the absence of Green functions in the integrand, only the zero mode $\hat {\rm G}_2$ of the
Fourier expansion (\ref{2.999}) of $V_2(i,j)$ contributes,
\begin{align}
  H_{12|34}^{(2,0)}[\emptyset] &=2{\rm \hat G}_{2} \, , \ \ \ \ \ \ 
  H_{12|34}^{(4,0)}[\emptyset]={\rm \hat G}_{2}^{2} \, .
  \label{npreg4}
\end{align}
At first order in $\ap$, the representation (\ref{2.999}) of $V_2(i,j)$
yields conditionally convergent lattice sums
\begin{align}
 H^{(2,0)}_{12|34}[G_{12}]&=- \cregform{2 &0 \\ 0 &0}  \, , \ \ \ \ \ \ 
 H^{(4,0)}_{12|34}[G_{12}] =- {\hat{\rm G}_2 } \cregform{2 &0 \\ 0 &0} \, .
   \label{npreg5}
\end{align}
The regularized value of $\cregform{2 &0 \\ 0 &0}$ and similar lattice sums 
will be determined by repeating the above calculations with
the representation (\ref{npreg1}) of $V_2(i,j)$ in terms of $f^{(w)}$ functions,
\begin{align}
H_{12|34}^{(2,0)}[\emptyset] = 2 \cregform{2 &0 \\ 0 &0} \, , \ \ \ \ \ \
  H^{(2,0)}_{12|34}[G_{12}] = - 2 \cform{3 &0 \\ 1 &0}- \cregform{1 &1 &1 \\ 0 &0 &1} \, .
  \label{npreg6}
\end{align}
By imposing these results to match (\ref{npreg4}) and (\ref{npreg5}), we can determine both $\cregform{2 &0 \\ 0 &0}$
and another regularized value $\cregform{1 &1 &1 \\ 0 &0 &1}$ relevant to a later step\footnote{Note that the
expressions in (\ref{npreg7}) cannot be reproduced from an extension of momentum conservation
(say $0= \cregform{1 &1 &1 \\ 0 &0 &1}+2\cregform{2 &1 &0 \\ 1 &0 &0}$) and (\ref{mgfrel3}) to 
regularized values.},
\begin{align}
 \cregform{2 &0 \\ 0 &0}  = \hat {\rm G}_2 \, , \ \ \ \ \ \
 \cregform{1 &1 &1 \\ 0 &0 &1} =  {\rm \hat G}_{2} - 2\cform{3 &0 \\ 1 &0} \, ,
   \label{npreg7}
\end{align}
see also \cite{Lerche:1987qk} for the first result.
A more general class of identities between regularized values can be generated by integrating 
the lattice sum $\sum_{p\neq 0} \frac{ e^{2\pi i \langle p,z_{ij} \rangle } }{p^a \bar p^b}$ against
the two representations (\ref{2.999}) and (\ref{npreg1}) of $V_2(i,j)$\footnote{Note that (\ref{npreg8})
is inconsistent with a naive extension of the holomorphic subgraph reduction (\ref{eq:42}) to $a_+ = a_-  = 1$
and regularized values.},
\begin{align}
\cregform{1 &1 &a \\ 0 &0 &b} =  \cregform{a+1 &0 \\ b-1 &0} - 2\cform{a+2 &0 \\ b &0} \, ,
   \label{npreg8}
\end{align}
where $\cregform{A \\ B}:=\cform{A \\ B}$ whenever the entries of $A$ and $B$ yield absolutely convergent sums.
With the above regularized values, the inequivalent heterotic graph forms at the first order in $\ap$ read
\begin{align}
 H^{(2,0)}_{12|34}[G_{12}] &= -{\rm \hat G}_{2}\, ,& H^{(2,0)}_{12|34}[G_{13}]& = 0  \label{npreg9} \\
  H^{(4,0)}_{12|34}[G_{12}] &= -{\rm \hat G}_{2}^{2}\, ,& H^{(4,0)}_{12|34}[G_{13}] &= 0 \, . \notag
\end{align}

\subsubsection{Subleading orders $\ap^2$, $\ap^3$ and beyond}

At the second order in $\ap$, we exploit the expressions for
 $\cregform{2 &0 \\ 0 &0}$ and $\cregform{1 &1 &1 \\ 0 &0 &1} $
in (\ref{npreg7}) to determine the inequivalent heterotic graph forms
\begin{align}
 H^{(2,0)}_{12|34}[G_{12}^{2}]&= 2 {\rm\hat G}_{2}{\rm E}_{2} +2 {\rm\hat G}_{2} - 4 \cform{3&0\\1&0} \notag\\
H^{(2,0)}_{12|34}[G_{13}^{2}]&=  2 {\rm\hat G}_{2}{\rm E}_{2} \notag\\
H^{(2,0)}_{12|34}[G_{13}G_{23}]&= - \cform{3&0\\1&0} \label{npreg10}\\
H^{(2,0)}_{12|34}[G_{12}G_{13}]&=H^{(2,0)}_{12|34}[G_{12}G_{34}]=H^{(2,0)}_{12|34}[G_{13}G_{24}]=0 \notag
\end{align}
and
\begin{align}
  H^{(4,0)}_{12|34}[G_{12}^{2}]&= {\rm\hat G}^{2}_{2} {\rm E}_{2} +2 {\rm\hat G}^{2}_{2} - 4 {\rm \hat G}_{2}  \cform{3&0\\1&0}
  \, , &
  H^{(4,0)}_{12|34}[G_{12}G_{13}]&=0 \notag\\
  H^{(4,0)}_{12|34}[G_{13}G_{24}]&={\rm G}_{4} \, ,
  &
  H^{(4,0)}_{12|34}[G_{13}^{2}]&=  {\rm\hat G}^{2}_{2}{\rm E}_{2}  \label{npreg11} \\
  H^{(4,0)}_{12|34}[G_{12}G_{34}]&={\rm\hat G}_{2}^{2} \, ,
  &
H^{(4,0)}_{12|34}[G_{13}G_{23}]&= - {\rm \hat G}_{2}  \cform{3&0\\1&0}  
  \ . \notag
\end{align}
On top of \eqref{npreg8}, the third order in $\ap$ involves the regularized value
\begin{align}
  \cregform{1 &1 &1 &1 \\ 0 &0 &1 &1} &= - 2 \cform{4 &0 \\ 2 &0}  +4 \cform{3 &0 \\ 1 &0} - 2 {\rm \hat G}_{2} - {\rm \hat G}_{2} {\rm E}_2\label{npreg12}
\end{align}
which has been inferred by evaluating $H^{(2,0)}_{12|34}[G_{12}^{2}]$ with both representations
(\ref{2.999}) and (\ref{npreg1}) of $V_2(i,j)$. The inequivalent heterotic graph forms at order $\ap^3$ from table 
\ref{npreg3} are given in appendix \ref{app:Y}. The four-point gauge amplitude only requires a specific combination of them that simplifies and is given below.

For higher orders in $\ap$, we speculate that the above
method can be used to assign a regularized value to all the conditionally convergent or divergent sums
in the non-planar heterotic graph forms: The idea is to integrate suitably chosen lattice sums 
(such as $\sum_{p\neq 0} \frac{ e^{2\pi i \langle p,z_{ij} \rangle } }{p^a \bar p^b}$ in case of (\ref{npreg8})) 
against $V_2(i,j)$ or $V_2(i,j)V_2(k,l)$ and equate the results that arise from different use of (\ref{2.999}) 
and (\ref{npreg1}). In particular, this technique has been checked to yield unique regularized values
for the most challenging contributions at order $\ap^{4}$.

\subsubsection{Assembling the results}

With the results for the non-planar heterotic graph forms at order $\ap^{\leq 3}$,
the low-energy expansions of the integrals ${\cal I}^{(w,0)}_{12|34}$ are found to be
\begin{align}
  {\cal I}^{(2,0)}_{12|34}&= 2{\rm \hat G}_{2}-2s_{12}{\rm \hat G}_{2}+2s_{12}^{2}\Big({\rm \hat G}_{2}(1+2{\rm E}_{2}) - \frac{\pi \nabla {\rm E}_2}{(\Im\tau)^2}\Big)-2s_{13}s_{23} \left(2{\rm \hat G}_{2} {\rm E}_{2}+\frac{\pi \nabla {\rm E}_2}{(\Im\tau)^2}\right)\notag\\
&\hphantom{=\ }\! \!   \! +2s_{12}^{3}\left(\frac{\pi \nabla {\rm E}_2}{(\Im\tau)^2}-\frac{2}{3}\frac{\pi \nabla {\rm E}_3}{(\Im\tau)^2}-(1+2 {\rm E}_{2}){\rm \hat G}_{2}\right) \label{sureB} \\
&\hphantom{=\ }\! \!   \! +2s_{12}s_{13}s_{23}\left((2 {\rm E}_{2}+5 {\rm E}_{3}+\zeta_{3}){\rm \hat G}_{2}+\frac{\pi \nabla {\rm E}_2}{(\Im\tau)^2}-\frac{2}{3}\frac{\pi \nabla {\rm E}_3}{(\Im\tau)^2}\right)+{\cal O}(\ap^{4})
 \notag \\
    {\cal I}^{(4,0)}_{12|34}&=  {\rm \hat G}_{2}^{2}-2s_{12}{\rm \hat G}_{2}^{2}\notag\\
&\hphantom{=\ } \! \!   \!    +s_{12}^{2}\Big({\rm G}_{4}+{\rm \hat G}_{2}^{2}(3{+}2 {\rm E}_{2})- \frac{2 {\rm \hat G}_{2}   \, \pi \nabla {\rm E}_2}{(\Im\tau)^2} \Big)-2s_{13}s_{23} \Big({\rm G}_{4}+{\rm \hat G}_{2}^{2}{\rm E}_{2}+ \frac{ {\rm \hat G}_{2} \, \pi \nabla {\rm E}_2}{(\Im\tau)^2}\Big)\notag\\
&\hphantom{=\ } \! \!   \! +s_{12}s_{13}s_{23}\left((4{\rm E}_{2}{+}5 {\rm E}_{3}{+}\zeta_{3}){\rm \hat G}_{2}^{2}-2 {\rm G}_{4}+\frac{( \pi \nabla)^2 {\rm E}_3 }{(\Im \tau)^4}+\frac{{\rm \hat G}_{2}\pi \nabla {\rm E}_2}{(\Im\tau)^2}-\frac{4}{3}\frac{{\rm \hat G}_{2}\pi \nabla {\rm E}_3}{(\Im\tau)^2}\right)
\label{sure2B}\\
 &\hphantom{=\ } \! \!   \! +s_{12}^{3}\left(-2 {\rm G}_{4}-4{\rm \hat G}_{2}^{2}(1{+}{\rm E}_{2})+\frac{( \pi \nabla)^2 {\rm E}_3 }{3 (\Im \tau)^4}+\frac{4{\rm \hat G}_{2}\pi \nabla {\rm E}_2}{(\Im\tau)^2}-\frac{4}{3} \frac{{\rm \hat G}_{2}\pi \nabla {\rm E}_3}{(\Im\tau)^2}\right) +{\cal O}(\ap^{4})\notag
\end{align}
after applying momentum conservation. Similar to the representations (\ref{goodrepB}) and
(\ref{eq:123B}) of the planar integrals, we have used the substitutions \eqref{eq:2}. Together with the
planar results and the expression (\ref{mgf10}) for ${\cal I}^{(0,0)}$, (\ref{sureB}) and (\ref{sure2B})
complete the ingredients for the $\tau$ integrand (\ref{3.1}) of the heterotic-string amplitude.

Note that (\ref{sureB}) and (\ref{sure2B}) have been reproduced from the alternative method in
appendix \ref{sec:3.6}, where the $\ap$-expansion of ${\cal I}^{(w,0)}_{12|34}$ is performed without
any conditionally convergent or divergent sums in intermediate steps. This adds further validation 
for the assignment of regularized values described above.


\subsection{Uniform transcendentality and $\ap$-expansions}
\label{sec:unif-transc}

As mentioned in the introduction, one of the remarkable features of type-II amplitudes is that they exhibit so-called uniform transcendentality at each order in $\ap$.  
In this section, we will study the transcendentality properties of the heterotic string by restricting to the salient points that require a rewriting of the basis integrals ${\cal I}^{(w,0)}_{\ldots}$; additional details can be found in appendix~\ref{sec:5}.

In analogy with the superstring, we associate transcendental weights to the various objects appearing in the low-energy expansion of the heterotic integrals over the punctures as follows. The Eisenstein series ${\rm G}_k$ and ${\rm E}_k$ as well as $\zeta_k$ are assigned transcendental weight $k$, i.e.\ $\pi$ has transcendental weight one, whereas $\tau$ and $\nabla$ have transcendental weight zero.  Accordingly, one finds transcendental weight one for both $\pi \nabla $ and $y=\pi  \Im \tau$, i.e.\ weight $k{+}p$ for $(\pi \nabla)^p {\rm E}_k$, and weight $4{+}p$ for $(\pi \nabla)^p{\rm E}_{2,2}$. A more general definition of transcendental weight in terms of iterated integrals is given in appendix~\ref{sec:5}, but the assignment above suffices for the discussion of this section.

Inspecting the $\ap$-expansions of the planar single-trace integrals ${\cal I}^{(0,0)}$ and ${\cal I}^{(2,0)}_{1234}$ of (\ref{mgf10}) and (\ref{goodrepB}), one sees that their $k^{\rm th}$ order consistently involves modular graph forms of weight $k$ and $k{+}2$, respectively. Thus, these two integrals are referred to as uniformly transcendental. 

By contrast, ${\cal I}^{(4,0)}_{1234}$ in \eqref{eq:123B} violates uniform transcendentality since the same type of transcendental object appears at different orders in the $\ap$-expansion. For instance, ${\rm G}_{4}$ of transcendentality four appears with $1+6s_{13}+\ldots$ and thus at different orders in $\ap$. Similarly, the integrals ${\cal I}^{(w,0)}_{12|34}$ in \eqref{sureB} and \eqref{sure2B} from the double-trace sector violate uniform transcendentality. This can for instance be seen from the terms $\sim(1{+}2{\rm E}_2)$ along with $s_{12}^2$  in ${\cal I}^{(2,0)}_{12|34}$ and $\sim(3{+}2{\rm E}_2)$ along with $s_{12}^2$ in ${\cal I}^{(4,0)}_{12|34}$, respectively.

This violation of uniform transcendentality can be traced back to the following phenomenon. For the planar integral ${\cal I}_{1234}^{(4,0)}$ we see from~\eqref{hetgraph2},~\eqref{eq:1} and~\eqref{hetgraph4} that there is a leading contribution with a closed cycle of the form $f^{(1)}_{12}f^{(1)}_{23}f^{(1)}_{34}f^{(1)}_{41}$. This cycle exhibits purely holomorphic modular weight
$(4,0)$ and is thus amenable to holomorphic subgraph reduction discussed in section~\ref{sec:hsr}. As is evident from the explicit examples in~\eqref{eq:28}, the reduction produces various terms of different transcendentality: $\cform{5&0\\1&0}$ carries transcendental weight 5 by \eqref{eq:2} whereas ${\rm G}_{4}$ carries weight 4. Moreover, the formula~\eqref{eq:42} for
dihedral holomorphic subgraphs generically produces explicit factors of ${\rm \hat G}_{2}={\rm G}_{2}-\frac{\pi}{\Im\tau}$ which are clearly not of uniform transcendental weight. We therefore expect that \emph{all} closed cycles $f^{(1)}_{12}f^{(1)}_{23}\ldots f^{(1)}_{k1}$ in the $n$-point integrand break uniform transcendentality, including those with $k{=}n$. This is in marked contrast with the genus-zero situation where only closed \emph{sub}cycles $(z_{12}z_{23}\ldots z_{k1})^{-1}$ in the integrand with $k\leq n{-}2$ violate uniform transcendentality~\cite{Oprisa:2005wu, Mafra:2011nw, Huang:2016tag, Schlotterer:2016cxa}.\footnote{This point can be illustrated by considering a four-point integral at genus zero over closed subcycles with an integrand of the form $(z_{12}z_{21})^{-1}(z_{34} z_{43})^{-1}$. Integrating by parts in $z$ leads to the cyclic factor $(z_{12}z_{23} z_{34} z_{41})^{-1}$ subtending all four punctures that is called a Parke--Taylor factor (see section~\ref{sec:4}). The integration by parts also generates the rational factor $\frac{s_{23} }{1+s_{12}}$ in Mandelstam invariants that mixes different orders in $\ap$. Genus-zero integrals over Parke--Taylor factors subtending all the punctures are known to be uniformly transcendental 
(which is for instance evident from their representation in terms of the Drinfeld associator \cite{Broedel:2013aza}). Hence,
the original subcycle expression with $(z_{12}z_{21})^{-1}(z_{34} z_{43})^{-1}$ must violate uniform transcendentality.}
The subcycles $f^{(1)}_{12}f^{(1)}_{21}$ in the integrands of the non-uniformly
transcendental integrals ${\cal I}_{12|34}^{(2,0)}$ and ${\cal I}_{12|34}^{(4,0)}$
in the non-planar sector confirm the general expectation.

At genus zero, any non-uniformly transcendental disk or sphere integral over $n$ punctures can be expanded in a basis of uniformly transcendental integrals, see \cite{Mizera:2017cqs} for a general argument and \cite{Mafra:2011nv, Huang:2016tag, Schlotterer:2016cxa} 
for examples. This basis, known as Parke--Taylor basis, consists of $(n{-}3)!$ elements \cite{Mafra:2011nv, Zfunctions} 
and spans the twisted cohomology defined by the Koba--Nielsen factor made out of $|z_{ij}|^{-s_{ij}}$~\cite{Mizera:2017cqs}.

At genus one, the classification of integration-by-parts inequivalent half-integrands -- i.e.\ chiral halves for
torus integrands -- is an open problem. While genus-one correlators of the open superstring exclude a variety 
of worldsheet functions by maximal supersymmetry \cite{Green:1982sw, Berkovits:2004px}, Kac--Moody correlators 
such as (\ref{halfper99}) give a more accurate picture of the problem: It remains to find a minimal set of
integrals that span all the above ${\cal I}^{(w,0)}_{\ldots}$ (and possibly other torus integrals from different
massless one-loop closed-string amplitudes) via integration by parts. These equivalence classes are again referred
to as twisted cohomologies, where the twist is defined by the Koba--Nielsen factor $\exp(\sum_{i<j} s_{ij}G_{ij})$ with the Green function in (\ref{mgf2}).

Therefore we shall now re-express the planar and non-planar integrands in a basis of uniformly transcendental integrals, hoping that this will also shed light on the question of a basis for twisted cohomologies at genus one. We present below candidate basis elements $\widehat {\cal I}^{(w,0)}_{\ldots}$ of conjectured uniform transcendentality that appear suitable for the four-current correlator (\ref{halfper99}). Our explicit expressions at leading orders in $\ap$ and their different modular weights can 
be used to exclude relations among the $\widehat {\cal I}^{(w,0)}_{\ldots}$. However, it is beyond the scope of this work to arrive at a reliable prediction for the basis dimension of uniform-transcendentality integrals at four points.

In the relation between the new quantities $\widehat {\cal I}^{(w,0)}_{\ldots}$ and the genus-one integrals ${\cal I}^{(4,0)}_{1234}$, ${\cal I}^{(2,0)}_{12|34}$ and ${\cal I}^{(4,0)}_{12|34}$ all terms that break uniform transcendentality are contained in simple explicit coefficients like ${\rm \hat G}_2$ or $(1{+}s_{12})^{-1}$.
The manipulations necessary to arrive at the $\widehat {\cal I}^{(w,0)}_{\ldots}$ are given in detail in appendices \ref{sec:3.5} and \ref{sec:3.6} and driven by integration by 
parts, resulting again in series in modular graph forms which
\begin{itemize}
\item bypass the need for holomorphic subgraph reduction
\item avoid the conditionally convergent or divergent lattice sums caused by
integration over $V_2(i,j)$, see section \ref{sec:3.3} for our regularization scheme.
\end{itemize}
Aspects of the computational complexity when using the ${\cal I}^{(w,0)}_{\ldots}$ versus the $\widehat {\cal I}^{(w,0)}_{\ldots}$ can be found in appendix \ref{sec:effic-new-repr}.

\subsubsection{The planar results in terms of uniform-transcendentality integrals}

As derived in appendix \ref{sec:3.5}, a decomposition of the single-trace part of the four-point gauge 
amplitude that exhibits uniform transcendentality is
\begin{align}
M_4(\tau) \, \big|_{ {\rm Tr}(t^{a_1}t^{a_2}t^{a_3}t^{a_4}) } = {\rm G}_4^2 \widehat {\cal I}^{(4,0)}_{1234} + {\rm G}_4 \Big( {\rm G}_4 \hat {\rm G}_2 - \frac{7}{2} {\rm G}_6  \Big) {\cal I}^{(2,0)}_{1234}
+ \Big(  \frac{49}{6} {\rm G}_{6}^2  - \frac{10}{3} {\rm G}_{4}^3 \Big) {\cal I}^{(0,0)}   \label{ut1} 
\, .
\end{align}
In this expression we introduced the following combination of modular weight $(4,0)$
\begin{align}
 \widehat {\cal I}^{(4,0)}_{1234}(s_{ij},\tau) &= {\cal I}^{(4,0)}_{1234}(s_{ij},\tau) - {\rm G}_4  {\cal I}^{(0,0)}_{1234}(s_{ij},\tau) -{\rm \hat G}_2 {\cal I}^{(2,0)}_{1234}(s_{ij},\tau) \nonumber\\
 &= 6s_{13} {\rm G}_4+(s_{13}^{2}+2s_{12}s_{23}) \frac{(\pi \nabla)^2{\rm E}_{3}}{6(\Im\tau)^{4}}+{\cal O}(\ap^{3})\ ,
\label{Ihat40}
\end{align}
that manifestly respects uniform transcendentality to the order given. In~\eqref{uniform21}, we provide a closed integral form of $\widehat {\cal I}^{(4,0)}_{1234}(s_{ij},\tau)$ that we conjecture to be uniformly transcendental at every order in $\ap$,
with weight $k{+}3$ at the order of $\ap^k$. As argued above, ${\cal I}^{(2,0)}_{1234}$ and ${\cal I}^{(0,0)} $ are uniformly transcendental and all non-uniformly transcendental terms in the above way of writing the planar amplitude are in the coefficients of the basis integrals. 
The coefficient of $s_{13}$ in (\ref{Ihat40}) may also be written as $6 {\rm G}_4=  \frac{(\pi \nabla)^2{\rm E}_{2}}{(\Im\tau)^{4}}$
to highlight the parallel with the modular graph form $\frac{(\pi \nabla)^2{\rm E}_{3}}{6(\Im\tau)^{4}}$
at the subleading order $\ap^2$.

Note that the coefficient of ${\cal I}^{(0,0)}$ in (\ref{ut1}) can be recognized as
\begin{align}
\label{nozeroagain}
 \frac{49}{6} {\rm G}_{6}^2  - \frac{10}{3} {\rm G}_{4}^3 =  - \frac{ 128 \pi^{12}}{2025} \eta^{24}
\, .
\end{align}
At the level of the integrated amplitude (\ref{3.0}), this
cancels the factor of $\eta^{-24}$ due to the partition function. Hence, one can import the
techniques of the type--II amplitude \cite{Green:2008uj, DHoker:2015foa} to perform the modular integrals
$\int_{\cal F}  \frac{\dd^2 \tau }{(\Im \tau)^2 } {\cal I}^{(0,0)}$
arising from the terms $\sim {\cal I}^{(0,0)}$ in (\ref{ut1}) as we shall see in section~\ref{sec:tauint}.

Similar to (\ref{nozeroagain}) the coefficient of ${\cal I}^{(2,0)}_{1234}$ in \eqref{ut1} exhibits
a special relative factor in the combination ${\rm G}_4 {\rm G}_2 - \frac{7}{2} {\rm G}_6$ 
that can therefore be written as a $\tau$-derivative\footnote{We are using the Ramanujan identities
$$
 q \frac{ \dd {\rm G}_2}{\dd q} = \frac{ {\rm G}_2^2- 5 {\rm G}_4 }{4\pi^2}  \, , \ \ \ \
 q \frac{ \dd {\rm G}_4}{\dd q} = \frac{2 {\rm G}_2 {\rm G}_4 - 7 {\rm G}_6 }{2\pi^2} \, , \ \ \ \ 
 q \frac{ \dd {\rm G}_6}{\dd q} = \frac{21 {\rm G}_2 {\rm G}_6 - 30 {\rm G}^2_4 }{14\pi^2} \, .
$$}
\begin{align}
\label{nozero}
{\rm G}_4 {\rm \hat G}_2 - \frac{7}{2} {\rm G}_6
= - \frac{\pi}{\Im \tau} {\rm G}_4 + \pi^2 q \frac{ \dd {\rm G}_4}{\dd q}\, .
\end{align}
Hence, the only contribution $\sim q^0$ to (\ref{nozero}) stems from the non-holomorphic term 
$-\frac{\pi}{\Im \tau}{\rm G}_{4}$.

\subsubsection{The non-planar results in terms of uniform-transcendentality integrals}

Similarly, we can also rewrite the non-planar part of the amplitude in terms of combinations that exhibit uniform transcendentality as follows
\begin{align}
&M_4(\tau)  \, \big|_{  {\rm Tr}(t^{a_1}t^{a_2}) {\rm Tr}(t^{a_3}t^{a_4})}   =  \frac{ {\rm G}_4^2 \, \big[  \widehat {\cal I}^{(4,0)}_{12|34}  + s_{13}^2 ( \widehat {\cal I}^{(4,0)}_{1243} {+} \hat {\rm G}_2 {\cal I}^{(2,0)}_{1243} )
+ s_{23}^2 (\widehat {\cal I}^{(4,0)}_{1234} {+} \hat {\rm G}_2 {\cal I}^{(2,0)}_{1234} ) \big]}{(1+s_{12})^2 }   \label{reallyuni} \\
 &\ + \Big( \frac{  {\rm G}_4^2 \hat {\rm G}_2 }{(1{+}s_{12})^2}    {-}   \frac{7 {\rm G}_4{\rm G}_6 }{2(1{+}s_{12})} \Big)
 \widehat   {\cal I}^{(2,0)}_{12|34} 
 +\Big(  \frac{ {\rm G}_4^2 \hat {\rm G}_2^2+{\rm G}^3_4 (s_{13}^2{+}s_{23}^2)  }{(1+s_{12})^2 }  {-} 7 \frac{ {\rm G}_4{\rm G}_6 \hat {\rm G}_2 }{1+s_{12}} {+} \frac{5}{3} {\rm G}_{4}^3 {+} \frac{ 49  }{6} {\rm G}_6^2 \Big) {\cal I}^{(0,0)}
 \, .\notag 
\end{align}
The details of the derivation of this result are given in appendix \ref{sec:3.6}. On top of a contribution from the planar integral given in~\eqref{Ihat40}, \eqref{reallyuni} contains the non-planar integrals of conjectured uniform transcendentality $\widehat  {\cal I}^{(w,0)}_{12|34}$ with modular weights $(w,0)$ and leading orders
\begin{align}
\widehat  {\cal I}^{(2,0)}_{12|34} &= \ -2(s_{12}^{2}+s_{13}s_{23})\frac{\pi\nabla {\rm E}_{2}}{(\Im\tau)^{2}}-4(
s_{12}^{3} + s_{12}s_{13}s_{23})\frac{\pi\nabla {\rm E}_{3}}{3(\Im\tau)^{2}}+{\cal O}(\ap^{4})
\notag\\
  \widehat  {\cal I}^{(4,0)}_{12|34} &= (s_{12}^{3}+3s_{12}s_{13}s_{23})\frac{( \pi \nabla)^2 {\rm E}_3 }{3(\Im \tau)^4}+{\cal O}(\ap^{4})\ .\label{eq:25}
\end{align}
For definitions of the $\widehat {\cal I}^{(w,0)}_{12|34}$ to all orders in $\ap$ via $z$-integrals, see \eqref{3.5NP} and \eqref{3.4NPNP}.

\subsection{The integrated amplitude and the low-energy effective action}
\label{sec:tauint}

Our main focus of this paper lies on the structure of the $\tau$-integrand $M_4(\tau)$ appearing in the four-point gauge amplitude~\eqref{3.0}. However, using the results of~\cite{MR656029, Green:1999pv, Lerche:1987qk, Green:2008uj, Angelantonj:2011br, DHoker:2015foa, Basu:2017nhs} on the integrals of certain combinations of modular graph forms and Eisenstein series, it is possible to 
perform the integral over $\tau$ analytically up to second order in $\ap$ both in the single-trace and the double-trace sector. We present the resulting values, keeping in mind that these have to be considered in our normalization~\eqref{3.0}, see also footnote~\ref{fn:norm}.

\subsubsection{Planar amplitude up to second order in $\ap$}

Upon collecting all terms of the same structure in Mandelstam invariants up to quadratic order from~\eqref{ut1} we have to evaluate the integrals appearing in
\begin{align}
{\cal M}_4\, &\big|_{  {\rm Tr}(t^{a_1}t^{a_2} t^{a_3}t^{a_4})}  \sim  \int_{\cal F} \frac{\dd^2 \tau }{(\Im \tau)^2 \, \eta^{24} }\left(\frac{49}{6} {\rm G}_6^2 -  \frac{10}{3} {\rm G}_4^3 \right)\nonumber\\
& + s_{13} \int_{\cal F} \frac{\dd^2 \tau }{(\Im \tau)^2 \, \eta^{24} } \left( 6 {\rm G}_4^3 + 3 {\rm G}_4^2 \hat{\rm G}_2\frac{ \pi \nabla {\rm E}_2}{(\Im \tau)^2} - \frac{21}2 {\rm G}_4 {\rm G}_6  \frac{\pi \nabla {\rm E}_2}{(\Im \tau)^2}\right)\nonumber\\
& + (s_{13}^2 - s_{12} s_{23})\int_{\cal F} \frac{\dd^2 \tau }{(\Im \tau)^2 \, \eta^{24} } \left(\frac{49}{3} {\rm G}_4 {\rm G}_6^2 {\rm E}_2 -\frac{20}{3} {\rm G}_4^3 {\rm E}_2 \right)\nonumber\\
&\quad +(s_{13}^2 + 2 s_{12} s_{23}) \int_{\cal F} \frac{\dd^2 \tau }{(\Im \tau)^2 \, \eta^{24} } \left(   \frac23 {\rm G}_4^2 {\rm \hat G}_{2} \frac{ \, \pi \nabla {\rm E}_3}{ (\Im \tau)^2}
-\frac{7}{3} {\rm G}_4{\rm G}_6 \frac{\pi \nabla  {\rm E}_3}{(\Im \tau)^2}
+\frac16 {\rm G}_4^2 \frac{(\pi \nabla)^2 {\rm E}_3}{(\Im \tau)^4}  \right)\nonumber\\
&+ {\cal O}(\ap^3)\,. \label{plantauints}
\end{align}
These integrals can be done using the following observations. The combination of ${\rm G}_4^3$ and ${\rm G}_6^2$ appearing in the first and third line is that of~\eqref{nozeroagain} leading to $\eta^{24}$, making the first line proportional to the volume of ${\cal F}$ that equals $\frac{\pi}{3}$ while the third line is proportional to the integral of ${\rm E}_2$ over ${\cal F}$ that vanishes~\cite{Green:1999pv}. Using furthermore the results of~\cite{Lerche:1987qk, Basu:2017nhs} for the 
remaining lines, we end up with the integrated planar amplitude
\begin{align}
{\cal M}_4\, \big|_{  {\rm Tr}(t^{a_1}t^{a_2} t^{a_3}t^{a_4})}  &\sim  - \frac{256\pi^{13}}{6075} + s_{13} \frac{32\pi^{13}}{6075} \left(\frac{25}{6} + \gamma_{\rm E} +\log \pi -2\frac{\zeta'_4}{\zeta_4}\right)\nonumber\\
&\quad\quad - (s_{13}^2 + 2 s_{12} s_{23}) \frac{32\pi^{13}}{30375} + {\cal O}(\ap^3)\,.
\label{patternPL}
\end{align}
The appearance of terms $\log\pi$ and $d\log\zeta$ is due to the method of cutting off the fundamental domain  
${\cal F}$ as in~\cite{Basu:2017nhs}. Note that these terms as well as the Euler--Mascheroni constant $ \gamma_{\rm E}$
cancel at the second order in $\ap$, a feature that we shall discuss in more detail in section~\ref{sec:intconst}.

From the point of view of the low-energy effective action, the terms above correspond to single-trace higher-derivative corrections of the schematic form ${\rm Tr} (F^4)$, ${\rm Tr} (D^2 F^4)$ and ${\rm Tr} (D^4 F^4)$, respectively. The lowest-order term in the one-loop scattering amplitude was already analyzed in~\cite{Sakai:1986bi, Lerche:1987qk, Lerche:1987sg, Lerche:1988zy, Ellis:1987dc}. The structure of higher-derivative invariants in super Yang--Mills theory was studied for example in~\cite{Drummond:2003ex,Howe:2010nu,Bossard:2010pk} and the three operators above are of $1/2$-, $1/4$- and non-BPS type, respectively. 
General references on the effective action of heterotic string theories include
\cite{Gross:1986mw, Kikuchi:1986cz, Bergshoeff:1989de, Tseytlin:1995bi, Stieberger:2002wk, Stieberger:2002fh, Bossard:2016zdx, Bossard:2017wum, Basu:2017nhs, Basu:2017zvt}.

\subsubsection{Non-planar amplitude up to second order in $\ap$}

The integrated contribution from the double-trace sector can be determined by similar methods by starting from~\eqref{reallyuni}. The $\tau$-integral to be performed to quadratic order in Mandelstam invariants is
\begin{align}
{\cal M}_4\, &\big|_{{\rm Tr}(t^{a_1}t^{a_2}) {\rm Tr}(t^{a_3}t^{a_4})}  \sim  \int_{\cal F} \frac{\dd^2 \tau }{(\Im \tau)^2 \, \eta^{24} }\left({\rm G}_4^2 \hat{{\rm G}}_2^2 -  7 {\rm G}_4 {\rm G}_6 \hat{\rm G}_2 + \frac53 {\rm G}_4^3 + \frac{49}{6} {\rm G}_6^2 \right)\nonumber\\
& + s_{12}  \int_{\cal F} \frac{\dd^2 \tau }{(\Im \tau)^2 \, \eta^{24}} \left( 7  {\rm G}_4 {\rm G}_6 \hat{\rm G}_2  -2 {\rm G}_4^2\hat{\rm G}_2^2\right)\nonumber\\
&+ s_{12}^2  \int_{\cal F} \frac{\dd^2 \tau }{(\Im \tau)^2 \, \eta^{24}} \Bigg( {\rm G}_4^3 + \hat{\rm G}_2^2 {\rm G}_4^2 (3+ 2{\rm E}_2) - \frac{2\hat{\rm G}_2 {\rm G}_4^2 \pi \nabla {\rm E}_2}{(\Im \tau)^2} 
-7 {\rm G}_4 {\rm G}_6  \hat{\rm G}_2 (1+2{\rm E}_2)\nonumber\\
&\hspace{40mm} +7 {\rm G}_4 {\rm G}_6 \frac{\pi \nabla {\rm E}_2}{(\Im \tau)^2}+\frac{10}{3} {\rm G}_4^3 {\rm E}_2 + \frac{49}{3}{\rm G}_6^2 {\rm E}_2 \Bigg)\nonumber\\
&+ s_{13} s_{23} \int_{\cal F} \frac{\dd^2 \tau }{(\Im \tau)^2 \, \eta^{24}} \Bigg( {-}2{\rm G}_4^3-2 {\rm G}_4^2 \hat{\rm G}_2^2 {\rm E}_2 -2 \frac{{\rm G}_4^2 \hat{\rm G}_2 \pi \nabla {\rm E}_2}{(\Im \tau)^2}
+14 {\rm G}_4 {\rm G}_6 \hat{\rm G}_2 {\rm E}_2 \nonumber\\
&\hspace{40mm} +7\frac{{\rm G}_4{\rm G}_6\pi\nabla {\rm E}_2}{(\Im \tau)^2} -\frac{10}3 {\rm G}_4^3 {\rm E}_2 - \frac{49}{3}  {\rm G}^2_6 {\rm E}_2\Bigg) + {\cal O}(\ap^3)\,.
\label{tauintsnp}
\end{align}
These integrals can again be performed using the results of~\cite{Lerche:1987qk,Basu:2017nhs,DHoker:2015foa}, and we obtain the integrated double-trace amplitude to second order in $\ap$ as 
\begin{align}
{\cal M}_4 \, \big|_{  {\rm Tr}(t^{a_1}t^{a_2}) {\rm Tr}(t^{a_3}t^{a_4})}   &\sim - \frac{128\pi^{13}}{6075} s_{12}
+s_{12}^2 \frac{64 \pi^{13}}{3645} \left( - \frac{91}{30} + \gamma_{\rm E} + \log \pi - 2\frac{\zeta'_4}{\zeta_4} \right)\nonumber\\
&\quad\quad -s_{13} s_{23}  \frac{256 \pi^{13}}{18225} \left( - \frac{11}{6} + \gamma_{\rm E} + \log \pi - 2\frac{\zeta'_4}{\zeta_4} \right) + {\cal O}(\ap^3) \,. \label{patternNP}
\end{align}
We note that there is no lowest-order term in the non-planar sector. As pointed out in~\cite{Stieberger:2002wk},
this is in agreement with the duality between the heterotic string and the type-I string \cite{Polchinski:1995df}, 
where $({\rm Tr} (F^2))^2$ is absent at tree level. The first non-trivial correction term for double-trace operators is then $({\rm Tr} (D F^2))^2$, and the eight-derivative order admits two independent kinematic structures.

\subsubsection{Consistency with tree-level amplitudes}
\label{sec:intconst}

Given the expressions (\ref{patternPL}) and (\ref{patternNP}) for the integrals over $\tau$,
the appearance of $\gamma_{\rm E} + \log \pi - 2\frac{\zeta'_4}{\zeta_4}$ signals an interplay
with the non-analytic momentum dependence of the respective amplitude at the same $\ap$-order, 
cf.\ \cite{Green:2008uj, DHoker:2015foa}.
The non-analytic part of the four-point gauge amplitude can be inferred to comprise factors of $\log s_{ij}$
\begin{itemize}
\item in the planar sector at the order of $\ap$ but not at the orders of $\ap^0$ or $\ap^2$, see (\ref{patternPL})
\item in the non-planar sector at the order of $\ap^2$ but not at the orders of $\ap^0$ or $\ap^1$, see (\ref{patternNP})
\end{itemize}
These patterns in the discontinuity of the one-loop amplitude are consistent with the $\ap$-expansion of
the respective tree amplitudes of the heterotic string \cite{Gross:1986mw}
\begin{align}
{\cal M}^{\rm tree}_4 \, \big|_{  {\rm Tr}(t^{a_1}t^{a_2}t^{a_3}t^{a_4})}  &\sim  
1 + 2 \zeta_3 s_{12}s_{13}s_{23} + {\cal O}(\ap^5) 
  \label{treeamps} \\
{\cal M}^{\rm tree}_4 \, \big|_{  {\rm Tr}(t^{a_1}t^{a_2}) {\rm Tr}(t^{a_3}t^{a_4})}  &\sim   s_{23} - s_{12}s_{23} + s_{12}^2s_{23} +  {\cal O}(\ap^4) \, , \notag
\end{align}
where a polarization-dependent factor $A^{\rm tree}_{\rm SYM}(1,2,3,4) $ has been subsumed
in the $\sim$. Unitarity relates the $(s_{ij})^w$-order of (\ref{treeamps}) to
the non-analytic terms in the one-loop amplitudes that are signaled by the
 $(s_{ij})^{w+1}$-order in (\ref{patternPL}) and (\ref{patternNP}). In particular, 
the absence of a subleading order $\ap^1$ in the planar sector of ${\cal M}^{\rm tree}_4 $ 
ties in with the absence of $\gamma_{\rm E} + \log \pi - 2\frac{\zeta'_4}{\zeta_4}$
at the $\ap^2$-order of (\ref{patternPL}). This is analogous to the discontinuity structure
of the massless type-II amplitude \cite{Green:2008uj, DHoker:2015foa}, where unitarity relates the 
$\ap^{w+1}$-order beyond the one-loop low-energy limit to the $\ap^w$-order of the tree amplitude 
beyond its supergravity limit.


\subsection{Modular graph forms in the massless $n$-point function}
\label{sec:3.8}

Although the main focus of this work is on the four-point amplitude involving gauge bosons,
we shall now explain how the above techniques can be extended to higher multiplicity 
and to external gravitons. As we will see, the integration over the punctures in $n$-point 
one-loop amplitudes of the heterotic string involving any combination of gauge bosons
and gravitons boils down to (heterotic) modular graph forms -- at any order in the $\ap$-expansion.

\subsubsection{$n$ external gauge bosons}

For the $n$-point generalization of the amplitude (\ref{presc1}) among four gauge bosons,
the structure of the correlation functions in the integrand is well known. The supersymmetric
chiral halves of the vertex operators (\ref{presc3}) exclusively contribute (complex conjugates of) 
$f^{(w)}_{ij}$ and holomorphic Eisenstein series ${\rm G}_k$ to the $n$-point correlators \cite{Broedel:2014vla}. 
This has been manifested in this reference by expressing RNS spin sums\footnote{Also see \cite{Mafra:2016nwr, Mafra:2017ioj}  for recent examples of $f^{(w)}$ functions in manifestly supersymmetric higher-point amplitudes in the pure-spinor formalism.} of the worldsheet fermions in terms 
of the $V_w$ functions (\ref{1.1a}) and ${\rm G}_k$, also see \cite{Lee:2017ujn} for analogous results with two external gauginos and \cite{Tsuchiya:1988va, Stieberger:2002wk}
for earlier work on the spin sums. The contributions from the worldsheet bosons $\partial_{\bar z} X(z,\bar z)$ 
are even simpler, they can be straightforwardly integrated out using Wick contractions that 
yield $\overline{f^{(1)}_{ij}}$ or $\partial_{\bar z_i} \overline{f^{(1)}_{ij}}$. Likewise, the Kac--Moody 
currents of (\ref{presc3}) exclusively contribute $V_w$ functions and ${\rm G}_k$ to the complementary 
chiral half of the $n$-point correlators~\cite{Dolan:2007eh}.

Given these results on the $n$-point integrands, it is important to note that ${\rm G}_k$ are modular graph forms 
and that $f^{(w)}_{ij}$ functions admit the same type of lattice-sum representation (\ref{1.5}) as the 
Green function (\ref{mgf4}). Then, the Fourier integrals over the $n$ punctures yield momentum-conserving 
delta functions similar to (\ref{2.14}), and one is left with the kinds of nested lattice sums that define 
modular graph forms \cite{DHoker:2016mwo}.

Starting from the five-point function, the singularities 
$\overline{f^{(1)}_{ij}} \sim \frac{1}{\bar z_{ij}}+ {\cal O}(z,\bar z)$
in the supersymmetric correlators introduce kinematic poles into the integrals over the punctures. 
Still, the residues of these kinematic poles reduce to lower-multiplicity results and therefore give modular
graph forms by an inductive argument. 

On these grounds, one-loop scattering of $n$ gauge bosons in the heterotic string boil down to
modular graph forms at each order in the $\ap$-expansion after integrating over the punctures at fixed $\tau$.

\subsubsection{Adjoining external gravitons}

The vertex operators of gravitons and gauge bosons in heterotic string theories have the same 
supersymmetric chiral half.
That is why for mixed $n$-point amplitudes involving external gauge bosons and gravitons, the supersymmetric 
half of the correlator is identical to that of $n$ gauge bosons. Only the non-supersymmetric chiral half 
of the correlators is sensitive to the species of massless states in the external legs since the graviton 
vertex operator involves the worldsheet boson $\partial_{z} X(z,\bar z)$ 
in the place of the Kac--Moody current \cite{Gross:1985rr}. 

These additional worldsheet bosons of the gravitons contribute (sums 
of products of) $f^{(1)}_{ij}$ and $\partial_{z_i}f^{(1)}_{ij}$ due to Wick contractions and decouple 
from the current correlators of the gauge bosons. Moreover, they admit zero-mode contractions
$\partial_{z_i} X(z_i,\bar z_i) \partial_{\bar z_j} X(z_j,\bar z_j) \rightarrow \frac{\pi}{\Im \tau}$ 
between left and right movers, known from type-II amplitudes \cite{Gregori:1997hi,BjerrumBohr:2008vc,Richards:2008jg,Green:2013bza}. These kinds of cross-contractions are specific to 
amplitudes involving gravitons, and the resulting factors of $\frac{\pi}{\Im \tau}$ have the same modular weight
$(1,1)$ as the contributions $f^{(1)}_{ij}\overline{f^{(1)}_{kl}}$ due to separate
Wick contractions of the left and right movers.

Hence, the additional contributions due to $\partial_{z} X(z,\bar z)$ in the graviton vertex operator 
boil down to $f^{(1)}_{ij}$, $\partial_{z_i}f^{(1)}_{ij}$ or $\frac{\pi}{\Im \tau}$. All of these factors
line up with the above statements on the correlators of the supersymmetric chiral half and the currents: 
The $n$-point correlators for mixed graviton- and gauge-boson amplitudes in heterotic string theories 
exclusively depend on the punctures via lattice sums
\begin{align}
\sum_{p\neq 0} \frac{ e^{2\pi i \langle p, z_{ij} \rangle} }{p^{m_{ij} } \bar p^{n_{ij}} } \, , \ \ \ \ \ \ m_{ij},n_{ij} \geq -1
\label{generalsum}
\end{align}
that may be accompanied by powers of $\frac{\pi}{\Im \tau}$ and yield simple Fourier 
integrals w.r.t.\ $z_2,\ldots,z_n$. Each term in the $\ap$-expansion of the
Koba--Nielsen factor is bound to yield modular graph forms upon integrating over the $z_j$. 
By the arguments given for $n$ gauge bosons, the kinematic poles do not alter this result.

Note that cases with $m_{ij}=n_{ij}=-1$ in \eqref{generalsum} are due to the spurious factors of $\partial_{z_i}f^{(1)}_{ij}$ and $\partial_{\bar z_i}\overline{f^{(1)}_{ij}}$ in the left- and right-moving contributions to the correlators.
One can always remove any appearance of $\partial_{z_i}f^{(1)}_{ij}$ and $\partial_{\bar z_i}\overline{f^{(1)}_{ij}}$ via integration by parts and thereby improve the bound in \eqref{generalsum} towards $m_{ij},n_{ij} \geq 0$.

The $\ap$-expansion of four-point amplitudes involving gravitons has been studied beyond the
leading order in \cite{Basu:2017nhs, Basu:2017zvt}. Moreover, selected terms in the five- and six-point 
gauge amplitudes that are relevant to ${\rm Tr}(F^5)$ and ${\rm Tr}(F^6)$ interactions have been 
studied in \cite{Stieberger:2002wk}. The comparison of $(n\geq 5)$-point $\ap$-expansions with 
open-string results is left for the future \cite{toappsoon}.

\section{Heterotic strings versus open superstrings}
\label{sec:4}

In this section, we point out new relations between open-superstring amplitudes
and the integral ${\cal I}^{(2,0)}_{1234}$ over the torus punctures in the planar sector 
of the heterotic-string amplitude, cf.\ (\ref{3.6}). The observations of this section can be viewed as generalizing
the genus-zero result that closed-string amplitudes are single-valued open-string amplitudes 
\cite{Schlotterer:2012ny, Stieberger:2013wea, Stieberger:2014hba, BrownDupont, Schlotterer:2018zce, Brown:2018omk}.
That is why we motivate the genus-one discussion by a brief review of the tree-level connection between
moduli-space integrals in open- and closed-string amplitudes.

\subsection{Review of the single-valued map in tree amplitudes}

The single-valued map ${\rm sv}: \, \zeta_{n_1,n_2,\ldots, n_r} \rightarrow \zeta^{\rm sv}_{n_1,n_2,\ldots, n_r}$ of MZVs \cite{Schnetz:2013hqa, Brown:2013gia},
\begin{align}
\zeta_{2k}^{\rm sv} &= 0 \, , \ \ \ \ 
\zeta_{2k+1}^{\rm sv} = 2 \zeta_{2k+1} \, , \ \ \ \ k \in \mathbb N \label{svmzv} \\
\zeta_{3,5}^{\rm sv} &= - 10 \zeta_3 \zeta_5 \, , \ \ \ \ 
\zeta_{3,5,3}^{\rm sv} = 2 \zeta_{3,5,3} - 2 \zeta_{3}\zeta_{3,5} - 10 \zeta_3^2 \zeta_5  
\, , \ \ \ \ \te{etc.} \,,
\notag
\end{align}
relates moduli-space integrals over punctured spheres to single-valued disk 
integrals at the level of their low-energy expansion. 
Open-string tree amplitudes boil down to disk integrals that can be characterized 
by an integration domain for the punctures on the boundary
\beq
D(1,2,\ldots,n) := \{(z_1,z_2,\ldots,z_n) \in \mathbb R^n\, : \ - \infty < z_1 < z_2 <\ldots < z_n < \infty\} 
\label{svmzv1}
\eeq
and a cyclic Parke--Taylor factor $(z_{12} z_{23} \ldots z_{n1})^{-1}$ in the integrand \cite{Mafra:2011nv, Zfunctions, Huang:2016tag, Azevedo:2018dgo},
\beq
Z_{\rm tree}(\rho(1,2,\ldots,n)\, | \, 1,2,\ldots, n) := \int \limits_{D(\rho(1,2,\ldots,n))} \frac{ \dd z_1 \, \dd z_2 \, \ldots \, \dd z_n}{ {\rm vol} \, {\rm SL}_2(\mathbb R) } \frac{ \prod_{i<j}^n |z_{ij}|^{-s_{ij}} }{z_{12} z_{23} \ldots z_{n-1,n} z_{n1}} \, .
\label{svmzv2}
\eeq
The ordering of the punctures on the boundary does not need to coincide with that of the contractions giving the Parke--Taylor factors and this is included in the formula above by the permutation $\rho\in S_n$.

The sphere integrals arising for the closed string do not admit a similar notion of cyclic ordering of the punctures but instead involve 
Parke--Taylor integrands in both the holomorphic and the antiholomorphic variables,
\beq
J_{\rm tree}(\rho(1,2,\ldots,n)\, | \, 1,2,\ldots, n) := \int\limits_{\mathbb C^n} \frac{ \dd^2 z_1 \, \dd^2 z_2 \, \ldots \, \dd^2 z_n}{\pi^{n-3} \, {\rm vol} \, {\rm SL}_2(\mathbb C) } \frac{ \prod_{i<j}^n |z_{ij}|^{-2s_{ij}} }{z_{12} z_{23} \ldots z_{n1} \, \rho(\bar z_{12} \bar z_{23} \ldots \bar z_{n1})} \, ,
\label{svmzv3}
\eeq
where the permutation $\rho \in S_n$ is understood to act as $\rho(\bar z_{12}\ldots) = (\bar z_{\rho(1)\rho(2)}\ldots)$.
The inverse volumes of ${\rm SL}_2$ instruct to fix any three punctures to $(0,1,\infty)$ with the standard Jacobian
\cite{Polchinski:1998rq, Blumenhagen:2013fgp}, and
the numerator factors $ |z_{ij}|^{-s_{ij}}$ and $|z_{ij}|^{-2s_{ij}}$ stem from the genus-zero analogues
of the plane-wave correlator (\ref{mgf0}).

For any choice of $\rho$, the low-energy expansions
of (\ref{svmzv2}) and (\ref{svmzv3}) are related by the single-valued map (\ref{svmzv}) 
\cite{Schlotterer:2012ny, Stieberger:2013wea, Stieberger:2014hba, BrownDupont, Schlotterer:2018zce, Brown:2018omk},
\beq
J_{\rm tree}(\rho(1,2,\ldots,n)\, | \, 1,2,\ldots, n) 
= {\rm sv} \, Z_{\rm tree}(\rho(1,2,\ldots,n)\, | \, 1,2,\ldots, n) \, .
\label{svmzv4}
\eeq
Hence, the cyclic ordering of the integration domain (\ref{svmzv1}) on the disk boundary 
is passed to the antiholomorphic Parke--Taylor factor $\rho(\bar z_{12} \bar z_{23} \ldots \bar z_{n1})^{-1}$
under the single-valued map, following a Betti--deRham duality\footnote{In the context of graphical 
hyperlogarithms in \cite{betti2}, inequalities $z_i<z_j$ and factors of $(\bar z_i-\bar z_j)^{-1}$ are 
represented by (directed) Betti edges and deRham edges, respectively.
By exchanging the roles of these edges, one converts the Parke--Taylor 
factor $(\bar z_{12} \bar z_{23} \ldots \bar z_{n1})^{-1}$ into a disk ordering 
$-\infty < z_1<z_2<\ldots < z_n<\infty$.} \cite{betti1, betti2}. 

The single-valued relation (\ref{svmzv4}) at tree level allows to extract the complete closed-string low-energy 
expansions $J_{\rm tree}$ from the open-string input $Z_{\rm tree}$ (see e.g.\
\cite{Broedel:2013aza, Mafra:2016mcc} for recent methods for open-string 
$\ap$-expansions at $n$ points).
In particular, the expressions for single-valued MZVs in (\ref{svmzv}) 
expose the absence of certain MZVs (including $\zeta_{2k}, \zeta_{3,5}$, also see \cite{Stieberger:2009rr}) 
in closed-string amplitudes which is obscured in the Kawai--Lewellen--Tye relations \cite{Kawai:1985xq}. 
Obtaining a loop-level analogue of the single-valued map (\ref{svmzv4}) would similarly allow extracting 
the closed-string one-loop amplitudes from open-string calculations.

A first hint for the existence of such loop-level relations has been given in \cite{DHoker:2015wxz}, 
where modular graph functions were identified as special values of infinite sums of single-valued multiple polylogarithms. The infinite sums
in the reference are proposed to be single-valued analogues of elliptic multiple polylogarithms.
Moreover, an explicit connection between open- and closed-string data at genus one has been found 
in \cite{Broedel:2018izr}, relating {\it symmetrized} integrals over open-string punctures\footnote{The 
symmetrized integrals of \cite{Broedel:2018izr} would correspond to scattering of abelian gauge bosons
which do not appear in the spectrum of the type-I superstring with gauge group $SO(32)$.
While the physical interpretation of the symmetrized open-string integrals is an open question,
the results of this section relate the esv-image of more general four-point open-string integrals
compatible with $SO(32)$ to closed-string quantities.} to the permutation-invariant
integral ${\cal I}^{(0,0)}$ in (\ref{mgf1}). The proposal for an elliptic single-valued map ``esv'' \cite{Broedel:2018izr} 
and the open-string setup will be reviewed below, and we will identify the closed-string counterpart 
to more general 
four-point open-string integrals.


\subsection{Open-superstring integrals at genus one}
\label{sec:4.1}

In the same way as the one-loop four-point amplitude of type-II superstrings boils 
down to a single integral ${\cal I}^{(0,0)}$, the single-trace sector of the corresponding 
open-string amplitude is governed by \cite{Green:1982sw}
\beq
I^{\te{open}}_{1234}(s_{ij},\tau) := \int \limits_{0\leq z_2 \leq z_3 \leq z_4\leq 1} \dd z_2 \, \dd z_3 \, \dd z_4 \, \exp\Big( \sum_{i<j}^4 s_{ij} G_{ij}^{\rm open}(\tau) \Big) \, .
\label{open0a}
\eeq
The integration domain for the punctures $z_2,z_3,z_4$ corresponds to a particular parametrization
of a cylinder worldsheet, where the unit interval $(0,1)$ represents one of the boundaries. We have
again fixed $z_1=0$ by translation invariance, and although the open-string amplitude only involves $\tau = i t, \ t \in \RR_+$, we will study the integral \eqref{open0a} for arbitrary $\tau$ in the upper half plane. The open-string counterpart to the closed-string Green function (\ref{mgf2}) 
can be chosen as
\beq
G_{ij}^{\rm open}(\tau) = -\int^{z_{ij}}_0 \dd w \, f^{(1)}(w,\tau)\, ,
\label{openGF}
\eeq
where all $z_{j}$ and $w$ are taken to be on the real axis, and the regularization of the
endpoint divergence is discussed in \cite{Broedel:2014vla}. Given that the Taylor expansion of
(\ref{open0a}) boils down to iterated integrals over the expansion coefficient $f^{(1)}$ in the 
Kronecker--Eisenstein series (\ref{1.1}), each order in $\ap$ is expressible in terms of elliptic 
multiple zeta values (eMZVs) \cite{Broedel:2014vla}. More precisely, the parametrization of the
cylinder boundary in (\ref{open0a}) gives rise to the A-cycle eMZVs of Enriquez \cite{Enriquez:Emzv}
\beq
\omega(n_1,n_2,\ldots,n_r|\tau) := \! \! \! \! \!   \int \limits_{0\leq z_1 \leq z_2 \leq \ldots \leq z_r\leq 1}  \! \! \! \! \! f^{(n_1)}(z_1,\tau) \,\dd z_1 \, f^{(n_2)}(z_2,\tau) \, \dd z_2 \, \ldots \, f^{(n_r)}(z_r,\tau) \, \dd z_r  \, , 
\label{open0b}
\eeq
where the number $r$ of arguments $n_i \in \NN_0$ and the sum $n_1+n_2+\ldots+n_r$ are referred to as
length and weight, respectively. The differential equations of eMZVs in $\tau$ and their degeneration
at the cusp can be used to infer a Fourier expansion \cite{Enriquez:Emzv, Broedel:2015hia}
\beq
\omega(n_1,n_2,\ldots,n_r|\tau) = \sum_{k=0}^\infty c_k(n_1,n_2,\ldots,n_r) q^k \, ,  
\label{emzvfourier}
\eeq
where the $\tau$-independent coefficients $c_k(n_1,\ldots,n_r)$ are $\mathbb Q[(2\pi i)^{-1}]$-linear combinations
of MZVs. The dependence of the eMZVs 
on $\tau$ will usually be suppressed for ease of notation. See 
\cite{Matthes:Thesis} for a comprehensive reference on eMZVs.

As exemplified by the leading low-energy behavior of (\ref{open0a}) \cite{Broedel:2014vla},
\begin{align}
 &I^{\te{open}}_{1234}(s_{ij},\tau)\, =\, \frac{1}{6} \, - \, 2 \, s_{13} \,\omega(0,1,0,0)  \, + \,  2 \,  \big( s_{12}^2  + s_{23}^2 \big)\,  \omega(0,1,1,0,0)  \label{open01} \\
&\, - \, 2\,s_{12}s_{23}  \, \omega(0,1,0,1,0) \, + \, 
s_{13}(s_{13}^2- s_{12} s_{23} ) \, \beta_5
\, + \, s_{12} s_{23}s_{13} \,\beta_{2,3} \,  + \, {\cal O}(\ap^4) \, ,
\notag
\end{align}
the order in $\ap$ is in one-to-one correspondence with the weight of the accompanying eMZVs, i.e.\
$I^{\te{open}}_{1234}(s_{ij},\tau)$ is said to be uniformly transcendental. We have introduced the following shorthands 
for the coefficients at the third order in $\ap$,
\begin{align}
\beta_5 &=  \frac{4}{3} \, \big[ \omega(0,0,1,0,0,2)+\omega(0,1,1,0,1,0) - \omega(2,0,1,0,0,0) - \zeta_2 \omega(0,1,0,0) \big] \label{open02}  \\
\beta_{2,3} &= \frac{ \zeta_3}{12} +
 \frac{8}{3} \zeta_2 \omega(0, 1, 0, 0) - \frac{5}{18} \omega(0,3,0,0) \ .
 \label{open03}
\end{align}

In the same way as MZVs descend from multiple polylogarithms, eMZVs are special values
of elliptic multiple polylogarithms \cite{BrownLev}. Various representations of elliptic multiple polylogarithms
have recently found appearance in the evaluation of Feynman integrals, see e.g.\ \cite{Bloch:2014qca, Adams:2017ejb, Remiddi:2017har, Bourjaily:2017bsb, Broedel:2017kkb, Broedel:2017siw, Broedel:2018iwv, Adams:2018bsn, Adams:2018kez,
Broedel:2018rwm, Blumlein:2018aeq, Blumlein:2018jgc, Broedel:2018qkq} and references therein.

\subsubsection{Decomposition into symmetry components}
By the properties of the integration cycle and the open-string Green function in (\ref{open0a}), the
open-string integral exhibits the same dihedral symmetries w.r.t.\ its labels $1,2,3,4$ as the $V_w$ functions at even values of $w$, 
\beq
I^{\te{open}}_{1234}(s_{ij},\tau)=I^{\te{open}}_{2341}(s_{ij},\tau) \, , \ \ \ \ \ \  
I^{\te{open}}_{4321}(s_{ij},\tau) = I^{\te{open}}_{1234}(s_{ij},\tau) \, ,
\label{opendih}
\eeq
cf.\ (\ref{1.1symm}). In order to explore further connections with the $V_w$ functions, we
decompose the integral $I^{\rm open}_{1234}=\frac{1}{6}Z^{(0)}+Z^{(2)}_{1234}$ into components with different symmetry properties in $1,2,3,4$,
\begin{align}
Z^{(0)}(s_{ij},\tau)&:= \sum_{\sigma\in S_3}  I^{\te{open}}_{1\sigma(234)}(s_{ij},\tau)  \label{open0e}\\
Z^{(2)}_{1234}(s_{ij},\tau) &:= \frac{1}{3} \big[2  I^{\te{open}}_{1234}(s_{ij},\tau) - I^{\te{open}}_{1342}(s_{ij},\tau)  - I^{\te{open}}_{1423}(s_{ij},\tau) \big]  \, .
\notag
\end{align}
While the permutation symmetric component $Z^{(0)}$ of the open-string integral has 
been studied in \cite{Broedel:2018izr}, we will here investigate the $\ap$-expansion of the
second component $Z^{(2)}_{\ldots}$ subject to
\begin{align}
Z^{(2)}_{1234}(s_{ij},\tau)+Z^{(2)}_{1342}(s_{ij},\tau)+Z^{(2)}_{1423}(s_{ij},\tau)
 = 0   \, .
\label{open0f}
\end{align}
The symmetry properties of the $Z^{(0)}$ and $Z^{(2)}_{\ldots}$ tie in with those of $V_0(1,2,3,4)=1$ and
$V_2(1,2,3,4)$, respectively, see (\ref{hetgraph22}).  

By inserting the $\ap$-expansion (\ref{open01}) along with momentum conservation
$s_{12}+s_{13}+s_{23}=0$ into (\ref{open0e}), we arrive at the following representation in terms of eMZVs\footnote{We have
used the following relations among eMZVs in simplifying (\ref{open0simp}) \cite{Broedel:2015hia}
$$
\omega(0, 1, 1,0, 0) = \frac{ \zeta_2}{12} + \omega(0, 0, 0, 0, 2) \, , \ \ \ \
\omega( 0, 1, 0, 1, 0) = \frac{ \zeta_2}{12}  +\frac{1}{2} \omega(0, 0, 2) - 4 \omega(0, 0, 0, 0, 2) 
\, .$$}
\begin{align}
Z^{(0)}(s_{ij},\tau)&= 1 + (s_{13}^2 - s_{12}s_{23}) \Big(2  \omega(0,0,2) + \frac{ 5\zeta_2}{3} \Big) + 6 s_{12}s_{23}s_{13} \beta_{2,3}  \, + \, {\cal O}(\ap^4)  \label{open0simp}\\
Z^{(2)}_{1234}(s_{ij},\tau)&= -2 s_{13} \omega(0,1,0,0) - \frac{2}{3} (s_{13}^2 +2 s_{12}s_{23}) \big[
 \omega(0,1,0,1,0) + \omega(0,1,1,0,0)  \big] \notag \\
 & \ \ \ +  s_{13}(s_{13}^2 - s_{12}s_{23})  \beta_5 \, + \, {\cal O}(\ap^4)  \ .
\label{open0g}
\end{align}
Note that the coefficient $\beta_{2,3}$ in (\ref{open01}) drops out from the definition of $Z_{1234}^{(2)}$ in (\ref{open0e}),
and we are only left with a specific linear combination of $\omega(0,1,0,1,0)$ and $\omega(0,1,1,0,0) $ at order~$\ap^{2}$.

\subsubsection{Modular transformation}

A connection between the symmetrized open-string integral  $Z^{(0)}$ in (\ref{open0e}) and 
closed-string integrals \cite{Broedel:2018izr} is based on the modular $S$ transformation 
$\tau \rightarrow -\frac{1}{\tau}$ of the contributing eMZVs. Otherwise, the $q$-series representation
(\ref{emzvfourier}) of the A-cycle eMZVs in (\ref{open0simp}) and (\ref{open0g}) would not exhibit any open-string analogue of
the expansion (\ref{crder9}) of modular graph forms around the cusp, more specifically of the 
Laurent polynomials $c^{\Gamma}_{m,n}(y)$ in $y=\pi \Im \tau$.

In order to determine the modular $S$ transformation of the $Z^{(w)}_{\ldots}$ integrals in (\ref{open0simp})
and (\ref{open0g}), we express
the A-cycle eMZVs in terms of iterated Eisenstein integrals (\ref{crder6}) \cite{Broedel:2015hia, JBOS} 
\begin{align}
Z^{(0)}(s_{ij},\tau)&= 1 + (s_{12}s_{23} - s_{13}^2 ) \big[ 12 {\cal E}_0(4,0) -  \zeta_2  \big] \notag \\
&- s_{12} s_{23} s_{13}\Big[ 12 {\cal E}_0(4,0,0) + 300 {\cal E}_0(6,0,0) -  \frac{5\zeta_3}{2}  \Big] \, + \, {\cal O}(\ap^4)  \label{open0jsymm}
\\
Z^{(2)}_{1234}(s_{ij},\tau)&= \frac{3 s_{13}}{2\pi^2} \,\big[ 6 {\cal E}_0(4,0,0)- \zeta_3  \big] + \frac{ s_{13}^2 +2 s_{12}s_{23}}{2\pi^2} \big[   120 {\cal E}_0(6,0,0,0) -  \zeta_4 \big]\notag \\
 & \! \! \!  \! \! \!  \! \! \!  \! \! \!   \! \! \!  \! \! \!  \! \! \!  \! \! \!    +\frac{s_{13}(s_{13}^2 - s_{12}s_{23})}{2\pi^2} \big[ 1296  {\cal E}_0(4, 4, 0, 0, 0) + 432 {\cal E}_0(4, 0, 4, 0, 0)   +  \tfrac{6}{5} {\cal E}_0(4, 0, 0, 0, 0) \label{open0j} \\
 &  \! \! \!  \! \! \!  \! \! \!  \! \! \!    
 + 4032 {\cal E}_0(8, 0, 0, 0, 0) - 216 {\cal E}_0(4, 0) {\cal E}_0(4, 0, 0) + 36 \zeta_3  {\cal E}_0(4,0)- 5 \zeta_5 
 \big]  \, + \, {\cal O}(\ap^4)  \ .
 \notag 
\end{align}
The modular properties of the holomorphic Eisenstein series give rise to $S$-transformations 
such as \cite{Brown:mmv, Broedel:2018izr}
\begin{align}
{\cal E}_0(4,0;-\tfrac{1}{\tau}) &= \frac{ T^2 }{1080} + \frac{ \pi^2 }{216} - \frac{ i \zeta_3}{6T} - \frac{ \pi^4 }{360 T^2} + 
{\cal E}_0(4,0;\tau)  + \frac{i}{T} {\cal E}_0(4,0,0;\tau)  \notag \\
{\cal E}_0(4,0,0;-\tfrac{1}{\tau}) &= \frac{i\pi^2 T }{540} + \frac{\zeta_3}{6} - \frac{ i \pi^4}{108 T} - \frac{ \pi^2 \zeta_3 }{6 T^2} + 
\frac{i \pi^6 }{540 T^3} + \frac{\pi^2}{T^2} {\cal E}_0(4,0,0;\tau)  \notag \\
{\cal E}_0(6,0,0;-\tfrac{1}{\tau}) &= - \frac{i T^3}{226800}
+ \frac{ i \pi^4}{21600 T}
+ \frac{ \zeta_5}{40 T^2}
- \frac{ i \pi^6}{22680 T^3}  \label{open05} \\
& \ \ \ \ +  {\cal E}_0(6, 0, 0;\tau)
 + \frac{ 3 i }{T}  {\cal E}_0(6, 0, 0, 0;\tau)
  - \frac{3 }{T^2} {\cal E}_0(6, 0, 0, 0, 0;\tau)
  \notag \\
{\cal E}_0(6,0,0,0;-\tfrac{1}{\tau}) &= \frac{\pi^2 T^2 }{226800} + \frac{ \pi^4 }{64800} + \frac{ \pi^6}{21600 T^2}  - \frac{ i \pi^2 \zeta_5}{60T^3} - \frac{ \pi^8 }{45360 T^4}  \notag \\
&\ \ \ \ + \frac{\pi^2}{T^2}
{\cal E}_0(6,0,0,0;\tau)  + \frac{2i \pi^2}{T^3} {\cal E}_0(6,0,0,0,0;\tau)  \, , \notag
\end{align}
and similar to (\ref{crder8}), we have absorbed powers of $\pi$ into\footnote{Note that this convention 
for $T$ agrees with the first arXiv version of \cite{Broedel:2018izr} 
but not with later versions of the reference, where another factor of $i$ will be 
absorbed into the definition of $T$.}
\beq
T:= \pi \tau  \, .
\eeq
The remaining modular transformations relevant to (\ref{open0j}) are displayed in
appendix \ref{appmod1}. These expressions for ${\cal E}_0(k_1,\ldots;-\tfrac{1}{\tau}) $
yield the following modular $\tau \rightarrow -\tfrac{1}{\tau}$ image of (\ref{open0jsymm}) \cite{Broedel:2018izr}
\begin{align}
Z^{(0)}(s_{ij},-\tfrac{1}{\tau}) &= 1+ (s_{13}^2 {-} s_{12}s_{23})
\Big(
{-} \frac{T^2}{90} +\frac{ \pi^2 }{ 9} + \frac{ 2 i \zeta_3}{T} +
\frac{ \pi^4}{30 T^2}
 - 12  {\cal E}_0(4, 0) -  \frac{12 i}{T}   {\cal E}_0(4, 0, 0)
 \Big) \notag \\
 & \  \ \hspace{-2.2cm}+ s_{12}s_{23}s_{13} \Big( 
 \frac{ i T^3}{756}  -   \frac{ i \pi^2 T }{45}  + \frac{ \zeta_3}{2}
 + \frac{ 7 i \pi^4}{72 T} + \frac{ 2 \pi^2 \zeta_3}{T^2} - \frac{ 15 \zeta_5}{2 T^2}
 - \frac{ 17 i \pi^6}{1890 T^3} - \frac{ 12 \pi^2 }{T^2} {\cal E}_0(4, 0, 0) \label{easyesv} \\
 & \ \ \ \ \ \ \ \ 
  - 300 {\cal E}_0(6, 0, 0)
  - \frac{900 i }{T}  {\cal E}_0(6, 0, 0, 0)
  + \frac{900 }{T^2}  {\cal E}_0(6, 0, 0, 0, 0)
 \Big)+ {\cal O}(\ap^4)  \notag
 \end{align}
and the following result for (\ref{open0j})
\begin{align}
Z^{(2)}_{1234}(s_{ij},-\tfrac{1}{\tau})&= s_{13} \Big(   \frac{i T}{60} - \frac{3 \zeta_3}{2 T^2}
- \frac{i \pi^2}{12 T}  +\frac{i \pi^4}{60 T^3}   + \frac{9}{T^2} 
  {\cal E}_0(4, 0, 0)   \Big) \notag \\
  & \ \ \ \ \hspace{-2.4cm}+  (s_{13}^2 +2 s_{12}s_{23})  \Big(  \frac{T^2}{3780} {-} \frac{i \zeta_5}{ T^3}
{-}\frac{\pi^2}{216 } {+} \frac{\pi^4}{360 T^2} {-} \frac{\pi^6}{756 T^4} {+} \frac{ 60}{T^2} {\cal E}_0(6, 0, 0, 0) {+} \frac{120 i}{T^3} {\cal E}_0(6, 0, 0, 0, 0)
\Big)  
\notag  \\
& \ \ \ \ \hspace{-2.4cm}+ s_{13}(s_{13}^2-s_{12}s_{23})  \beta_5\big({-} \tfrac{1}{\tau} \big)  \, + \, {\cal O}(\ap^4)\, ,
\label{open0k}
\end{align}
where the modular S transformation of $\beta_5(\tau)$ is given by
\begin{align}
\beta_5\big({-} \tfrac{1}{\tau} \big)  &=  - \frac{ i T^3}{7560} + \frac{ i \pi^2 T }{540 }
 - \frac{ \zeta_3}{20} - \frac{ i \pi^4 }{ 120 T}  - \frac{ 5 \zeta_5}{2 T^2} + \frac{ \pi^2 \zeta_3}{ 12 T^2} 
+ \frac{ 29 i \pi^6}{11340 T^3} + \frac{ \pi^4 \zeta_3}{60 T^4}  + \frac{ 3 \zeta_7}{T^4} -\frac{ i \pi^8}{1800 T^5} \notag \\
&+ \Big( {-}\frac{ i T}{5}     + \frac{ i \pi^2 }{T}
 + \frac{ 18  \zeta_3}{T^2}  - \frac{ i \pi^4  }{ 5 T^3}
 \Big) {\cal E}_0(4, 0) + \Big( \frac{3}{10}    - \frac{ \pi^2  }{2 T^2}
 - \frac{ \pi^4  }{10 T^4}
 \Big) {\cal E}_0(4, 0, 0) \label{beta5trf} \\
 & - \frac{ 108 }{T^2} {\cal E}_0(4, 0) {\cal E}_0(4, 0, 0) + \frac{ 216}{T^2} \Big( {\cal E}_0(4, 0, 4, 0, 0)+ 3  {\cal E}_0(4, 4, 0, 0, 0)+ \frac{ {\cal E}_0(4, 0, 0, 0, 0)}{360  } \Big) \notag \\
 & + \frac{2016 }{T^2} {\cal E}_0(8, 0, 0, 0, 0)
 + \frac{10080 i }{T^3} {\cal E}_0(8, 0, 0, 0, 0, 0)
 - \frac{15120}{T^4}   {\cal E}_0(8, 0, 0, 0, 0, 0, 0) \, .\notag
\end{align}
The ${\cal E}_0(\ldots)$ on the right-hand sides
of (\ref{easyesv}) to (\ref{beta5trf}) are understood to be evaluated at argument $\tau$ rather than 
$-\frac{1}{\tau}$. Note that from the results of \cite{Broedel:2014vla}, any order in the $\ap$-expansion 
of (\ref{easyesv}) and (\ref{open0k}) is expressible in terms of the B-cycle eMZVs of 
Enriquez \cite{Enriquez:Emzv}. A general discussion of the asymptotic expansion of B-cycle eMZVs
around the cusp can be found in \cite{Enriquez:Emzv, Broedel:2018izr, Zerbini:2018sox, Zerbini:2018hgs}.


\subsection{A proposal for a single-valued map at genus one}
\label{sec:4.2}

After modular transformation, the open-string expressions (\ref{easyesv}) to (\ref{beta5trf}) 
resemble the expansion (\ref{crder9}) of modular graph forms around the cusp:
The Laurent polynomials of (\ref{easyesv}) in $T= \pi \tau$ parallel the Laurent polynomials
of modular graph forms in $y= \pi \Im \tau$.
For instance, with the representations (\ref{nholo3}) of ${\rm E}_k$ in terms of iterated Eisenstein integrals,
the $\ap$-expansion (\ref{mgf10}) of the closed-string integral ${\cal I}^{(0,0)}$ takes the following form,
\begin{align}
{\cal I}^{(0,0)}(s_{ij},\tau)&= 1+2 (s_{13}^2 -s_{12}s_{23})
\Big( \frac{ y^2}{45} + \frac{ \zeta_3 }{y} - 12  \Re[ {\cal E}_0(4, 0) ]- \frac{6}{y}  \Re[{\cal E}_0(4, 0, 0)] \Big) \notag \\
&+ s_{12}s_{23}s_{13} \Big(  
\frac{ 2 y^3}{189} 
+  \zeta_3
+ \frac{ 15   \zeta_5  }{4 y^2}
-  600 \Re[ {\cal E}_0(6, 0, 0)]  \label{typeIIexp}  \\
& \ \ \ \  \ \ \ \
 - \frac{900 }{y}  \Re [{\cal E}_0(6, 0, 0, 0)]
 - \frac{ 450 }{y^2}  \Re[ {\cal E}_0(6, 0, 0, 0, 0)]
\Big) + {\cal O}(\ap^4) \, .
\notag
\end{align}
Moreover, the leading low-energy orders ${\cal I}^{(0,0)}=1+{\cal O}(s_{ij}^2)$ 
line up with the symmetry component $Z^{(0)}$ of the open-string integral, cf.\ (\ref{open0simp}), whereas
the expansion of $Z^{(2)}_{1234}$ starts at ${\cal O}(s_{ij})$. 
That is why the expression (\ref{typeIIexp}) was compared with the modular S transformation
of the {\it symmetrized} open-string integral in (\ref{easyesv}) \cite{Broedel:2018izr}. In fact, the
coefficients of the Mandelstam polynomials $s_{13}^2 {-}s_{12}s_{23}$ and
$s_{12}s_{23}s_{13}$ on the open- and closed-string side were observed to be related via \cite{Broedel:2018izr}
\beq
{\rm esv} \, : \ \ \  T \rightarrow 2iy \, , \ \ \ \ {\cal E}_0(k_1,k_2,\ldots,k_r) \rightarrow 2 \Re{\cal E}_0(k_1,k_2,\ldots,k_r) \, , \ \ \ \ \zeta_{n_1,\ldots,n_r} \rightarrow \zeta_{n_1,\ldots,n_r}^{\rm sv} 
\label{open0m}
\eeq
with $k_1 \neq 0$.
The single-valued MZVs of depth one can be found in (\ref{svmzv}), and the first part $\tau \rightarrow 2i \Im \tau$ of the esv 
map in (\ref{open0m}) does not apply to the exponents in the series 
representation (\ref{crder7}) of iterated Eisenstein integrals in $q=e^{2\pi i \tau}$. The factors of $2i$ and $2$
in (\ref{open0m}) ensure that the holomorphic derivatives
of $\tau$ and ${\cal E}_0(\ldots;\tau)$ are preserved under esv, and 
they were engineered in \cite{Broedel:2018izr} to obtain
\begin{align}
{\rm esv}  \, Z^{(0)}(s_{ij},-\tfrac{1}{\tau}) &= 1+2 (s_{13}^2 -s_{12}s_{23} )
\Big( \frac{ y^2}{45} + \frac{ \zeta_3 }{y} - 12  \Re[ {\cal E}_0(4, 0) ]- \frac{6}{y}  \Re[{\cal E}_0(4, 0, 0)] \Big) \notag \\
&+ s_{12}s_{23}s_{13} \Big(  
\frac{ 2 y^3}{189} 
+  \zeta_3
+ \frac{ 15   \zeta_5  }{4 y^2}
-  600 \Re[ {\cal E}_0(6, 0, 0)]    \\
& \ \ \ \  \ \ \ \
 - \frac{900 }{y}  \Re [{\cal E}_0(6, 0, 0, 0)]
 - \frac{ 450 }{y^2}  \Re [{\cal E}_0(6, 0, 0, 0, 0)]
\Big) + {\cal O}(\ap^4) 
\notag
\end{align}
which exactly matches (\ref{typeIIexp}) to the orders shown,
\beq
{\cal I}^{(0,0)}(s_{ij},\tau) =
{\rm esv}  \, Z^{(0)} (s_{ij}, -\tfrac{1}{\tau})   \, .
\label{esvZI}
\eeq
In fact, this correspondence has been checked to persist up to and including the order of $\ap^6$ and
is conjectural at higher orders \cite{Broedel:2018izr}.

The relation (\ref{esvZI}) between open- and closed-string $\ap$-expansions at genus one strongly resembles the 
tree-level relation (\ref{svmzv4}) between disk and sphere integrals. Hence, the rules in (\ref{open0m}) were proposed
\cite{Broedel:2018izr} to implement an elliptic analogue of the single-valued map (\ref{svmzv}) of MZVs. 

However, the formulation of the esv rules in (\ref{esvZI}) is in general ill-defined as it is not 
compatible with the shuffle multiplication of iterated Eisenstein integrals (\ref{crder6}) \cite{Broedel:2018izr}. 
For instance, applying the esv rules to the right-hand side of
\beq
{\cal E}_0(4,0,0)^2 = 
2{\cal E}_0(4,0,0,4,0,0)
+6{\cal E}_0(4,0,4,0,0,0)
+12{\cal E}_0(4,4,0,0,0,0)
\label{esvprob1}
\eeq
yields a different result than the square of ${\rm esv}  \,  {\cal E}_0(4,0,0)= {\cal E}_0(4,0,0) + \overline{{\cal E}_0(4,0,0)}$.
The ambiguity in applying esv to (\ref{esvprob1}) is proportional to the cross-term
${\cal E}_0(4,0,0) \overline{{\cal E}_0(4,0,0)} $ which is of order ${\cal O}(q \, \bar q)$
by the $q$-expansion (\ref{crder7}). More generally, terms of the form $q^n \bar q^0$ and $q^0\bar q^n$ with
$n \in \mathbb N_0$ in the output of the esv rules (\ref{open0m}) are well-defined, i.e.\ independent of the 
order of applying shuffle multiplication and esv. We will later on encounter a similar restriction on the powers
of $q$ and $\bar q$ in the expansion (\ref{crder9}) of modular graph forms that can be reliably predicted from
open-string input.

For the iterated Eisenstein integrals of depth one in (\ref{easyesv}), 
the convergent ${\cal E}_0(k,0,\ldots,0)$ with $k\geq 4$ cannot be rewritten 
via shuffle multiplication without introducing divergent examples ${\cal E}_0(0,\ldots)$. By restricting
the esv rules (\ref{open0m}) to convergent iterated Eisenstein integrals, the ambiguities due to shuffle
multiplication are relegated to depth two. The relation (\ref{esvZI}) has been established to the
order of $\ap^6$ (including ${\cal E}_0$ of depth three) by picking an ad hoc convention for the use 
of shuffle-multiplication in the open-string input, see section 4.3.3 of \cite{Broedel:2018izr} for details.

Note that a reformulation of the esv action on ${\cal E}_0(\ldots)$ that bypasses the
issues with the shuffle multiplication and should make contact with the equivariant
Eisenstein integrals of \cite{Brown:2017qwo2} will be discussed in \cite{futureref}.


\subsection{The closed-string integral over $V_2(1,2,3,4)$ versus ${\rm esv} \, Z^{(2)}_{1234}$}
\label{sec:4.3}

Given the above significance of the symmetrized open-string integral $Z^{(0)}$, we
will next apply the esv rules (\ref{open0m}) to the symmetry component $Z^{(2)}_{1234}$ in (\ref{open0e}).
Starting from the low-energy expansion (\ref{open0k}) and (\ref{beta5trf}) of its modular S transformation,
one arrives at 
\begin{align}
{\rm esv}  \, &Z^{(2)}_{1234}(s_{ij}, -\tfrac{1}{\tau}) = s_{13} \Big(   {-}\frac{y}{30} + \frac{3 \zeta_3}{4 y^2}
 - \frac{9}{2 y^2} 
 \Re[ {\cal E}_0(4, 0, 0)]   \Big) \notag \\
 & \! \! \!  \! \! \! \! \! +  (s_{13}^2 +2s_{12}s_{23})  \Big(  {-}\frac{y^2}{945} + \frac{ \zeta_5}{4 y^3}
- \frac{ 30}{y^2} \Re[ {\cal E}_0(6, 0, 0, 0)]  - \frac{30}{y^3} \Re[ {\cal E}_0(6, 0, 0, 0, 0)]
\Big) \! \!   \notag \\
& \! \! \!  \! \! \! \! \! + s_{13}(s_{13}^2 - s_{12}s_{23}) \Big({-}
\frac{y^3}{945}  -\frac{ \zeta_3}{10} + \frac{5 \zeta_5}{4 y^2} + \frac{3 \zeta_7}{8 y^4}
+\Big[ \frac{4y}{5} {-} \frac{ 18 \zeta_3}{y^2}  \Big] \Re[ {\cal E}_0(4, 0)]   + \frac{3}{5} \Re[ {\cal E}_0(4, 0, 0)] \notag \\
& 
 -\frac{108}{y^2} \Re[ {\cal E}_0(4, 0, 4, 0, 0) + 3 {\cal E}_0(4, 4, 0, 0, 0)  +\tfrac{1}{360} {\cal E}_0(4, 0, 0, 0, 0)]
 \notag \\
 & + \frac{ 108}{y^2} \Re[ {\cal E}_0(4, 0)] \Re[ {\cal E}_0(4, 0, 0)]
  - \frac{1008}{y^2} \Re[ {\cal E}_0(8, 0, 0, 0, 0)]  \notag\\
 &  - \frac{2520 }{y^3} \Re[ {\cal E}_0(8, 0, 0, 0, 0, 0)] 
 - \frac{1890 }{y^4} \Re[ {\cal E}_0(8, 0, 0, 0, 0, 0, 0)] 
\Big) \, + \, {\cal O}(\ap^4)  \, ,\label{open0p}
\end{align}
which will now be related to the $\ap$-expansion of closed-string integrals.


\subsubsection{The closed-string expansion in terms of iterated Eisenstein integrals}

Given that the symmetry properties (\ref{open0f}) of $Z^{(2)}_{1234}$ have been tailored to match
those of $V_2(1,2,3,4)$, it is natural to compare (\ref{open0p}) with the integral ${\cal I}_{1234}^{(2,0)}$
from the heterotic string. The low-energy expansion (\ref{goodrepB}) of ${\cal I}_{1234}^{(2,0)}$ is written in 
terms of Cauchy--Riemann derivatives of modular graph functions and can therefore be expressed in terms of iterated 
Eisenstein integrals. More precisely, the representations (\ref{nholo3}) and (\ref{newCR8})
of ${\rm E}_k$ and ${\rm E}_{2,2}$ yield \cite{Broedel:2018izr}
\begin{align}
\pi \nabla {\rm E}_{2} &=  \frac{ 2 y^3}{45} -
\zeta_3 + 24 y^2  {\cal E}_0(4) + 12 y {\cal E}_0(4, 0) + 6 \Re[ {\cal E}_0(4, 0, 0) ] 
\notag \\
\pi \nabla {\rm E}_{3} &= \frac{2 y^4}{315} - \frac{3 \zeta_5}{2 y} + 240 y^2 {\cal E}_0(6, 0) + 360 y {\cal E}_0(6, 0, 0) + 
 180 {\cal E}_0(6, 0, 0, 0)   \notag  \\
 &+ 180 \Re[ {\cal E}_0(6, 0, 0, 0) ] + 
 \frac{180 \Re[ {\cal E}_0(6, 0, 0, 0, 0) ]}{y}\notag \\
\pi \nabla {\rm E}_{4} &= \frac{4 y^5}{4725}- \frac{15 \zeta_7}{8 y^2} + 3360 y^2 {\cal E}_0(8, 0^2) + 10080 y {\cal E}_0(8, 0^3) + 
 12600 {\cal E}_0(8, 0^4)  \label{service4}  \\
 &+ \frac{6300 {\cal E}_0(8, 0^5)}{y} + 
 5040 \Re[ {\cal E}_0(8, 0^4) ] + \frac{12600 \Re[ {\cal E}_0(8, 0^5) ]}{y} + 
 \frac{9450 \Re[ {\cal E}_0(8, 0^6) ]}{y^2} \notag \\
\pi \nabla {\rm E}_{2,2} &=
- \frac{2 y^5}{10125} +  \frac{y^2 \zeta_3 }{45} - \frac{5 \zeta_5}{12}  + \frac{\zeta_3^2}{2 y}
+\Big( \frac{ 4  y^3}{15}   - 6   \zeta_3 \Big){\cal E}_0(4, 0) -  \Big( \frac{2y^2 }{15}    + \frac{ 6 \zeta_3}{y} \Big) \Re[ {\cal E}_0(4, 0, 0)]  \notag \\
&  +  \frac{ 2 y^2}{5}  {\cal E}_0(4, 0, 0) 
+ 36 y {\cal E}_0(4, 0)^2 + 36 {\cal E}_0(4, 0) \Re[ {\cal E}_0(4, 0, 0)] + \frac{18 \Re[ {\cal E}_0(4, 0, 0)]^2}{y}  \notag  \\
& + 144 y^2 {\cal E}_0(4, 4, 0) + 72 y \big( {\cal E}_0(4, 4, 0, 0) + \tfrac{1}{360} {\cal E}_0(4, 0, 0, 0)  \big) \notag \\
&  + 36 \Re[ {\cal E}_0(4, 0, 4, 0, 0) + 3 {\cal E}_0(4, 4, 0, 0, 0) + \tfrac{1}{360} {\cal E}_0(4, 0, 0, 0, 0) ] \,,\notag
 \end{align}
which can be used to cast the leading 
orders of (\ref{goodrepB}) into the following form
\begin{align}
&{\cal I}_{1234}^{(2,0)}(s_{ij},\tau) = \pi^2 s_{13} \Big(  
\frac{2y}{15} - \frac{ 3 \zeta_3}{y^2} + \frac{18}{y^2}  \Re[ {\cal E}_0(4, 0, 0)] + 72 {\cal E}_0(4) + \frac{36}{y} {\cal E}_0(4,0)  
\Big) \label{open0r} \\
&\ \ \ + \pi^2 (s_{13}^2+2s_{12}s_{23}) \Big( 
\frac{ 4 y^2 }{945} - \frac{ \zeta_5}{y^3} + \frac{120}{y^2} \Re[ {\cal E}_0(6,0, 0, 0)]  + \frac{120}{y^3} \Re[ {\cal E}_0(6,0,0, 0, 0)] 
 \notag \\
 &\hspace{30mm} + 160 {\cal E}_0(6, 0)  + \frac{240}{y} {\cal E}_0(6,0, 0)  + \frac{120}{y^2} {\cal E}_0(6,0,0, 0) 
\Big)
+ {\cal O}(\ap^3) \ .
\notag
\end{align}
A similar expression for the $\ap^3$-order is displayed in appendix \ref{appmod3}. 
There is a notable difference between the terms involving real parts of iterated Eisenstein integrals $\Re[ {\cal E}_0]$ and the terms without real parts. The real parts $\Re[ {\cal E}_0]$ and the pure $y$-terms match the esv image of the open-string integral in (\ref{open0p}) up to a global rescaling of ${\rm esv} \, Z^{(2)}_{1234}(s_{ij},-\tfrac{1}{\tau})$. By contrast, the iterated Eisenstein integrals without real part -- specifically, the above
$72 {\cal E}_0(4) $ and $\frac{36}{y} {\cal E}_0(4,0)$ as well as the last line of (\ref{open0r}) -- do not have any open-string counterpart in (\ref{open0p}). The same mismatch also arises at the third order in $\ap$, cf.~\eqref{eq:10}.
%


\subsubsection{The $P_{\rm Re}$ projection}

We shall now give a more precise description of the commonalities and differences of the expressions (\ref{open0p})
and (\ref{open0r}) for ${\rm esv}  \, Z^{(2)}_{1234}(s_{ij}, -\tfrac{1}{\tau}) $ and ${\cal I}_{1234}^{(2,0)}(s_{ij},\tau)$. 
The contributions to (\ref{open0r}) which do not have any obvious open-string correspondent will be 
isolated by defining a formal projection $P_{\rm Re}$ via
\beq
P_{\rm Re}\big(\overline{{\cal E}_0(k_1,\ldots,k_r) } \big) := 2  \Re[ {\cal E}_0(k_1,\ldots,k_r) ]  \, , \ \ \ \
P_{\rm Re}\big( {\cal E}_0(k_1,\ldots,k_r)   \big) := 0
\label{defPRE}
\eeq
with $k_1 \neq0$ which acts factor-wise on a product. The projection $P_{\rm Re}$ is designed to only keep the real parts of iterated Eisenstein 
integrals, i.e.\ the cases where holomorphic and antiholomorphic terms pair up. Moreover,
Laurent polynomials in $y$ and MZVs are taken to be inert
\beq
P_{\rm Re}\big(y^m \zeta_{n_1,n_2,\ldots,n_r} \big) := y^m \zeta_{n_1,n_2,\ldots,n_r} \, .\label{defPRE2}
\eeq
Similar to the esv rule (\ref{open0m}), the action of $P_{\rm Re}$ on $\overline{{\cal E}_0(k_1,\ldots,k_r) } $
is incompatible with shuffle multiplication and necessitates ad-hoc conventions for the presentation
of its input when two or more of the entries $k_j$ are non-zero. By the 
expansions ${\cal E}_0(k_1,\ldots)  {=} {\cal O}(q)$ and $\overline{{\cal E}_0(k_1,\ldots)}  {=} {\cal O}(\bar q)$
for $k_1\neq 0$, the ambiguity in evaluating $P_{\rm Re}$ has again at least one factor of both $q$ and $\bar q$.
Based on their representation given in (\ref{service4}), the
modular graph forms relevant to the expansion (\ref{goodrepB}) are mapped to
\begin{align}
P_{\rm Re}(\pi \nabla {\rm E}_{2}) &=  \frac{ 2 y^3}{45} -
\zeta_3+ 6 \Re[ {\cal E}_0(4, 0, 0) ] 
\notag \\
P_{\rm Re}(\pi \nabla {\rm E}_{3}) &= \frac{2 y^4}{315} - \frac{3 \zeta_5}{2 y} 
+ 180 \Re[ {\cal E}_0(6, 0, 0, 0) ] + 
 \frac{180 \Re[ {\cal E}_0(6, 0, 0, 0, 0) ]}{y} \notag \\
P_{\rm Re}(\pi \nabla {\rm E}_{4})&= \frac{4 y^5}{4725}- \frac{15 \zeta_7}{8 y^2} 
 + 5040 \Re[ {\cal E}_0(8, 0,0,0,0) ]  \label{drop3}
 \\
& \! \! \! \! \! \! \!  
 + \frac{12600 \Re[ {\cal E}_0(8, 0,0,0,0,0) ]}{y} + 
 \frac{9450 \Re[ {\cal E}_0(8, 0,0,0,0,0,0) ]}{y^2} \notag
 \\
P_{\rm Re}(\pi \nabla {\rm E}_{2,2} )&=
- \frac{2 y^5}{10125} +  \frac{y^2 \zeta_3 }{45} - \frac{5 \zeta_5}{12}  + \frac{\zeta_3^2}{2 y}
 -  \Big( \frac{2y^2 }{15}    + \frac{ 6 \zeta_3}{y} \Big) \Re[ {\cal E}_0(4, 0, 0)] \notag \\
& \! \! \! \! \! \! \!  + \frac{18 \Re[ {\cal E}_0(4, 0, 0)]^2}{y}  + 36 \Re[ {\cal E}_0(4, 0, 4, 0, 0) + 3 {\cal E}_0(4, 4, 0, 0, 0) + \tfrac{1}{360} {\cal E}_0(4, 0, 0, 0, 0) ] \, .\notag
 \end{align}
The expression (\ref{nholo3}) for ${\rm E}_{2}$ is  invariant under the projection (\ref{defPRE}), so it naturally extends to the product
\begin{align}
P_{\rm Re}({\rm E}_2 \pi \nabla {\rm E}_{2} )&=
\frac{2 y^5}{2025} +  \frac{ y^2 \zeta_3 }{45} - \frac{ \zeta_3^2}{y}
 + \Big( 12 \zeta_3 - \frac{ 8 y^3}{15} \Big) \Re[ {\cal E}_0(4, 0)] + \Big( \frac{12 \zeta_3}{y} - \frac{ 2 y^2 }{15} \Big)  \Re[ {\cal E}_0(4, 0, 0)]   \notag \\
 &
 -  72 \Re[ {\cal E}_0(4, 0)]  \Re[ {\cal E}_0(4, 0, 0)] - \frac{36}{y}   \Re[ {\cal E}_0(4, 0, 0)]^2 \ .
 \label{drop5}
\end{align}
%


\subsubsection{The relation between $Z^{(2)}_{1234}$ and ${\cal I}_{1234}^{(2,0)}$}

When applied to the low-energy expansion (\ref{open0r}) and \eqref{eq:10} of 
${\cal I}^{(2,0)}$, the projection $P_{\rm Re}$ removes all standalone instances of 
${\cal E}_0$ but preserves the real parts $P_{\rm Re}\Re[ {\cal E}_0]=\Re[ {\cal E}_0]$:
\begin{align}
&P_{\rm Re}\big({\cal I}_{1234}^{(2,0)}(s_{ij},\tau)\big) = \pi^2 s_{13} \Big(  
\frac{2y}{15} - \frac{ 3 \zeta_3}{y^2} + \frac{18}{y^2}  \Re[ {\cal E}_0(4, 0, 0)] 
\Big) \notag \\
&\ \ + \pi^2 (s_{13}^2 + 2s_{12}s_{23}) \Big( 
\frac{ 4 y^2 }{945} - \frac{ \zeta_5}{y^3} + \frac{120}{y^2} \Re[ {\cal E}_0(6,0, 0, 0)]  + \frac{120}{y^3} \Re[ {\cal E}_0(6,0,0, 0, 0)] 
\Big) \notag \\
& \ \ + \pi^2 s_{13}(s_{13}^2 - s_{12}s_{23}) \Big(
\frac{ 4 y^3}{945} + \frac{2 \zeta_3}{5}  - \frac{ 5 \zeta_5}{y^2} - \frac{3 \zeta_7}{2 y^4} + \Big(  \frac{72   \zeta_3}{y^2} - \frac{16 y}{5}    \Big)      \Re[ {\cal E}_0(4, 0)] \notag \\
& \ \  \ \  \ \  \ \  \ \  \ \  \ \  \ \  \ \  \ \  \ \   -  \frac{12}{5} \Re[ {\cal E}_0(4, 0, 0)]    - \frac{432 }{y^2}  \Re[ {\cal E}_0(4, 0)] \Re[ {\cal E}_0(4, 0, 0)]
\label{open0s} \\
& \ \  \ \  \ \  \ \  \ \  \ \  \ \  \ \  \ \  \ \  \ \    +\frac{432}{y^2}  \Re[  {\cal E}_0(4, 0, 4, 0, 0)+3{\cal E}_0(4, 4, 0, 0, 0)+ \tfrac{1}{360} {\cal E}_0(4, 0, 0, 0, 0)]
 \notag \\
 & \ \  \ \  \ \  \ \  \ \  \ \  \ \  \ \  \ \  \ \  \ \   + \frac{4032}{y^2}   \Re[ {\cal E}_0(8, 0, 0, 0, 0)]
  + \frac{10080}{y^3}   \Re[ {\cal E}_0(8, 0, 0, 0, 0, 0)] 
  \notag \\
  &  \ \  \ \  \ \  \ \  \ \  \ \  \ \  \ \  \ \  \ \  \ \   + \frac{ 7560 }{y^4}  \Re[ {\cal E}_0(8, 0, 0, 0, 0, 0, 0)]   \Big) 
+ {\cal O}(\ap^4) \, .
\notag
\end{align}
Up to a global prefactor $(2 \pi i)^2$, this expression agrees with the 
esv image (\ref{open0p}) of the open-string integral $Z^{(2)}_{1234}$. 
Hence, we have checked to the order of $\ap^3$ that
\begin{align}
  P_{\rm Re}\big({\cal I}_{1234}^{(2,0)}(s_{ij},\tau)\big) = (2\pi i)^2 \, {\rm esv}  \, &Z^{(2)}_{1234}(s_{ij}, -\tfrac{1}{\tau})   \, , \label{open0t}
\end{align}
and conjecture this relation between open- and closed-string integrals to hold at higher orders as well.
In the order-$\ap^3$ contribution (\ref{beta5trf}) to $Z^{(2)}_{1234}(s_{ij}, -\tfrac{1}{\tau}) $, 
the product in the third line is understood to be mapped to ${\rm esv} \, ({\cal E}_0(4, 0) {\cal E}_0(4, 0, 0))
=({\rm esv} \, {\cal E}_0(4, 0))({\rm esv} \,  {\cal E}_0(4, 0, 0))$, see (\ref{open0p}), i.e.\ without shuffle multiplication 
prior to the application of esv. Similar ad-hoc convention are expected to be possible at higher orders of 
${\cal I}_{1234}^{(2,0)}$ and $Z^{(2)}_{1234}$ such as to satisfy (\ref{open0t}).

Given that the $\ap$-expansion of ${\cal I}_{1234}^{(2,0)}$ is expressible in terms of modular graph forms,
its expansion around the cusp is expected to be of the type (\ref{crder9}),
\beq
{\cal I}_{1234}^{(2,0)}(s_{ij},\tau) = \sum_{m,n=0}^\infty j_{m,n}(s_{ij},y) q^m \bar q^n \, .
\label{cuspex}
\eeq
The coefficients $j_{m,n}(s_{ij},y)$ are series in $s_{ij}$ such that each $\ap$-order comprises Laurent polynomials in $y$.
Since the ambiguities in the evaluation of esv and $P_{\rm Re}$ were pointed out to be ${\cal O}(q  \, \bar q)$,
one can turn (\ref{open0t}) into a well-defined conjecture by dropping terms $\sim q$,
\begin{align}
{\cal I}_{1234}^{(2,0)}(s_{ij},\tau) = (2\pi i)^2 \, {\rm esv}  \, &Z^{(2)}_{1234}(s_{ij}, -\tfrac{1}{\tau})  + {\cal O}(q) \, . \label{welldef}
\end{align}
This form of our conjecture predicts all the coefficients  $j_{0,n}(s_{ij},y)$ of $q^0 \bar q^n$ in (\ref{cuspex})
with $n\in \mathbb N_0$ including the zero mode $j_{0,0}(s_{ij},y)$ from the open-string 
quantity $Z^{(2)}_{1234}$. The omission of ${\cal O}(q)$-contributions 
in (\ref{welldef}) bypasses both the need for the $P_{\rm Re}$ projection in (\ref{open0t}) and the 
incompatibility of esv with the shuffle multiplication.

The modular weight $(2,0)$ of ${\cal I}_{1234}^{(2,0)}(s_{ij},\tau)$ is not at all evident from the relations
(\ref{open0t}) and (\ref{welldef}) with open-string integrals.
Hence, it should be possible to infer the coefficients $j_{m,n}(s_{ij},y)$ in (\ref{cuspex}) with $m\geq 1$ 
that do not have any known open-string counterpart from $j_{0,n}(s_{ij},y)$ via modular properties. 
This approach is particularly tractable as long as an ansatz of modular graph forms of suitable transcendental
weight is available for a given order in $\ap$: For instance, suppose the $\ap^3$ order of ${\cal I}_{1234}^{(2,0)}$ is known 
to involve a $\mathbb Q$-linear combination of $\pi\nabla {\rm E}_{4}, \ {\rm E}_{2} \pi\nabla {\rm E}_{2}$ 
and $\pi\nabla {\rm E}_{2,2}$, cf.\ (\ref{goodrepB}). Then, the coefficients $c_1,c_2,c_3 \in \mathbb Q$ in an ansatz
\beq
{\cal I}_{1234}^{(2,0)} \big|_{\ap^3} = \frac{s_{13}(s_{13}^2 - s_{12}s_{23})}{(\Im \tau)^2} 
(c_1 \pi\nabla {\rm E}_{4}+ c_2 {\rm E}_{2} \pi\nabla {\rm E}_{2} + c_3 \pi\nabla {\rm E}_{2,2})
\label{Qcomb}
\eeq
are uniquely determined to be $(c_1 ,c_2 ,c_3)=(\frac{4}{5},6,12)$ by (\ref{open0t}) and (\ref{welldef}).
At the $\ap^4$-order of ${\cal I}_{1234}^{(2,0)}$, one could envision a $(4+4)$-parameter ansatz 
comprising $\pi\nabla {\rm E}_{5}, \ {\rm E}_{2} \pi\nabla {\rm E}_{3}, \ {\rm E}_{3} \pi\nabla {\rm E}_{2}$ 
and $\pi\nabla {\rm E}_{2,3}$ (see section 4.2 of \cite{Broedel:2018izr} for the modular graph function ${\rm E}_{2,3}$)
along with both $s_{12}s_{23}s_{13}^2$ and $s_{12}^4 - 4 s_{12}^2 s_{23}^2 + s_{23}^4$.
A more systematic single-valued map extending our present approach involving $P_{\rm Re}$
will be discussed in \cite{futureref}.


\subsubsection{Integration cycles versus elliptic functions}

It is amusing to compare the single-valued relation between genus-zero integrals in (\ref{svmzv4})
with our present evidence for an elliptic single-valued correspondence between open and closed strings.
At tree level, the single-valued map of MZVs was found to relate integration cycles (\ref{svmzv1}) on a disk boundary
to Parke--Taylor factors $(z_{12}z_{23}\ldots z_{n1})^{-1}$. At genus one, the two links (\ref{esvZI})
and (\ref{open0t}) between open- and closed-string $\ap$-expansions suggest that integration cycles 
on a cylinder boundary translate into combinations of the elliptic functions $V_w$ in (\ref{1.1a}).

It would be interesting to explain the correspondence between symmetrized open-string cycles and
$V_0(1,2,\ldots,n)=1$ as well as the four-point cycles of $Z^{(2)}_{1234}$ and $V_2(1,2,3,4)$ from the viewpoint
of Betti-deRham duality \cite{betti1, betti2}. The general dictionary between $V_w(1,2,\ldots,n)$ functions
in a closed-string integrand and formal sums of integration cycles $\{(z_1,\ldots, z_n)\in \mathbb R^n ,$ $ 0 {<}z_1{<}z_2{<}\ldots{<}z_n{<}1 \}$ 
on the open-string side will be explored in a sequel of this work \cite{toappsoon}.

One might wonder if the integral ${\cal I}_{1234}^{(4,0)}$ over the elliptic function $V_4(1,2,3,4)$ also
admits an open-string correspondent along the lines of (\ref{esvZI}) and (\ref{open0t}). However, the
independent permutations of $V_0(1,2,3,4)=1$ and $V_2(1,2,3,4)$ already exhaust the three combinations
of four-point cycles that share the invariance under reflection $z_j \rightarrow 1-z_j$ of the even-weight 
$V_{2k}(1,2,3,4)$. Moreover, since ${\cal I}_{1234}^{(4,0)}= {\rm G}_4(1+6s_{13})+\frac{3 s_{13} \hat {\rm G}_2 \pi \nabla {\rm E}_2}{(\Im \tau)^2} +{\cal O}(\ap^2)$ violates uniform transcendentality, it might be hard
to identify a suitable open-string integral with the same property. But it might be a more
tractable problem to identify open-string counterparts for the conjecturally uniformly transcendental
integrals $\widehat  {\cal I}^{(4,0)}_{1234}$, $\widehat  {\cal I}^{(2,0)}_{12|34}$ and 
$\widehat  {\cal I}^{(4,0)}_{12|34}$ in (\ref{uniform21}), (\ref{3.5NP}) and (\ref{3.4NPNP}), respectively.


\section{Conclusions}
\label{sec:concl}

In this work, we have developed techniques to systematically determine low-energy expansions of one-loop
heterotic-string amplitudes. At each order in $\ap$, the integrations over the punctures are performed
at the level of the Fourier modes in a lattice-sum representation of the correlation functions.
Consequently, the coefficients in the $\ap$-expansion are identified to be modular graph forms for any
combination of external gravitons and gauge bosons. Explicit results are given for the $\tau$-integrand of the four-point amplitude of non-abelian gauge bosons up to the third subleading order in $\ap$ -- both in the planar and in the non-planar sector. The integral over the modular parameter up to the second order in $\ap$ have also been carried out in both cases.

We have furthermore pointed out a striking relation between a specific integral in the planar 
four-gauge-boson amplitude and a corresponding open-string integral over cylinder punctures. 
In this way, we provide another incarnation of the idea that closed-string integrals follow from
open-string quantities via suitable formal operations -- single-valued maps of the periods in the 
respective $\ap$-expansions. Our results support a recent proposal for an elliptic single-valued map which arose from
inspecting the eMZVs in open-string $\ap$-expansions and modular graph functions in closed-string $\ap$-expansions
at one loop \cite{Broedel:2018izr}. The proposal in the reference only applies to situations where the open-string punctures
are symmetrized, i.e.\ independently integrated over the entire cylinder boundary. The heterotic-string integral in this work,
by contrast, relates to generic (i.e.\ non-symmetrized) cyclic orderings on the open-string side under the tentative elliptic 
single-valued map.

These new connections between four-point open- and closed-string integrals point towards the 
$n$-point systematics of relating Koba--Nielsen integrals over cylinder and torus punctures. Integration
cycles on the cylinder boundaries should translate into elliptic functions in closed-string integrands via
Betti-deRham duality \cite{betti1, betti2}. Since we have identified the open-string integration cycles 
dual to the elliptic $V_2(1,2,3,4)$ function (\ref{hetgraph4}) of modular weight (2,0), one can expect
a similar dictionary for various elliptic $V_w$ functions at higher multiplicity \cite{toappsoon}.

Still, the proposal for an elliptic single-valued map \cite{Broedel:2018izr} raises several open questions: First, the 
empirical prescription (\ref{open0m}) to replace iterated Eisenstein integrals by twice their real parts is
incompatible with shuffle multiplication and awaits some reformulation to be 
given in~\cite{futureref}. Second, it remains to make
contact with Brown's construction of single-valued eMZVs via equivariant iterated Eisenstein integrals 
\cite{Brown:2017qwo, Brown:2017qwo2}. Third, the conjectural relation between open- and closed-string integrals
in this work only predicts certain subsectors when expanding closed-string data around the cusp, see (\ref{welldef}) for details.
Given that the main results in this work concern modular graph forms of non-zero modular weights, we hope that they
harbor new input on the above open questions.

Some of the Koba--Nielsen integrals in the four-gauge-boson amplitude are found to violate uniform transcendentality 
in their $\ap$-expansion. These integrals are rewritten using integration-by-parts relations such as to alleviate
certain technical challenges in their $\ap$-expansion -- repeated holomorphic subgraph reduction or 
regularization of non-absolutely convergent lattice sums. Our integration-by-parts manipulations yield 
combinations of integrals which are individually believed to have a uniformly transcendental $\ap$-expansion beyond the lowest orders where we have checked this explicitly. 
On these grounds, our rewritings should be instrumental in
constructing a uniform-transcendentality basis for the twisted cohomology for four-point Koba--Nielsen integrals at genus one.

It would be particularly exciting to explore higher-genus generalizations of the present results. 
Recent progress in the low-energy expansion at genus $g\geq 2$ has been initiated by the identification \cite{DHoker:2013fcx} 
and the computation \cite{DHoker:2014oxd} of the Zhang--Kawazumi invariant in two-loop amplitudes of type-II superstrings
at the subleading order in $\ap$. Moreover, a higher-genus definition of modular graph functions has been recently proposed 
in \cite{DHoker:2017pvk, DHoker:2018mys}, along with a comprehensive investigation of their degeneration limits.
It would be interesting to extend the analysis of the references to higher-genus modular graph forms, to establish
their appearance in heterotic-string amplitudes at $g\geq 2$ loops and to connect with the respective open-string 
$\ap$-expansions.


\section*{Acknowledgments}

We are grateful to the Hausdorff Center for Mathematics Bonn for providing a stimulating atmosphere, 
support, and hospitality through the Hausdorff Trimester Program ``Periods in Number Theory,
Algebraic Geometry and Physics'' and the workshop ``Amplitudes and Periods''.
Moreover, we would like to thank Guillaume Bossard, Johannes Broedel, Justin Kaidi, Carlos Mafra, Erik Panzer,
Oliver Schnetz and Federico Zerbini for combinations of inspiring discussions
and collaboration on related topics. We would like to thank Justin Kaidi, Sebastian Mizera 
and Federico Zerbini for valuable comments on the manuscript.
JG thanks the Perimeter Institute for hospitality during the initial stages of this work. Moreover, JG is supported by the International Max Planck Research School for
Mathematical and Physical Aspects of Gravitation, Cosmology and Quantum Field Theory. This research was supported in part by Perimeter Institute 
for Theoretical Physics. Research at Perimeter Institute is supported by the Government of Canada through the Department 
of Innovation, Science and Economic Development Canada and by the Province of Ontario through 
the Ministry of Research, Innovation and Science.


\appendix


\section{Relations among modular graph forms}
\label{app:B}

In this appendix, we present the relations among modular graph forms which are necessary to simplify the heterotic graph forms in the sections \ref{sec:3.1} and \ref{sec:3.3} and comment on some general properties of identities between modular graph forms.

Note that up to the first order in $\ap$, the identities \eqref{mgfrel4} and \eqref{eq:28} given in the main text are sufficient to simplify the planar heterotic graph forms to single lattice sums. These are also sufficient to simplify the order $\ap^{2}$ of $\mathcal{I}^{(2,0)}_{1234}$. For the non-planar heterotic graph forms one needs also the identity \eqref{mgf9} on top of \eqref{mgfrel4} and \eqref{eq:28}, but these relations are then sufficient for all simplifications up to third order in $\ap$.


\subsection{Relations among dihedral graphs at weight $(4,0)$ and order $\ap^{2}$}
\label{app:B1}
The following identities are instrumental in evaluating the $\ap^2$-order of the 
integral ${\cal I}^{(4,0)}_{1234}$ in (\ref{3.2}) and can be derived by applying the momentum-conservation identities
(\ref{mgfrel1}) along with (\ref{mgfrel3}),
\begin{align}
\cform{4&1&1\\1&1&0} &=   - \frac{3}{2} \cform{6&0\\2&0}  +\frac{1}{2} \left(\cform{3&0\\1&0} \, \right)^2 \, ,  &  \cform{4&1&1\\0 &1&1} &=   3 \cform{6&0\\2&0}  - \left(\cform{3&0\\1&0}\, \right)^2\notag\\
\cform{2&2&2\\1&1&0} &=0 \, ,  &  \cform{3&2&1\\1&1&0}&= \frac{1}{2}  \cform{6&0\\2&0}   - \frac{1}{2} \left( \cform{3&0\\1&0}\, \right)^2 \label{manyrels}\\
\cform{3&2&1\\1&0&1}&= - \frac{1}{2}  \cform{6&0\\2&0}   + \frac{1}{2}  \left(\cform{3&0\\1&0}\, \right)^2  \, , &  \cform{3&2&1\\0&1&1}&=0\ .\notag
\end{align}

\subsection{Relations among dihedral graphs at weight $(2,0)$ and order $\ap^{3}$}
In order to perform simplifications at the third order in $\ap$ of $\mathcal{I}^{(2,0)}_{1234}$, one needs the following identities,
\begin{align}
  \cform{2&1&1&1\\1&1&1&0}&=-{\rm E}_{2}\cform{3&0\\1&0}-2 \cform{3&1&1\\1&1&1}+2 \cform{5&0\\3&0}\notag\\
  \cform{2&1&1&1\\0&1&1&1}&=3{\rm E}_{2}\cform{3&0\\1&0}+6 \cform{3&1&1\\1&1&1}-6 \cform{5&0\\3&0}\notag\\
  \cform{2&2&1\\1&1&1}&=0\label{eq:14}\\
  \cform{2&2&1\\2&0&1}&=-\frac{1}{2}\cform{3&1&1\\1&1&1}+\cform{5&0\\3&0}\notag\\
  \cform{2&2&1\\2&1&0}&=\frac{1}{2}\cform{3&1&1\\1&1&1}-\cform{5&0\\3&0}\notag\\
  \cform{3&1&1\\2&1&0}&=-\frac{1}{2}\cform{3&1&1\\1&1&1}\notag\ .
\end{align}
The first one of these can be verified up to a function of $\bar\tau$ of modular weight $(0,-2)$ by acting on both sides with the Cauchy--Riemann operator $\nabla$ in \eqref{crder1}. 
The vanishing of the constant follows from the corollary $H^{(2,0)}_{1234}[G_{12}G_{13}G_{23}]=0$ of \eqref{hetgraph21} as can be seen by inserting
a trihedral identity from \eqref{eq:12} and one of the other identities of \eqref{eq:14} into \eqref{eq:7}.
The remaining identities of \eqref{eq:14} can all be derived from the momentum-conservation identities \eqref{mgfrel1} and \eqref{mgfrel2}.


\subsection{Eisenstein regularized sums}
\label{app:Eisreg}

The procedure of holomorphic subgraph reduction~\cite{DHoker:2016mwo} is based on performing partial-fraction decomposition on the summand of the lattice sum using the momentum of the holomorphic edge. This produces sums of the form
\begin{align}
  \sum_{\substack{p\neq0\\p\neq q}} \frac{1}{p^{n}} \, ,\quad n\geq 1\ .\label{eq:17}
\end{align}
These can be performed explicitly and hence the original graph is simplified. For $n\geq3$, the sums \eqref{eq:17} are absolutely convergent and evaluate to ${\rm G}_{n}-\frac{1}{q^{n}}$. For $n=1,2$, however, \eqref{eq:17} is not absolutely convergent and hence ill-defined. This ambiguity arises because the original, absolutely convergent, sum was illegally distributed over terms which are not absolutely convergent. One can make this step well-defined by first committing to a summation prescription for which all the sums \eqref{eq:17} are convergent and then distributing the sum. A summation prescription which satisfies this criterion is the so-called Eisenstein summation~$\esum$ which is defined by~\cite{Schoen:1974, nettequelle}
\begin{align}
    \esum_{\substack{p\neq 0\\p \neq q}}f(p):=\lim_{M\rightarrow\infty}\sum_{\substack{m=-M\\m\neq0}}^{M}\left(\lim_{N\rightarrow\infty}\sum_{n=-N}^{N}f(m\tau+n)\right)+\lim_{N\rightarrow\infty}\sum_{\substack{n=-N\\n\neq0}}^{N}f(n)-f(q)\ .\label{eq:19}
\end{align}
Under Eisenstein summation we obtain (cf.\ (\ref{1.6BB}))
\begin{align}
  \esum_{\substack{p\neq0\\p\neq q}} \frac{1}{p}&=-\frac{1}{q}\, ,&\esum_{\substack{p\neq0\\p\neq q}} \frac{1}{p^{2}}&=-\frac{1}{q^{2}}+{\rm \hat G}_{2}+\frac{\pi}{\Im \tau}\ .
    \label{someqqsB}
\end{align}
Note that the right-hand side for the sum over $p^{-2}$ is not modular covariant since the 
contributions $\frac{\pi}{\Im\tau}$ and ${\rm \hat G}_{2}$ have different modular weights. As explained in the following, these cancel from the final expressions for the present dihedral modular graph forms \cite{DHoker:2016mwo} and also for more general cases of holomorphic subgraph reduction \cite{Gerken:2018zcy}.
In the partial-fraction decomposition of modular graph forms, also shifted sums of the form
\begin{align}
  \esum_{\substack{p\neq0\\p\neq q}} \frac{1}{(q-p)^{n}} \, ,\quad n\geq 1
\end{align}
appear. For $n\geq2$, the shift (and sign flip) of the summation variable is irrelevant. For $n=1$, however, we obtain
\begin{align}
  \esum_{\substack{p\neq0\\p\neq q}} \frac{1}{q-p}=-\frac{1}{q}-\frac{\pi}{\Im\tau}(q-\bar q)\ .
    \label{someqqsA}
\end{align}
The right-hand side is again not modular covariant due to the $\frac{\pi}{\Im\tau}q$ term. When
assembling the partial-fraction decomposition of modular graph forms, all
the terms of the wrong modular weight are found to cancel out, and we are left with modular covariant 
results. The terms in curly brackets in \eqref{eq:42} are due to the contributions from $n=1,2$.

Note that in deriving the general expression \eqref{eq:42}, it is convenient to use the functions
\begin{align}
  Q_{1}(q)&:=-\frac{1}{q}-\frac{\pi}{2\Im\tau}(q-\bar q) \, ,& Q_{2}(q)&:=-\frac{1}{q^{2}}+{\rm \hat G}_{2}+\frac{\pi}{\Im\tau}
  \label{someqqs}
\end{align}
and then to replace
\begin{align}
  \sum_{\substack{p\neq0\\p\neq q}}\frac{1}{p}&\rightarrow Q_{1}(q) \, , & \sum_{\substack{p\neq0\\p\neq q}}\frac{1}{q-p}&\rightarrow Q_{1}(q) \, ,& \sum_{\substack{p\neq0\\p\neq q}}\frac{1}{p^{2}}&\rightarrow Q_{2}(q)\ .
  \label{noneisen}
\end{align}
In the expressions obtained form partial-fraction decomposition of modular graph forms, 
this is equivalent to performing the Eisenstein summation as outlined above.


\subsection{Relations among trihedral graphs}
\label{app:B3}
Also for trihedral modular graph forms there exist similar simplification identities as for dihedral modular graph forms. The trihedral momentum-conservation identity is very similar to its dihedral analogue \eqref{mgfrel1} and given by~\cite{DHoker:2016mwo}
\begin{align}
  \sum_{i=1}^{R_{1}}\cformtri{A_{i}\\B\vphantom{A_{i}}}{C\vphantom{A_{i}}\\D\vphantom{A_{i}}}{E\vphantom{A_{i}}\\F\vphantom{A_{i}}}-\sum_{j=1}^{R_{2}}\cformtri{A\vphantom{C_{j}}\\B\vphantom{C_{j}}}{C_{j}\\D\vphantom{C_{j}}}{E\vphantom{C_{j}}\\F\vphantom{C_{j}}}&=0\notag\\
  \sum_{i=1}^{R_{1}}\cformtri{A\vphantom{B_{i}}\\B_{i}}{C\vphantom{B_{i}}\\D\vphantom{B_{i}}}{E\vphantom{B_{i}}\\F\vphantom{B_{i}}}-\sum_{j=1}^{R_{2}}\cformtri{A\vphantom{D_{j}}\\B\vphantom{D_{j}}}{C\vphantom{D_{j}}\\D_{j}\vphantom{D_{j}}}{E\vphantom{D_{j}}\\F\vphantom{D_{j}}}&=0\ , \label{eq:13}
\end{align}
where $A=(a_{1},\dots,a_{R_{1}})$ and similarly for $B$, $C$, $D$, $E$ and $F$ as in figure \ref{fig:trihedralMGF}.
Moreover, we are using the shorthand $A_{i}=(a_{1},\dots,a_{i-1},a_{i}{-}1,a_{i+1},\dots,a_{R_{1}})$ and similarly for $B_{i}$, $C_{i}$ and $D_{i}$. Since the order of the blocks in the argument of $\mathcal{C}$ is irrelevant, \eqref{eq:13} straightforwardly generalizes to any pair of blocks, i.e.\ with $(C,D)\leftrightarrow(E,F)$
or $(A,B)\leftrightarrow(E,F)$ interchanged.

As in the dihedral case \eqref{mgfrel2}, whenever an edge carries the trivial decoration $(0,0)$, one summation variable can be removed~\cite{DHoker:2016mwo}
\begin{align}
  \cformtri{A&0\\B&0}{C\\D}{E\\F}=(-1)^{|C|+|D|}\cform{C&E\\D&F}\prod_{i=1}^{R_{1}} \cform{a_{i}&0\\b_{i}&0}-\cformtri{A\\B}{C\\D}{E\\F}\ .\label{eq:15}
\end{align}
For trihedral graphs of the form $\cformtri{0\\0}{C\\D}{E\\F}$ one can use \eqref{eq:15} with the empty product set to one and the replacement
\begin{align}
  \cformtri{}{C\\D}{E\\F}:=\cform{C\\D}\cform{E\\F}\ .
\end{align}
If two blocks carry just one edge, the trihedral graph simplifies to a dihedral one,\footnote{Note that when specializing \eqref{eq:15} to $\cformtri{a&0\\b&0}{c\\d}{e_{1}&e_{2}\\f_{1}&f_{2}}$ and using \eqref{eq:16}, there appears to be an overall $(-1)^{c+d}$ difference compared to (7.12) of \cite{DHoker:2016mwo}. This is due to the reversed sign in the $(c,d)$ column used in this section of \cite{DHoker:2016mwo}, cf.\ (7.8) therein.}
\begin{align}
  \cformtri{a\\b}{c\\d}{E\\F}=(-1)^{a+b+c+d}\cform{a+c&E\\b+d&F}\ .\label{eq:16}
\end{align}
To simplify the third order in $\ap$ of $H^{(2,0)}_{1234}$, we need the trihedral identities
\begin{align}
  \cformtri{1&1\\1&1}{1&1\\1&0}{1\\0}&=\cform{3&1&1\\1&1&1}+\cform{2&1&1&1\\1&1&1&0}+{\rm E}_{2}\cform{3&0\\1&0}\notag\\
  \cformtri{1&1\\1&0}{1&1\\1&0}{1\\1}&=\frac{1}{2}\cform{3&1&1\\1&1&1}+\frac{1}{2}\cform{2&1&1&1\\1&1&1&0}+\frac{1}{2}{\rm E}_{2}\cform{3&0\\1&0}\label{eq:12}\ ,
\end{align}
which can be derived from the identities \eqref{eq:13} and \eqref{eq:15}.

As in the dihedral case, holomorphic subgraph reduction is also possible for trihedral graphs~\cite{DHoker:2016mwo}. In this case, the holomorphic subgraph can be either two-valent (i.e.\ the lattice sum is of the form $\cformtri{a_{1}&a_{2}&A\\0&0&B}{C\\D}{E\\F}$) or three-valent (so that the lattice sum has the form $\cformtri{a_{1}&A\\0&B}{a_{2}&C\\0&D}{a_{3}&E\\0&F}$). While the reduction of two-valent subgraphs is a straightforward generalization of dihedral holomorphic subgraph reduction (\ref{eq:42}) and given explicitly in \cite{Gerken:2018zcy},
the decomposition of three-valent holomorphic subgraphs yields new sums of the form
\begin{align}
    \sum_{\substack{p\neq0\\p\neq q_{1},q_{2}}} \frac{1}{p^{n}} \, ,\quad n\geq 1
\end{align}
generalizing \eqref{eq:17}. As in the dihedral case, one can perform a replacement
\begin{align}
  \sum_{\substack{p\neq0\\p\neq q_{1},q_{2}}}\frac{1}{p}&\rightarrow Q_{1}(q_{1},q_{2}) \, ,& \sum_{\substack{p\neq0\\p\neq q_{1},q_{2}}}\frac{1}{q_{1}-p}&\rightarrow Q_{1}(q_{1},q_{1}{-}q_{2})  \, , & \sum_{\substack{p\neq0\\p\neq q_{1},q_{2}}}\frac{1}{p^{2}}&\rightarrow Q_{2}(q_{1},q_{2})
\end{align}
of the conditionally convergent sums by suitably defined functions \cite{Gerken:2018zcy}
\begin{align}
  Q_{1}(q_{1},q_{2})&:=-\frac{1}{q_{1}}-\frac{1}{q_{2}}-\frac{\pi}{3\Im\tau}(q_{1}+q_{2}-\bar q_{1}-\bar q_{2})\notag\\
  Q_{2}(q_{1},q_{2})&:=-\frac{1}{q_{1}^{2}}-\frac{1}{q_{2}^{2}}+{\rm \hat G}_{2}+\frac{\pi}{\Im\tau}\, .\label{eq:18}
\end{align}
Once the expressions \eqref{eq:18} are known, an explicit (though combinatorially involved) formula for trihedral 
three-valent holomorphic subgraph reduction can be obtained, see \cite{Gerken:2018zcy} for details. Again, the
terms of incorrect modular weight due to $Q_{1}$ and $Q_{2}$ in (\ref{eq:18})  cancel out in the final expression.

Using this explicit formula for the trihedral modular graph forms appearing at order $\ap^{2}$ in $\mathcal{I}^{(4,0)}_{1234}$, we obtain
\begin{align}
  \cformtri{2&1\\0&1}{1&1\\1&0}{1\\0}&=-{\rm \hat G}_{2} \cform{ 4 & 0 \\ 2 & 0 }-{\rm \hat G}_{2} \cform{ 2 & 1 & 1 \\ 1 & 1 & 0 }-\cform{ 5 & 0 \\ 1 & 0 }\notag\\
             &\hphantom{=}\ +4 \cform{ 6 & 0 \\ 2 & 0 }-\cform{ 3 & 1 & 1 \\ 0 & 0 & 1 }-\cform{ 3 & 2 & 1 \\ 0 & 1 & 1 }+3 \cform{ 4 & 1 & 1 \\ 1 & 1 & 0 }\notag\\
  \cformtri{ 1 & 1 \\ 1 & 0 }{ 1 & 1 \\ 1 & 0 }{ 2 \\ 0 }&=2 {\rm \hat G}_{2} \cform{ 4 & 0 \\ 2 & 0 }+2 {\rm \hat G}_{2} \cform{ 2 & 1 & 1 \\ 1 & 1 & 0 }+2 \cform{ 5 & 0 \\ 1 & 0 }\\
  &\hphantom{=}\ -6 \cform{ 6 & 0 \\ 2 & 0 }+2 \cform{ 3 & 1 & 1 \\ 0 & 0 & 1 }+2 \cform{ 3 & 2 & 1 \\ 1 & 0 & 1 }-6 \cform{ 4 & 1 & 1 \\ 1 & 1 & 0 }\ .\notag
\end{align}
Simplifying these using the identities from section \ref{app:B1} as well as the dihedral holomorphic subgraph relation \eqref{eq:28} yields the trihedral decompositions of \eqref{greatrels}.


\subsection{Tetrahedral holomorphic subgraph reduction}
\label{sec:tet-hsr}
The tetrahedral graph $\mathcal{C}_{\rm tet}$ defined in \eqref{tetgraph} can be decomposed using tetrahedral holomorphic subgraph reduction. It is convenient to employ the following
partial-fraction decomposition of the sum \eqref{tetgraph} w.r.t.\ $p_{3}$,
\begin{align}
  {\cal C}_{\rm tet}  &=\left(\frac{\Im\tau}{\pi}\right)^{2}\hspace{-1em}\sum_{\substack{p_1,p_2,p_3\neq0\\p_{1}+p_{3}\neq0\\p_{2}+p_{3}\neq0\\p_{1}+p_{2}+p_{3}\neq0\\p_{1}\neq\pm p_{2}}}\frac{1}{|p_1|^2|p_{2}|^{2}p_{3}(p_{1}{+}p_{3})(p_{2}{+}p_{3})(p_{1}{+}p_{2}{+}p_{3})}+\mathcal{C}_{\rm tet}^{p_{1}=p_{2}}+\mathcal{C}_{\rm tet}^{p_{1}=-p_{2}}\notag\\
  &= \left(\frac{\Im\tau}{\pi}\right)^{2}\hspace{-1em}  \sum_{\substack{p_1,p_2,p_3\neq0\\p_{1}+p_{3}\neq0\\p_{2}+p_{3}\neq0\\p_{1}+p_{2}+p_{3}\neq0\\p_{1}\neq\pm p_{2}}}\frac{1}{|p_1|^2|p_{2}|^{2}}\left[\frac{1}{p_1 \left(p_1{-}p_2\right) p_2 \left(p_1{+}p_3\right)}-\frac{1}{p_1 \left(p_1{-}p_2\right) p_2 \left(p_2{+}p_3\right)}\right.\label{eq:23}\\[-2.5em]
                      &\hphantom{\left(\frac{\Im\tau}{\pi}\right)^{2}\hspace{-1em}\sum_{\substack{p_1,p_2,p_3\neq0\\p_{1}+p_{3}\neq0\\p_{2}+p_{3}\neq0\\p_{1}+p_{2}+p_{3}\neq0}}}\hspace{1.5em}\left.-\frac{1}{p_1 p_2 \left(p_1{+}p_2\right) \left(p_1{+}p_2{+}p_3\right)}+\frac{1}{p_1 p_2 \left(p_1{+}p_2\right) p_3}\right]+\mathcal{C}_{\rm tet}^{p_{1}=p_{2}}+\mathcal{C}_{\rm tet}^{p_{1}=-p_{2}}\ ,\notag
\end{align}
where  $\mathcal{C}_{\rm tet}^{p_{1}=\pm p_{2}}$ is the sum \eqref{tetgraph} with $p_{1}$ set to $\pm p_{2}$. 
The terms $\mathcal{C}_{\rm tet}^{p_{1}=\pm p_{2}}$ involving fewer momenta are of lower complexity and can be simplified using the techniques of section \ref{app:B3}. The $p_{3}$-sum in the remaining 
contributions to (\ref{eq:23}) can be performed by means of the replacements
\begin{align}
  \sum_{\substack{p\neq0\\p\neq q_{1},q_{2},q_{3}}}\frac{1}{p}&\rightarrow Q_{1}(q_{1},q_{2},q_{3})  \, ,& \sum_{\substack{p\neq0\\p\neq -q_{1},q_{2},q_{3}}}\frac{1}{p+q_{1}}&\rightarrow Q_{1}(q_{1},q_{1}{+}q_{2},q_{1}{+}q_{3})\, , \label{eq:21}
\end{align}
where \cite{Gerken:2018zcy}
\begin{align}
  Q_{1}(q_{1},q_{2},q_{3})&:=-\frac{1}{q_{1}}-\frac{1}{q_{2}}-\frac{1}{q_{3}}-\frac{\pi}{4\Im\tau}(q_{1}+q_{2}+q_{3}-\bar q_{1}-\bar q_{2}-\bar q_{3})\ .\label{eq:22}
\end{align}
Putting everything together, \eqref{eq:23} can be written in terms of dihedral modular graph forms,
\begin{align}
  \mathcal{C}_{\rm tet}=-2 {\rm \hat G}_{2} \cform{ 4 & 0 \\ 2 & 0 }+2 \cform{ 6 & 0 \\ 2 & 0 }+4 \cform{ 2 & 1 & 2 \\ 0 & 0 & 1 }+4 \cform{ 2 & 2 & 2 \\ 1 & 1 & 0 }-8 \cform{ 3 & 2 & 1 \\ 1 & 1 & 0 }\ .
\end{align}
After using \eqref{eq:28} and the identities from section \ref{app:B1}, this simplifies to the last line in \eqref{greatrels}.


\section{The $V_2(1,2,3,4)$ integral at the third order in $\ap$}
\label{app:Z}
In this appendix, we list the heterotic graph forms contributing to $\mathcal{I}^{(2,0)}_{1234}$ at order $\ap^{3}$.

\subsection{The twelve inequivalent heterotic graph forms}
\label{sec:Z.1}
Table \ref{tableplan} lists the twelve inequivalent heterotic graph forms at order $\ap^{3}$ in the planar sector. The  contributions of modular weight $(2,0)$ are explicitly given by
\begin{align}
  H^{(2,0)}_{1234}[G_{12}^3]&= -\cform{2&1&1&1\\0&1&1&1}\notag\\
  H^{(2,0)}_{1234}[G_{13}^3]&=2\cform{2&1&1&1\\0&1&1&1}\notag\\
  H^{(2,0)}_{1234}[G_{12}^2 G_{13}]&=-\cform{3&1&1\\1&1&1}-\cform{2&1&1&1\\1&1&1&0}+{\rm E}_{2}\cform{3&0\\1&0}\notag\\
  H^{(2,0)}_{1234}[G_{12}^2 G_{14}]&=\cform{3&1&1\\1&1&1}- {\rm E}_{2}\cform{3&0\\1&0}\notag\\
  H^{(2,0)}_{1234}[G_{12}^2 G_{34}]&=\cform{3&1&1\\1&1&1}- {\rm E}_{2}\cform{3&0\\1&0}\notag\\
  H^{(2,0)}_{1234}[G_{13}^2G_{12}]&=-\cform{3&1&1\\1&1&1}+\cformtri{1&1\\1&1}{1&1\\1&0}{1\\0}-{\rm E}_{2}\cform{3&0\\1&0}\label{eq:7}\\
  H^{(2,0)}_{1234}[G_{13}^2 G_{24}]&= -2\cform{3&1&1\\1&1&1}+2{\rm E}_{2}\cform{3&0\\1&0}\notag\\
  H^{(2,0)}_{1234}[G_{12} G_{13} G_{14}]&=-\cform{2&2&1\\1&1&1}\notag\\
  H^{(2,0)}_{1234}[G_{12} G_{13} G_{23}]&=-\cform{2&2&1\\2&0&1}+\cformtri{1&1\\1&0}{1&1\\1&0}{1\\1}\notag\\
  H^{(2,0)}_{1234}[G_{12} G_{13} G_{24}]&=-\cform{5&0\\3&0}+\cform{2&2&1\\1&1&1}-2\cform{2&2&1\\2&1&0}+\cform{3&1&1\\2&1&0}\notag\\
  H^{(2,0)}_{1234}[G_{12} G_{13} G_{34}]&=-\cform{2&2&1\\1&1&1}\notag\\
  H^{(2,0)}_{1234}[G_{12} G_{14} G_{23}]&=-\cform{5&0\\3&0}-3\cform{3&1&1\\2&1&0}\notag\ .
\end{align}


\subsection{Identities among modular graph forms at order $\ap^3$}
\label{sec:Z.2}
Using the relations from appendix \ref{app:B}, the heterotic graph forms \eqref{eq:7} can be simplified. We obtain e.g.
\begin{align}
  H^{(2,0)}_{1234}[G_{12}^3]&=-6 \cform{3&1&1\\1&1&1}+6 \cform{5&0\\3&0}-3 {\rm E}_{2}\cform{3&0\\1&0}\notag\\
  H^{(2,0)}_{1234}[G_{12}^2 G_{13}]&=\cform{3&1&1\\1&1&1}-2 \cform{5&0\\3&0}+2{\rm E}_{2}\cform{3&0\\1&0}\label{eq:9}\\
  H^{(2,0)}_{1234}[G_{12}^2 G_{14}]&=\cform{3&1&1\\1&1&1}-{\rm E}_{2}\cform{3&0\\1&0}\notag\\
  H^{(2,0)}_{1234}[G_{12} G_{14} G_{23}]&=\frac{3}{2}\cform{3&1&1\\1&1&1}-\cform{5&0\\3&0}\notag\ .
\end{align}
Once this simplification is performed for all heterotic graph forms in \eqref{eq:7}, it is clear that the remaining contributions in \eqref{eq:7} are either related to those in \eqref{eq:9} or zero,
\begin{align}
  H^{(2,0)}_{1234}[G_{13}^3]&=-2H^{(2,0)}_{1234}[G_{12}^3]\notag\\
  H^{(2,0)}_{1234}[G_{12}^2 G_{34}]&=H^{(2,0)}_{1234}[G_{12}^2 G_{14}]\notag\\
  H^{(2,0)}_{1234}[G_{13}^2 G_{12}]&=-H^{(2,0)}_{1234}[G_{12}^2 G_{13}]-H^{(2,0)}_{1234}[G_{12}^2 G_{14}]\notag\\
  H^{(2,0)}_{1234}[G_{13}^2 G_{24}]&=-2H^{(2,0)}_{1234}[G_{12}^2 G_{14}]\label{eq:8}\\
  H^{(2,0)}_{1234}[G_{12} G_{13} G_{24}]&=-H^{(2,0)}_{1234}[G_{12} G_{14} G_{23}]\notag\\
  H^{(2,0)}_{1234}[G_{12} G_{13} G_{14}]&=H^{(2,0)}_{1234}[G_{12} G_{13} G_{23}]=H^{(2,0)}_{1234}[G_{12} G_{13} G_{34}]=0\notag\ .
\end{align}
Note that the expressions \eqref{eq:8} are compatible with the symmetries \eqref{hetgraph21}.


\section{The $V_2(i,j)$ integrals at the third order in $\ap$}
\label{app:Y}

In this appendix, we list the heterotic graph forms contributing to the double-trace sector at order $\ap^{3}$.

\subsection{The inequivalent heterotic graph forms of weight $(2,0)$}
\label{sec:Y.1}
Table \ref{npreg3} lists the twelve inequivalent heterotic graph forms at order $\ap^{3}$ in the 
non-planar sector. Using the regularized values given in \eqref{npreg8} and \eqref{npreg12},
the weight-$(2,0)$ contributions are given by
\begin{align}
  H^{(2,0)}_{12|34}[G_{12}^3]&={\rm \hat G}_{2} (2 ({\rm E}_{3}+\zeta_{3}-3)-3 {\rm E}_{2})+12 \cform{ 3 & 0 \\ 1 & 0 }-6 \cform{ 4 & 0 \\ 2 & 0 }\notag\\
  H^{(2,0)}_{12|34}[G_{13}^3]&=2 {\rm \hat G}_{2} ({\rm E}_{3}+\zeta_{3})\notag\\
  H^{(2,0)}_{12|34}[G_{12}^2G_{34}]&=-{\rm E}_{2} {\rm \hat G}_{2}\notag\\
  H^{(2,0)}_{12|34}[G_{12}G_{13}^{2}]&=-{\rm E}_{2} {\rm \hat G}_{2}\notag\\
  H^{(2,0)}_{12|34}[G_{13}^2G_{14}]&=-\cform{ 4 & 0 \\ 2 & 0 }\\
  H^{(2,0)}_{12|34}[G_{12}G_{13}G_{24}]&=-\cform{ 4 & 0 \\ 2 & 0 }\notag\\
  H^{(2,0)}_{12|34}[G_{13}G_{14}G_{34}]&=2 {\rm E}_{3} {\rm \hat G}_{2}+\cform{ 3 & 0 \\ 1 & 0 }-\frac{5}{2} \cform{ 4 & 0 \\ 2 & 0 }\notag\\
  H^{(2,0)}_{12|34}[G_{12}^2G_{13}]&=H^{(2,0)}_{12|34}[G_{12}G_{13}G_{34}]=H^{(2,0)}_{12|34}[G_{12}G_{13}G_{14}]\notag\\
                             &=H^{(2,0)}_{12|34}[G_{13}^{2}G_{24}]= H^{(2,0)}_{12|34}[G_{13}G_{14} G_{23}]=0\notag\ .
\end{align}
Note that the appearance of $\zeta_{3}$ is due to the modular graph function $C_{1,1,1}$ 
in intermediate steps, see \eqref{mgf9}.


\subsection{The inequivalent heterotic graph forms of weight $(4,0)$}
\label{sec:Y.2}
Similarly to section \ref{sec:Y.1}, the regularized values of the weight-$(4,0)$ contributions are given by
\begin{align}
  H^{(4,0)}_{12|34}[G_{12}^3]&={\rm \hat G}_{2}^{2} ({\rm E}_{3}-3 {\rm E}_{2}+\zeta_{3}-6)+12{\rm \hat G}_{2} \cform{ 3 & 0 \\ 1 & 0 }-6{\rm \hat G}_{2} \cform{ 4 & 0 \\ 2 & 0 }\notag\\
  H^{(4,0)}_{12|34}[G_{13}^3]&={\rm \hat G}_{2}^2 ({\rm E}_{3}+\zeta_{3})\notag\\
  H^{(4,0)}_{12|34}[G_{12}^2G_{34}]&=-(2+{\rm E}_{2}) {\rm \hat G}_{2}^{2}+4{\rm \hat G}_{2} \cform{ 3 & 0 \\ 1 & 0 }\notag\\
  H^{(4,0)}_{12|34}[G_{12}G_{13}^{2}]&=-{\rm E}_{2} {\rm \hat G}_{2}^2\notag\\
  H^{(4,0)}_{12|34}[G_{13}^2G_{14}]&=-{\rm \hat G}_{2}\cform{ 4 & 0 \\ 2 & 0 }\\
  H^{(4,0)}_{12|34}[G_{12}G_{13}G_{14}]&={\rm \hat G}_{2} \cform{ 3 & 0 \\ 1 & 0 }\notag\\
  H^{(4,0)}_{12|34}[G_{13}^{2}G_{24}]&=-2 {\rm G}_{4}-2 {\rm \hat G}_{2} \cform{ 3 & 0 \\ 1 & 0 }+8 \cform{ 5 & 0 \\ 1 & 0 }\notag\\
  H^{(4,0)}_{12|34}[G_{12}G_{13}G_{24}]&=-2 {\rm G}_{4}-{\rm \hat G}_{2} \cform{ 3 & 0 \\ 1 & 0 }-{\rm \hat G}_{2} \cform{ 4 & 0 \\ 2 & 0 }+6 \cform{ 5 & 0 \\ 1 & 0 }\notag\\
  H^{(4,0)}_{12|34}[G_{13}G_{14}G_{34}]&={\rm E}_{3} {\rm \hat G}_{2}^2+{\rm \hat G}_{2} \cform{ 3 & 0 \\ 1 & 0 }-\frac{5}{2} {\rm \hat G}_{2} \cform{ 4 & 0 \\ 2 & 0 }\notag\\
  H^{(4,0)}_{12|34}[G_{13}G_{14} G_{23}]&=2 {\rm G}_{4}+2 {\rm \hat G}_{2} \cform{ 3 & 0 \\ 1 & 0 }-6 \cform{ 5 & 0 \\ 1 & 0 }\notag\\
  H^{(4,0)}_{12|34}[G_{12}^2G_{13}]&=  H^{(4,0)}_{12|34}[G_{12}^2G_{13}]=0\notag\ .
\end{align}


\section{Decomposition into uniform-transcendentality integrals}
\label{sec:5}

In this appendix, we give more details on the rearrangements of the integrals in section \ref{sec:unif-transc} and in particular derive the expressions for the conjecturally uniformly transcendental integrals $\mathcal{\widehat I}^{(w,0)}_{\ldots}$ via integrations by parts.

The notion of transcendental weight for modular graph forms can be derived from the terminology for eMZVs (\ref{open0b}), where $\omega(n_1,n_2,\ldots,n_r)$ is said to have
weight $n_1{+}n_2{+}\ldots{+}n_r$. This convention implies weight one for $\pi$, weight $n_1{+}n_2{+}\ldots{+}n_r$
for $\zeta_{n_1,n_2,\ldots,n_r}$ and weight $r$ for iterated Eisenstein integrals ${\cal E}_0(k_1,k_2,\ldots,k_r)$
in (\ref{crder6}). This means that both types of Eisenstein series ${\rm G}_k$ and ${\rm E}_k$ have weight $k$ 
and that $\pi \nabla $ as well as $y=\pi  \Im \tau$ have weight one. Similarly, $(\pi \nabla)^p {\rm E}_k$
and $(\pi \nabla)^p {\rm E}_{2,2}$ are found to carry weight $k{+}p$ and $4{+}p$, respectively.


\subsection{Planar integrals}
\label{sec:3.5}
In this section, we provide a decomposition of ${\cal I}^{(4,0)}_{1234}$ defined in (\ref{3.2}) into the integrals
of uniform transcendentality. In particular, we derive \eqref{ut1}.

\subsubsection{Integration-by-parts manipulations}

The idea is to exploit the fact that total derivatives $\partial_i = \frac{ \partial }{\partial z_i}$ 
of the Koba--Nielsen factor (with $\partial_i G_{ij} = - f^{(1)}_{ij}$)
\begin{align}
{\cal J}_n :=
\prod_{i<j}^n \exp \Big( s_{ij} G_{ij}(\tau) \Big) \, , \ \ \ \ \ \
\partial_i {\cal J}_n = - {\cal J}_n \sum_{j\neq i}^n s_{ij} f^{(1)}_{ij}
\label{uniform08}
\end{align}
integrate to zero. In order to relate this to the constituents $f^{(w)}$ of the integrand $V_4(1,2,3,4)$ 
of ${\cal I}^{(4,0)}_{1234}$, cf.\ (\ref{hetgraph4}), the total $z_i$-derivatives have to furthermore act 
on suitably chosen functions of modular weight three.
As we will see, the identities of interest involve the combination\footnote{OS would like to thank Carlos Mafra for
collaboration on related topics, where the meromorphic version of (\ref{uniform06}) with all the
$\Im z_{j}$ removed has been studied.}
\begin{align}
X^{(3)}_{1234} &= f^{(1)}_{12}f^{(1)}_{23}f^{(1)}_{34} +{1\over 6}( f^{(3)}_{12} + f^{(3)}_{34}) + {2\over 3} f^{(3)}_{23} +{1\over 3} f_{23}^{(1)}( f^{(2)}_{12} + f^{(2)}_{34})  \label{uniform06}\\
& \ \ \ \ \ +  {2\over 3} f^{(2)}_{23} ( f^{(1)}_{12} + f^{(1)}_{34}) + {1\over 2} ( f^{(1)}_{12}  f^{(2)}_{34} + f^{(2)}_{12}  f^{(1)}_{34}) \, ,
\notag
\end{align}
whose derivatives in $z_1,\ldots,z_4$ can be evaluated using
\begin{align}
\partial_z f^{(1)}(z) &= 2 f^{(2)}(z) - \big( f^{(1)}(z) \big)^2 - {\rm \hat G}_2 \notag \\
\partial_z f^{(2)}(z) &= 3 f^{(3)}(z) -  f^{(1)}(z)   f^{(2)}(z)  - {\rm \hat G}_2  f^{(1)}(z) 
\label{uniform04}
\\
\partial_z f^{(3)}(z) &= 4 f^{(4)}(z) -  f^{(1)}(z)   f^{(3)}(z)  - {\rm \hat G}_2  f^{(2)}(z) - {\rm G}_4 \, .\notag
\end{align}
The virtue of the combination (\ref{uniform06}) is that it allows generating $V_4(1,2,3,4)$ and simpler 
elliptic functions by means of total derivatives in the punctures as detailed below. Indeed, by the 
symmetries $X^{(3)}_{1234} = - X^{(3)}_{4321}$ and $X^{(3)}_{1234}+X^{(3)}_{2134}+X^{(3)}_{2314}
+X^{(3)}_{2341}=0$ due to Fay identities \cite{BrownLev, Broedel:2014vla}, one can show that
\begin{align}
\partial_4 X^{(3)}_{1234} +\partial_2 X^{(3)}_{1432}  - \partial_3( X^{(3)}_{1423} +X^{(3)}_{1243} )&= {\rm  G}_4+{\rm \hat G}_2 V_2(1,2,3,4) - V_4(1,2,3,4) \ . \label{eq:133}
\end{align}
In order to relate this to the Koba--Nielsen integral ${\cal I}^{(4,0)}_{1234}$,
we extend (\ref{eq:133}) to the following total derivative via (\ref{uniform08}) 
and $V_4(1,2,3,4)+{\rm cyc}(2,3,4) = 3 {\rm G}_4$,
\begin{align}
&\partial_4 (X^{(3)}_{1234} {\cal J}_4) 
+\partial_2 (X^{(3)}_{1432} {\cal J}_4)
-\partial_3 (X^{(3)}_{1243} {\cal J}_4)
-\partial_3 (X^{(3)}_{1423} {\cal J}_4)   \label{uniform15} \\
&= {\cal J}_4 \big[  {\rm \hat G}_2 V_2(1,2,3,4)-(1+s_{1234}) V_4(1,2,3,4)  + {\rm G}_4 + 3(s_{13}+s_{24}) {\rm G}_4  +\widehat V_4(1,2,3,4) \big]  \ ,
\notag
\end{align}
using the shorthands $s_{1234}:=s_{12}+s_{13}+s_{14}+s_{23}+s_{24}+s_{34}$ and
\begin{align}
\widehat V_4(1,2,3,4)&= s_{12} R_{2|34|1}+s_{23} R_{3|41|2}+s_{34} R_{4|12|3}+s_{14} R_{1|23|4}  \notag
\\
& \ \  -s_{13} (R_{1|24|3}+R_{1|42|3}) -s_{24} (R_{2|13|4}+R_{2|31|4})  \label{uniform16} 
\end{align}
comprising several permutations of
\begin{align}
R_{1|23|4} = 
f_{14}^{(1)} X^{(3)}_{1234} + V_4(1,2,3,4) \, . \label{uniform16B}
\end{align}
In the remainder of this subsection, we elaborate on some of the intermediate steps in (\ref{uniform15}): When the derivatives act 
on the Koba--Nielsen factor ${\cal J}_4$, they generate the terms
\begin{align}
&X^{(3)}_{1234} (s_{14}f^{(1)}_{14} + s_{24}f^{(1)}_{24} + s_{34}f^{(1)}_{34} )
+X^{(3)}_{1432} (s_{12}f^{(1)}_{12} - s_{23}f^{(1)}_{23} - s_{24}f^{(1)}_{24} ) \notag
\\
& \ \ - (X^{(3)}_{1243}+X^{(3)}_{1423})(s_{13}f^{(1)}_{13} + s_{23}f^{(1)}_{23} - s_{34}f^{(1)}_{34} )   \label{uniform17}
\\
&= \big[ s_{14} f^{(1)}_{14} X^{(3)}_{1234}  + {\rm cyc}(1,2,3,4) \big] - s_{13} f^{(1)}_{13} (X^{(3)}_{1243}+X^{(3)}_{1423})
- s_{24} f^{(1)}_{24} (X^{(3)}_{2134}+X^{(3)}_{2314}) \, , \notag
\end{align}
where the symmetries $X^{(3)}_{1234} = - X^{(3)}_{4321}$ and $X^{(3)}_{1234}+X^{(3)}_{2134}+X^{(3)}_{2314}
+X^{(3)}_{2341}=0$ have been used in passing to the last line. However, we have refrained from using
momentum conservation in (\ref{uniform15}) or (\ref{uniform17}) so far. The term $s_{1234}V_4(1,2,3,4)$ in the 
second line of (\ref{uniform15}) has been generated by rewriting each term in (\ref{uniform17}) 
via permutations of (\ref{uniform16B}). The coefficients of $s_{13}$ and $s_{24}$ in 
(\ref{uniform17}) require the additional intermediate step
\begin{align}
-f^{(1)}_{13} (X^{(3)}_{1243}+X^{(3)}_{1423}) &= V_4(1,2,4,3) + V_4(1,4,2,3) - R_{1|24|3} - R_{1|42|3}
 \notag \\
&= 3 {\rm G}_4 -V_4(1,2,3,4) - R_{1|24|3} - R_{1|42|3} 
\label{uniform18}
\end{align}
in reproducing (\ref{uniform15}).

\subsubsection{Uniform transcendentality decomposition}

Using the representation (\ref{uniform06}) and (\ref{hetgraph4}) of its constituents $X^{(3)}_{1234}$ and $V_4(1,2,3,4)$, $R_{1|23|4}$ can be expanded as
\begin{align}
R_{1|23|4} &= f^{(1)}_{ 2 3} f^{(1)}_{ 3 4} f^{(2)}_{ 1 2} 
    +  f^{(1)}_{ 1 2} f^{(1)}_{ 3 4} f^{(2)}_{ 2 3} 
 + f^{(1)}_{ 1 2} f^{(1)}_{ 2 3} f^{(2)}_{ 3 4} 
+ f^{(1)}_{ 1 2} f^{(1)}_{ 2 3} f^{(2)}_{ 1 4} + 
 f^{(1)}_{ 1 2} f^{(1)}_{ 3 4} f^{(2)}_{ 1 4} + 
 f^{(1)}_{ 2 3} f^{(1)}_{ 3 4} f^{(2)}_{ 1 4}  \notag \\
&
 - \frac{1}{3} f^{(1)}_{ 1 2} f^{(1)}_{ 1 4} f^{(2)}_{ 2 3}
 -  \frac{1}{3} f^{(1)}_{ 1 4} f^{(1)}_{ 3 4} f^{(2)}_{ 2 3} 
- \frac{2}{3} f^{(1)}_{ 1 4} f^{(1)}_{ 2 3} f^{(2)}_{ 1 2} 
- \frac{ 2}{3} f^{(1)}_{ 1 4} f^{(1)}_{ 2 3} f^{(2)}_{ 3 4}
   - \frac{1}{2} f^{(1)}_{ 1 4} f^{(1)}_{ 3 4} f^{(2)}_{ 1 2}
 \notag \\
& - \frac{1}{2} f^{(1)}_{ 1 2} f^{(1)}_{ 1 4} f^{(2)}_{ 3 4} + f^{(2)}_{ 1 2} f^{(2)}_{ 1 4} + f^{(2)}_{ 1 2} f^{(2)}_{ 2 3} + 
 f^{(2)}_{ 1 4} f^{(2)}_{ 2 3}   + f^{(2)}_{ 1 2} f^{(2)}_{ 3 4} + 
 f^{(2)}_{ 1 4} f^{(2)}_{ 3 4} + f^{(2)}_{ 2 3} f^{(2)}_{ 3 4} \hspace{-0.2em}
\label{eq:135} \\
& - \frac{5}{6} f^{(1)}_{ 1 4} f^{(3)}_{ 1 2} 
 - \frac{5}{6} f^{(1)}_{ 1 4} f^{(3)}_{ 3 4} 
- \frac{1}{3} f^{(1)}_{ 1 4} f^{(3)}_{ 2 3} 
 - f^{(1)}_{ 1 2} f^{(3)}_{ 1 4} - f^{(1)}_{ 2 3} f^{(3)}_{ 1 4}  - f^{(1)}_{ 3 4} f^{(3)}_{ 1 4} 
 + f^{(1)}_{ 2 3} f^{(3)}_{ 1 2} \notag \\
 &+ f^{(1)}_{ 2 3} f^{(3)}_{ 3 4} 
 +  f^{(1)}_{ 3 4} f^{(3)}_{ 1 2}  + f^{(1)}_{ 1 2} f^{(3)}_{ 3 4} 
+ f^{(1)}_{ 1 2} f^{(3)}_{ 2 3}   + f^{(1)}_{ 3 4} f^{(3)}_{ 2 3}
 + f^{(4)}_{ 1 2} +f^{(4)}_{ 1 4} + f^{(4)}_{ 2 3} + f^{(4)}_{ 3 4} \ .\notag
\end{align}
Hence, all the terms of $R_{1|23|4} $ involve at most three factors of $f^{(n)}_{ij}$, and none of them exhibits a subcycle $f^{(m)}_{ij}f^{(n)}_{ji}$
or $f^{(m)}_{ij}f^{(n)}_{jk}f^{(p)}_{ki}$. This means none of the resulting graphs in the order-by-order integration against monomials in $G_{ij}$ contains closed holomorphic subgraphs and hence the need for holomorphic subgraph reduction is removed.

Apart from $R_{1|23|4}=R_{4|32|1}$, there are no further relations among the 12 reflection-independent 
permutations of $R_{1|23|4}$ in (\ref{uniform16B}). The reflection property is sufficient to show that
$\widehat V_4(1,2,3,4)+ {\rm cyc}(2,3,4)=0$. 

In the momentum phase-space of four particles with $s_{13}=s_{24}$ and $s_{1234}=0$,
one can solve (\ref{uniform15}) for
\begin{align}
V_4(1,2,3,4) \cong
{\rm \hat G}_2 V_2(1,2,3,4) + {\rm G}_4 + 6s_{13} {\rm G}_4  +\widehat V_4(1,2,3,4)
\, ,
\label{uniform19}
\end{align}
where the equivalence relation $\cong$ indicates that total derivatives $\partial_i(\ldots {\cal J}_4)$
have to be discarded in equating (${\cal J}_4$ times) the two sides of (\ref{uniform19}).
At the level of the integrals, this implies
\begin{align}
 {\cal I}^{(4,0)}_{1234}(s_{ij},\tau) = {\rm G}_4  {\cal I}^{(0,0)}_{1234}(s_{ij},\tau) 
 +{\rm \hat G}_2 {\cal I}^{(2,0)}_{1234}(s_{ij},\tau) 
 +\widehat {\cal I}^{(4,0)}_{1234}(s_{ij},\tau) \, ,
\label{uniform20}
\end{align}
cf.\ \eqref{Ihat40}, where we have introduced a new Koba--Nielsen integral
\begin{align}
&\widehat{ {\cal I}}^{(4,0)}_{1234}(s_{ij},\tau) =  \int \dd \mu_4 \, \exp \Big( \sum_{1\leq i<j}^4 s_{ij} G_{ij}(\tau) \Big)
\, \Big\{  6s_{13} {\rm G}_4 + s_{12} R_{2|34|1}+s_{23} R_{3|41|2} \label{uniform21} \\
& \ \ +s_{34} R_{4|12|3}+s_{14} R_{1|23|4} -s_{13} (R_{1|24|3}+R_{1|42|3}) -s_{24} (R_{2|13|4}+R_{2|31|4}) \Big\}\, .
\notag
\end{align}
Given that the $R_{i|jk|l}$ boil down to the $f^{(w)}$ with a lattice-sum representation (\ref{1.5}), the coefficients
in the $\ap$-expansion of (\ref{uniform21}) are guaranteed to be modular graph forms by the arguments of section \ref{sec:3.0}.
By the absence of subcycles in the constituents (\ref{uniform16B}), the $\ap$-expansion of $\widehat{ {\cal I}}^{(4,0)}_{1234} $
is expected to exhibit uniform transcendentality with weight $k{+}3$ at the order of $\ap^k$. This is confirmed
by the leading orders in $\ap$,
\begin{align}
\widehat{ {\cal I}}^{(4,0)}_{1234}(s_{ij},\tau) =6s_{13}{\rm G}_{4}+2(s_{13}^{2}+2s_{12}s_{23})\cform{5&0\\1&0}+{\cal O}(\ap^{3})\ ,\label{eq:137B}
\end{align}
as can be seen from the form given in \eqref{Ihat40}.

By inserting the decomposition (\ref{uniform20}) of the integral ${\cal I}_{1234}^{(4,0)}$ into (\ref{3.6}), we arrive at the decomposition of the complete planar $\tau$-integrand as stated in \eqref{ut1}.

\subsubsection{Consistency check of the leading contributions to $\mathcal{\widehat I}^{(4,0)}_{1234}$}
\label{sec:consistency-check-ut}
The leading-order result \eqref{eq:137B} can not only been obtained from (\ref{uniform20}) but can also be checked in an independent calculation based on an $\ap$-expansion of (\ref{uniform21}) as follows. We use the notation 
\begin{align}
{\cal R}_{a|bc|d} \Big[ \prod_{i<j} G_{ij}^{n_{ij}} \Big] = \int \dd \mu_4 \, R_{a|bc|d} \, \prod_{i<j} G_{ij}^{n_{ij}} 
\end{align}
analogous to (\ref{eq:1}). In the absence of closed subcycles in the expression (\ref{eq:135}) 
for $R_{1|23|4}$, the leading order evidently vanishes, 
\beq
{\cal R}_{a|bc|d}[\emptyset]=0 \, .
\eeq
At the subleading order in $\ap$, the same representation of $R_{1|23|4}$ yields
\begin{align}
  {\cal R}_{1|23|4}[G_{12}]&=-3\cform{5&0\\1&0} \, ,&
  {\cal R}_{1|23|4}[G_{13}]&=\frac{35}{6}\cform{5&0\\1&0}\, ,&
  {\cal R}_{1|23|4}[G_{14}]&=-4\cform{5&0\\1&0} \notag \\
  {\cal R}_{1|23|4}[G_{23}]&=-3\cform{5&0\\1&0}\, ,&
  {\cal R}_{1|23|4}[G_{24}]&=\frac{35}{6}\cform{5&0\\1&0}\, ,&
  {\cal R}_{1|23|4}[G_{34}]&=-3\cform{5&0\\1&0}\ , \label{eq:139}
\end{align}
and any other $ {\cal R}_{a|bc|d}[G_{ij}]$ can be obtained by relabeling. Finally, the contribution 
of $6s_{13} {\rm G}_4$ in (\ref{uniform21}) integrates to $6s_{13} {\rm G}_4{\cal I}^{(0,0)}$ 
with ${\cal I}^{(0,0)}=1+{\cal O}(\ap^2)$, completing the verification of \eqref{eq:137B} to the orders shown.


\subsection{Non-planar integrals}
\label{sec:3.6}

The non-planar integrals ${\cal I}^{(2,0)}_{12|34}$ and ${\cal I}^{(4,0)}_{12|34}$ 
in (\ref{3.4}) and (\ref{3.5}) admit integration-by-parts 
manipulations analogous to (\ref{uniform19}) to be rewritten in terms of (conjecturally) uniformly transcendental 
integrals $\widehat{{\cal I}}^{(w,0)}_{12|34}$. In this section we derive these decompositions, given in \eqref{reallyuni} in the main text.

In the non-planar cases, the total derivatives are simpler and boil down to iterations of
\begin{align}
\partial_2 (f^{(1)}_{12} {\cal J}_4) &= {\cal J}_4 \big[ {\rm \hat G}_2 - (1+s_{12}) V_2(1,2) + 2 s_{12}f^{(2)}_{12} - f^{(1)}_{12}(s_{23}f_{23}^{(1)} +s_{24}f_{24}^{(1)} ) \big] \,,
\label{uniform10}
\end{align}
which can be used to solve for
\begin{align}
(1+s_{12}) V_2(1,2) \cong {\rm \hat G}_2+
2s_{12}f^{(2)}_{12} - f_{12}^{(1)} (s_{23} f^{(1)}_{23}+s_{24} f^{(1)}_{24}) 
\, .
\label{uniform10B}
\end{align}
Again, $ \cong$ indicates that total derivatives $\partial_i(\ldots {\cal J}_4)$ have been discarded
in passing to the right-hand side. A similar identity can be derived by taking a $z_1$-derivative 
in (\ref{uniform10}), so the right-hand side of (\ref{uniform10B}) turns out to be symmetric under the
simultaneous exchange of $(z_1,s_{1j}) \leftrightarrow (z_2,s_{2j})$, at least up to
total derivatives.


\subsubsection{Rewriting the integral ${\cal I}^{(2,0)}_{12|34}$}

By applying (\ref{uniform10B}) to both summands $V_2(1,2)$ and $V_2(3,4)$ of (\ref{3.5}),
one arrives at a decomposition
\begin{align}
{\cal I}^{(2,0)}_{12|34}(s_{ij},\tau)  &= \frac{ 2\hat {\rm G}_2 {\cal I}^{(0,0)} (s_{ij},\tau)
+\widehat {\cal I}^{(2,0)}_{12|34} (s_{ij},\tau) }{1+s_{12}}
\label{npdecA}
\end{align}
involving a new integral that should be uniformly transcendental,
\begin{align}
\widehat {\cal I}^{(2,0)}_{12|34}(s_{ij},\tau)  &= \int \dd \mu_4 \, \exp \Big( \sum_{1\leq i<j}^4 s_{ij} G_{ij}(\tau) \Big)
\notag\\
&\hspace{-2.4em}\times \,
 \big[ 2s_{12}f^{(2)}_{12} - f_{12}^{(1)} (s_{23} f^{(1)}_{23}+s_{24} f^{(1)}_{24})
    + 2s_{34}f^{(2)}_{34} - f_{34}^{(1)} (s_{41} f^{(1)}_{41}+s_{42} f^{(1)}_{42})\big]\label{3.5NP} \\
  &=2 \int \dd \mu_4 \, \exp \Big( \sum_{1\leq i<j}^4 s_{ij} G_{ij}(\tau) \Big)
 \big[ 2s_{12}f^{(2)}_{12} - f_{12}^{(1)} (s_{23} f^{(1)}_{23}+s_{24} f^{(1)}_{24})\big]\, . \notag
\end{align}
In passing to the last line, we have used Mandelstam identities and the symmetries of the 
Koba--Nielsen factor to obtain identities such as
\begin{align}
  \int \dd \mu_4 \, \exp \Big( \sum_{1\leq i<j}^4 s_{ij} G_{ij}(\tau) \Big) s_{23}f^{(1)}_{12}f^{(1)}_{23}=\int \dd \mu_4 \, \exp \Big( \sum_{1\leq i<j}^4 s_{ij} G_{ij}(\tau) \Big) s_{14}f^{(1)}_{34}f^{(1)}_{41}\ .\label{eq:24}
\end{align}
The expansion of $\widehat {\cal I}^{(2,0)}_{12|34}$ up to the third order in $\ap$ is given in \eqref{eq:25} 
and verified to be uniformly transcendental.


\subsubsection{Rewriting the integral ${\cal I}^{(4,0)}_{12|34}$}

Repeated application of (\ref{uniform10}) leads to a total-derivative relation
for the integrand of ${\cal I}_{12|34}^{(4,0)}$,
\begin{align}
&(1+s_{12})(1+s_{34}) V_2(1,2) V_2(3,4) \cong {\rm \hat G}_2^2+ s_{13}^2 f^{(1)}_{12} f^{(1)}_{24}f^{(1)}_{43} f^{(1)}_{31}
+ s_{14}^2 f^{(1)}_{12} f^{(1)}_{23}f^{(1)}_{34} f^{(1)}_{41} \notag\\
& \ \ \ \ \  + {\rm \hat G}_2 \, \Big[ 2s_{12}f^{(2)}_{12}+2s_{34}f^{(2)}_{34} + \frac{1}{2} f^{(1)}_{12}(s_{13}f^{(1)}_{13}
+s_{14}f^{(1)}_{14}-s_{23}f^{(1)}_{23}-s_{24}f^{(1)}_{24}) \notag \\
& \ \ \ \ \  \ \ \ \ \  \   +\frac{1}{2} f^{(1)}_{34}(s_{24}f^{(1)}_{24}
+s_{14}f^{(1)}_{14}-s_{23}f^{(1)}_{23}-s_{13}f^{(1)}_{13}) \Big] + 4 s_{12}s_{34} f_{12}^{(2)} f^{(2)}_{34} \label{uniform31} \\
&  \ \ \ \ \  + s_{34} f_{34}^{(2)} f_{12}^{(1)} (s_{13}f^{(1)}_{13}
+s_{14}f^{(1)}_{14}-s_{23}f^{(1)}_{23}-s_{24}f^{(1)}_{24}) \notag \\
&  \ \ \ \ \   + s_{12} f_{12}^{(2)} f_{34}^{(1)} (s_{24}f^{(1)}_{24}
+s_{14}f^{(1)}_{14}-s_{23}f^{(1)}_{23}-s_{13}f^{(1)}_{13})  
 \ ,\notag
\end{align}
and therefore to a similar decomposition for the integral as in (\ref{npdecA}):
\begin{align}
{\cal I}^{(4,0)}_{12|34}(s_{ij},\tau) &= \frac{ \hat {\rm G}_2^2 {\cal I}^{(0,0)}(s_{ij},\tau) 
 + \hat {\rm G}_2 \widehat {\cal I}^{(2,0)}_{12|34}(s_{ij},\tau)
+ \widecheck {\cal I}^{(4,0)}_{12|34}(s_{ij},\tau)   }{(1+s_{12})^2} \, .
\label{newnpdecA}
\end{align}
Here, we have introduced the integral
\begin{align}
\widecheck{\cal I}^{(4,0)}_{12|34}(s_{ij},\tau)  = \int \dd \mu_4 \,
 \exp \Big( \sum_{1\leq i<j}^4 s_{ij} G_{ij}&(\tau) \Big)  \Big[  4 s_{12}s_{34} f_{12}^{(2)} f^{(2)}_{34}+ 4s_{34} f_{34}^{(2)} f_{12}^{(1)} (s_{13}f^{(1)}_{13}{+}s_{14}f^{(1)}_{14}) \notag \\
&+ s_{13}^2 f^{(1)}_{12} f^{(1)}_{24}f^{(1)}_{43} f^{(1)}_{31}+ s_{14}^2 f^{(1)}_{12} f^{(1)}_{23}f^{(1)}_{34} f^{(1)}_{41}\Big] \label{3.4NP}
 \end{align}
over the terms without $\hat {\rm G}_2$ on the right-hand side of (\ref{uniform31}), where relabeling identities 
of the form \eqref{eq:24} were used to simplify the result.
Again, we have independently verified (\ref{npdecA}) and (\ref{newnpdecA}) to the order of $\ap^3$ by expanding
the integrals in (\ref{3.5NP}) and (\ref{3.4NP}) along the lines of section \ref{sec:consistency-check-ut}.
Upon insertion into (\ref{3.6}), we arrive at an alternative representation of the non-planar sector
of the four-point amplitude,
\begin{align}
M_4(\tau)  \, \big|_{  {\rm Tr}(t^{a_1}t^{a_2}) {\rm Tr}(t^{a_3}t^{a_4})}   &=  \frac{ {\rm G}_4^2 \widecheck  {\cal I}^{(4,0)}_{12|34} }{(1+s_{12})^2 } + \Big( \frac{  {\rm G}_4^2 \hat {\rm G}_2 }{(1+s_{12})^2}    - \frac{7}{2} \frac{ {\rm G}_4{\rm G}_6 }{1+s_{12}} \Big)
 \widehat   {\cal I}^{(2,0)}_{12|34}  \label{almostuni} \\
 &+\Big(  \frac{ {\rm G}_4^2 \hat {\rm G}_2^2 }{(1+s_{12})^2 }  - 7 \frac{ {\rm G}_4{\rm G}_6 \hat {\rm G}_2 }{1+s_{12}} + \frac{5}{3} {\rm G}_{4}^3 + \frac{ 49  }{6} {\rm G}_6^2 \Big) {\cal I}^{(0,0)}
 \, .\notag 
\end{align}
%


\subsubsection{Towards a uniform-transcendentality basis}

However, this is not yet the desired uniform-transcendentality decomposition since the
last line in the integrand (\ref{3.4NP}) of $\widecheck {\cal I}^{(4,0)}_{12|34}$ exhibits closed subcycles
$f^{(1)}_{ij}f^{(1)}_{jk}f^{(1)}_{kl}f^{(1)}_{li}$. One can isolate a piece of uniform transcendentality from (\ref{3.4NP})
by subtracting these subcycles via $V_4(i,j,k,l)$, i.e.\ the integral
\begin{align}
\widehat  {\cal I}^{(4,0)}_{12|34}(&s_{ij},\tau)  = \int \dd \mu_4 \,
 \exp \Big( \sum_{1\leq i<j}^4 s_{ij} G_{ij}(\tau) \Big)  \Big[  4 s_{12}s_{34} f_{12}^{(2)} f^{(2)}_{34}+ 4s_{34} f_{34}^{(2)} f_{12}^{(1)} (s_{13}f^{(1)}_{13}{+}s_{14}f^{(1)}_{14}) \notag \\
 & + s_{13}^2 \big( f^{(1)}_{12} f^{(1)}_{24}f^{(1)}_{43} f^{(1)}_{31}- V_4(1,2,4,3) \big)+ s_{14}^2  \big(f^{(1)}_{12} f^{(1)}_{23}f^{(1)}_{34} f^{(1)}_{41}- V_4(1,2,3,4) \big)\Big]\label{3.4NPNP}
 \end{align}
is claimed to be uniformly transcendental. Then, by the decomposition (\ref{uniform20})
of integrals over $V_4(i,j,k,l)$, we can relate this to (\ref{3.4NP}) via
\begin{align}
  \widecheck {\cal I}^{(4,0)}_{12|34}  &= \widehat  {\cal I}^{(4,0)}_{12|34}
+(s_{13}^2+s_{23}^2) {\rm G}_4 {\cal I}^{(0,0)} 
+ s_{13}^2 ( \widehat {\cal I}^{(4,0)}_{1243} + \hat {\rm G}_2 {\cal I}^{(2,0)}_{1243} )
+ s_{23}^2 (\widehat {\cal I}^{(4,0)}_{1234} + \hat {\rm G}_2 {\cal I}^{(2,0)}_{1234} )\, .
\end{align}
Upon insertion into (\ref{almostuni}), we obtain admixtures of the planar integrals
$\widehat {\cal I}^{(4,0)}_{1234}$ and ${\cal I}^{(2,0)}_{1234}$ defined by (\ref{uniform21})
and (\ref{3.3}), respectively, and arrive at \eqref{reallyuni}. Similar to section \ref{sec:consistency-check-ut},
the results for ${\cal I}^{(w,0)}_{12|34}$ in (\ref{sureB}) and (\ref{sure2B}) have been confirmed
by performing an independent $\ap$-expansion of (\ref{3.5NP}) and (\ref{3.4NPNP}) and inserting
into the above integration-by-parts relations.


\subsection{Efficiency of the new representations for higher-order expansions}
\label{sec:effic-new-repr}

Given the Mandelstam invariants in the integrands (\ref{uniform21}), (\ref{3.5NP}) and (\ref{3.4NPNP}) 
of $\widehat {\cal I}^{(4,0)}_{1234}$, $\widehat {\cal I}^{(2,0)}_{12|34}$ and $\widehat {\cal I}^{(4,0)}_{12|34}$, 
the $k^{\rm th}$ order in their $\ap$-expansion can be computed from less than $k$ factors of $G_{ij}$ from the Koba--Nielsen factor.
However, the variety of $f^{(w)}_{ij}$ along with the different $s_{kl}$ in the integrands increases the number of independent
calculations w.r.t.\ relabeling the punctures and momenta at a fixed order of the Koba--Nielsen expansion. 
Hence, the combinatorial efficiency of the
new representations (\ref{ut1}) and (\ref{reallyuni}) of $M_4(\tau)$ for higher-order $\ap$-expansion should be
comparable to the old one in (\ref{3.6}).

Instead, the main advantages of the integrals $\widehat {\cal I}^{(4,0)}_{1234}$, $\widehat {\cal I}^{(2,0)}_{12|34}$ and 
$\widehat {\cal I}^{(4,0)}_{12|34}$ are the following:
\begin{itemize}
\item The integrand of $\widehat {\cal I}^{(4,0)}_{1234}$ in (\ref{uniform21}) does not share the 
term $f^{(1)}_{12}f^{(1)}_{23}f^{(1)}_{34}f^{(1)}_{41}$ of the $V_4(1,2,3,4)$ function.
Like this, expansion of $\widehat {\cal I}^{(4,0)}_{1234}$ bypasses numerous holomorphic subgraph 
reductions that introduced a spurious complexity into the calculations of section \ref{sec:3.1}.
\item The $\ap$-expansions of ${\cal I}^{(w,0)}_{12|34}$ in section \ref{sec:3.3} were plagued by
conditionally convergent or divergent lattice sums. Lack of absolute convergence is caused by the terms $(f^{(1)}_{ij})^2$ or
$\sum_{p\neq 0}\frac{ \bar p}{p}  e^{2\pi i \langle p,z_{ij} \rangle}$ in the representations (\ref{npreg1}) or (\ref{2.999}) 
of $V_2(i,j)$. Both of them are manifestly absent in the integrands (\ref{3.5NP}) and (\ref{3.4NPNP})
of $\widehat {\cal I}^{(2,0)}_{12|34}$ and $\widehat {\cal I}^{(4,0)}_{12|34}$.
\end{itemize}


\section{The elliptic single-valued map at the third order in $\ap$}
\label{appmod}

In this appendix, we give some further explicit expressions necessary to relate the $(\ap)^3$-order
of open- and closed-string integrals through the tentative elliptic single-valued map in 
section~\ref{sec:4}.


\subsection{Modular transformation of iterated Eisenstein integrals}
\label{appmod1}
Modular transformations of the iterated Eisenstein integrals at the orders $\ap$ and $\ap^{2}$ 
of \eqref{open0j} were given in \eqref{open05}. The order $\ap^{3}$ also necessitates the 
modular transformations \cite{Brown:mmv, Broedel:2018izr}
\begin{align}
&{\cal E}_0(8,0,0,0,0;-\tfrac{1}{\tau}) =
 - \frac{ i \pi^2 T^3}{ 95256000}
 + \frac{ i \pi^6}{13608000 T}
 + \frac{i \pi^8}{  2857680 T^3}  
+ \frac{\pi^2 \zeta_7 }{  672 T^4}
- \frac{i \pi^{10}}{4536000 T^5}
\notag \\
 & \ \ \ \ \ + \frac{\pi^2}{T^2}{\cal E}_0(8, 0, 0, 0, 0)
  + \frac{5 i \pi^2 }{T^3}  {\cal E}_0(8, 0, 0, 0, 0, 0)
  - \frac{15 \pi^2 }{2 T^4} {\cal E}_0(8, 0, 0, 0, 0, 0, 0)
 \\
&{\cal E}_0(4,0,4,0,0;-\tfrac{1}{\tau}) +3{\cal E}_0(4,4,0,0,0;-\tfrac{1}{\tau}) +  \tfrac{1}{360} {\cal E}_0(4,0,0,0,0;-\tfrac{1}{\tau})  =  
\frac{i \pi^2 T^3}{2916000}
+ \frac{i \pi^4 T}{ 116640 } \notag \\
& \ \ \ \ \
 + \frac{5 \zeta_5}{432}- \frac{\pi^2 \zeta_3}{  6480} - \frac{91 i \pi^6}{1458000 T}
 - \frac{  5 \pi^2 \zeta_5}{432 T^2}  - \frac{ \pi^4 \zeta_3}{ 1296 T^2}  + \frac{ i \pi^8}{38880 T^3}
  + \frac{ i \pi^2 \zeta_3^2}{72 T^3} 
  \\
 &  \ \ \ \ \ + \frac{\pi^6 \zeta_3}{2160 T^4} - \frac{ i \pi^{10}}{324000 T^5} +\Big( \frac{ \pi^2 }{ 1080 }+ \frac{ \pi^4  }{ 216 T^2}  - \frac{ i \pi^2   \zeta_3}{6 T^3}
 - \frac{ \pi^6  }{360 T^4} \Big)     {\cal E}_0(4, 0, 0)
   \notag \\
& \ \ \ \ \ + \frac{ i \pi^2 }{ 2 T^3}{\cal E}_0(4, 0, 0)^2 + \frac{ \pi^2}{T^2} \Big({\cal E}_0(4, 0, 4, 0, 0)+ 3  {\cal E}_0(4, 4, 0, 0, 0)+ \frac{ {\cal E}_0(4, 0, 0, 0, 0)}{360  }  \Big)
    \, .  \notag
\end{align}
%


\subsection{The third $\ap$-order of ${\cal I}_{1234}^{(2,0)}$ and iterated Eisenstein integrals}
\label{appmod3}

The first and second order of $\mathcal{I}^{(2,0)}_{1234}$ in $\ap$ were written in terms of iterated Eisenstein integrals in \eqref{open0r}. Similarly, the third order of the expansion in \eqref{goodrepB} can be written as
\begin{align}
&{\cal I}_{1234}^{(2,0)}\, \big|_{\ap^3} = \pi^2 s_{13}(s_{12}^2+s_{12}s_{23}+s_{23}^2) \Big(
\frac{ 4 y^3}{945} + \frac{2 \zeta_3}{5}  - \frac{ 5 \zeta_5}{y^2} - \frac{3 \zeta_7}{2 y^4} + \Big(  \frac{72   \zeta_3}{y^2} - \frac{16 y}{5}    \Big)      \Re[ {\cal E}_0(4, 0)] \notag \\
&  -  \frac{12}{5} \Re[ {\cal E}_0(4, 0, 0)]     +\frac{432}{y^2}  \Re[  {\cal E}_0(4, 0, 4, 0, 0)+3{\cal E}_0(4, 4, 0, 0, 0)+ \tfrac{1}{360} {\cal E}_0(4, 0, 0, 0, 0)]
 \notag \\
 & + \frac{4032}{y^2}   \Re[ {\cal E}_0(8, 0, 0, 0, 0)]
  + \frac{10080}{y^3}   \Re[ {\cal E}_0(8, 0, 0, 0, 0, 0)]
   + \frac{ 7560 }{y^4}  \Re[ {\cal E}_0(8, 0, 0, 0, 0, 0, 0)]
  \notag \\
  &   - \frac{432 }{y^2}  \Re[ {\cal E}_0(4, 0)] \Re[ {\cal E}_0(4, 0, 0)]  
  + \Big(  \frac{16y^2}{5}   + \frac{ 144   \zeta_3}{y}    \Big)  {\cal E}_0(4)
+ \frac{24}{5} y {\cal E}_0(4, 0) + \frac{24}{5} {\cal E}_0(4, 0, 0) \notag \\
&+ \frac{ 432 }{y} {\cal E}_0(4, 0)^2 -  1728 {\cal E}_0(4) \Re[ {\cal E}_0(4, 0)] 
- \frac{864 }{y}   {\cal E}_0(4, 0) \Re[ {\cal E}_0(4, 0)]
 - \frac{864 }{y} {\cal E}_0(4) \Re[ {\cal E}_0(4, 0, 0)] \notag \\
 &+  1728 {\cal E}_0(4, 4, 0)   + \frac{864}{y}  \big[  {\cal E}_0(4, 4, 0, 0) + \frac{1}{360}  {\cal E}_0(4, 0, 0, 0) \big]  + 2688 {\cal E}_0(8, 0, 0)
 \notag \\
 &+ \frac{8064 }{y}  {\cal E}_0(8, 0, 0, 0)
 + \frac{10080 }{y^2}  {\cal E}_0(8, 0, 0, 0, 0)
 + \frac{5040 }{y^3}  {\cal E}_0(8, 0, 0, 0, 0, 0)
\Big)\ .\label{eq:10}
\end{align}
As in \eqref{open0r}, not all of these terms are reproduced by $(2\pi i)^2 \, {\rm esv}  \, Z^{(2)}_{1234}(s_{ij}, -\tfrac{1}{\tau})|_{\ap^{3}}$, but only the ones which are left invariant under the projection $P_{\rm Re}$ defined
in (\ref{defPRE}). Specifically, $P_{\rm Re}$ maps the terms $\left(\frac{16y^2}{5}+\frac{ 144   \zeta_3}{y} \right) {\cal E}_0(4)
+ \frac{24}{5} y {\cal E}_0(4, 0) + \frac{24}{5} {\cal E}_0(4, 0, 0)$ in the fourth line and all terms in the following lines to zero. 



\providecommand{\href}[2]{#2}\begingroup\raggedright\endgroup

\end{document}